\documentclass[12pt,a4paper]{article}
\usepackage[OT1]{fontenc}
\usepackage[utf8]{inputenc}
\usepackage[a4paper,tmargin=3truecm,bmargin=3truecm,rmargin=2.2truecm,lmargin=2.2truecm]{geometry}
\usepackage{amssymb, amsmath}
\usepackage[amsmath, hyperref, thmmarks]{ntheorem}
\usepackage{calc} 
\usepackage{ifthen}
\usepackage[all]{xy}
\usepackage[english]{babel}
\usepackage{hyperref}

\theoremnumbering{arabic}
\theoremstyle{break}
\theoremsymbol{}
\theorembodyfont{\slshape}
\theoremheaderfont{\normalfont\bfseries}
\theoremseparator{}
\newtheorem{Theorem}{Theorem}[section]

\newtheorem{Proposition}[Theorem]{Proposition}

\theorembodyfont{\upshape}
\theoremsymbol{\ensuremath{\blacklozenge}}
\newtheorem{Example}[Theorem]{Example}

\newtheorem{Remark}[Theorem]{Remark}

\newtheorem{Definition}[Theorem]{Definition}

\newcommand\algnotation[1]{{\mathbf{#1}}}
\newcommand\algA{{\algnotation{A}}}
\newcommand\algB{{\algnotation{B}}}

\newcommand\algI{{\algnotation{I}}}

\newcommand\algzero{{\grnotation{0}}}

\newcommand\grnotation[1]{{\mathsf{#1}}}
\newcommand\grA{{\grnotation{A}}}

\newcommand\grG{{\grnotation{G}}}

\newcommand\grV{{\grnotation{V}}}
\newcommand\Gr{{\grnotation{Gr}}} 

\newcommand\evnotation[1]{{\mathnormal{#1}}}

\newcommand\evV{{\evnotation{V}}}

\newcommand\ehnotation[1]{{\mathcal{#1}}}

\newcommand\ehH{\ehnotation{H}}

\newcommand\modnotation[1]{{\boldsymbol{#1}}}

\newcommand\modM{{\modnotation{M}}}
\newcommand\modN{{\modnotation{N}}}

\newcommand\tncA{\mathcal{A}}

\newcommand\gR{{\mathbb R}}
\newcommand\gQ{{\mathbb Q}}
\newcommand\gK{{\mathbb K}}

\newcommand\gF{{\mathbb F}}
\newcommand\gC{{\mathbb C}}
\newcommand\gZ{{\mathbb Z}}
\newcommand\gN{{\mathbb N}}
\newcommand\gS{{\mathbb S}}
\newcommand\gT{{\mathbb T}}

\newcommand\gG{{\mathbb G}}
\newcommand\gRP{{\mathbb{RP}}}
\newcommand\gCP{{\mathbb{CP}}}

\newcommand\caB{{\mathcal B}}

\newcommand\caK{{\mathcal K}}
\newcommand\caL{{\mathcal L}}

\newcommand\caO{{\mathcal O}}
\newcommand\caP{{\mathcal P}}
\newcommand\caQ{{\mathcal Q}}

\newcommand\caS{{\mathcal S}}
\newcommand\caT{{\mathcal T}}
\newcommand\caU{{\mathcal U}}

\newcommand\caZ{{\mathcal Z}}

\newcommand\kg{{\mathfrak g}}

\newcommand\kC{{\mathfrak C}}

\newcommand\kS{{\mathfrak S}}

\newcommand\ksl{{\mathfrak{sl}}}
\newcommand\ko{{\mathfrak{o}}}
\newcommand\ku{{\mathfrak{u}}}

\newcommand\kso{{\mathfrak{so}}}
\newcommand\Ug{{\caU(\kg)}}

\newcommand\bbbone{{ \mathchoice {1\mskip-4mu\mathrm{l} } {1\mskip-4mu\mathrm{l} }{1\mskip-4.5mu\mathrm{l} } {1\mskip-5mu\mathrm{l}} }}

\newcommand{\omi}[1]{\buildrel { \buildrel{#1}\over{\vee} } \over .}
\newcommand{\spmatrix}[1]{\left( \begin{smallmatrix}#1\end{smallmatrix}\right)} 

\newcommand\ensvide{{\varnothing}} 
\newcommand\exter{{\textstyle\bigwedge}} 
\newcommand\otimeshat{\mathrel{\widehat\otimes}} 
\newcommand\symes{{\mathchoice{\textstyle\mathsf{S}}{\textstyle\mathsf{S}}%
{\scriptstyle\mathsf{S}}{\scriptscriptstyle\mathsf{S}}}} 

\newcommand\der{{\text{Der}}}

\newcommand\gr{{\text{gr}}}
\newcommand\Id{{\text{Id}}}

\newcommand\lie{{\text{Lie}}}

\newcommand\sign{{\text{sign}}}

\newcommand{\grast}{\bullet}
\newcommand{\indast}{\sharp}

\newcommand{\espast}{\boldsymbol{\ast}}

\DeclareMathOperator{\ch}{ch} 
\DeclareMathOperator{\cs}{cs} 
\DeclareMathOperator{\End}{End} 
\DeclareMathOperator{\Hom}{\mathsf{Hom}} 
\DeclareMathOperator{\ind}{Ind} 
\DeclareMathOperator{\Ker}{Ker} 
\renewcommand{\ker}{\Ker} 
\DeclareMathOperator{\Pf}{Pf} 
\DeclareMathOperator{\Ran}{Ran} 

\DeclareMathOperator{\rank}{rank}

\DeclareMathOperator{\tr}{Tr} 

\begin{document}

\title{An informal introduction to the ideas and concepts of noncommutative geometry}
\author{T. Masson$^a$}
\date{}
\maketitle
\begin{center}
$^a$ Laboratoire de Physique Th\'eorique (UMR 8627)\\
B\^at 210, Universit\'e Paris-Sud Orsay\\
F-91405 Orsay Cedex
\end{center}

\bigskip
\begin{abstract}
This informal introduction is an extended version of a three hours lecture given at the 6th Peyresq meeting ``Integrable systems and quantum field theory''. In this lecture, we make an overview of some of the mathematical results which motivated the development of what is called noncommutative geometry. The first of these results is the theorem by Gelfand and Neumark about commutative $C^\ast$-algebras; then come some aspects of the $K$-theories, first for topological spaces, then for $C^\ast$-algebras and finally the purely algebraic version. Cyclic homology is introduced, keeping in mind its relation to differential structures. The last result is the construction of the Chern character, which shows how these developments are related to each other.
\end{abstract}

\vfill
\begin{flushleft}
LPT-Orsay/06-87
\end{flushleft}
\newpage

\tableofcontents

\section{Introduction}

Once upon a time, in a perfect land, the idea of point was conceived.  

This was a beautiful concept, full of potentiality, especially in Natural Science: how easy is it to say where objects are when one has introduced such a precise definition of localization! How easy is it to describe the kinematics of bodies when one assigns to them a point at each time\dots Well, at least if time is there too! And then laws were found for the interactions of moving bodies, and then predictions were formulated: Pluto, the former planet, was \emph{where} it was calculated to be! Better: generalized geometries where conceived, in which parallels can meet. And you know it: physicists (one of them at least!) where fool enough to show us how useful these geometries can be to describe gravitation.

But nature seems often more subtle than human mind. And the dream ceased when quantum mechanics entered the game. We are no more allowed to say where an electron is exactly located on its ``orbit'' around the proton in the hydrogen atom. What is the photon trajectory in the Young's double slit experiment? It is forbidden to know! Knowing destroys the diffraction pattern on the target screen. 

The main feature of quantum mechanics which exposes us to this annoying situation is the non-commutativity of observables.

How can we accommodate this? Well, one of the reasonable answers can be found in mathematics. Surprisingly enough, mathematicians discovered, not so long ago, that we can speak about spaces without even mentioning them. The trick is to use algebraic objects, and the surprise is that spaces (some of them at least, miracles are not the prerogative of mathematicians!) can be \emph{reconstructed} from them. In the language of mathematics, one has an equivalence of categories\dots

The algebraic objects we need to deal with are associative algebras, not only with their friendly product but also with other structures, like involutions and norms. There, quantum mechanics is at home: observables are special operators on a Hilbert space, so that they live in such an algebra! Where are the ``points'' which were mentioned? Take a normal operator (it commutes with its adjoint) in such an operator algebra, consider the smallest subalgebra it generates. This subalgebra is a \emph{commutative} algebra, which can be shown to be the algebra of continuous functions on the spectrum of this normal operator. Associate to this element as many other normal operators as you can find, on the condition that they commute among themselves, and you get another algebra of continuous functions on a topological space. Yes, we get it: a topological space from pure algebraic objects! 

This is one of the main results behind \emph{noncommutative geometry}. The idea is the following: if commutative algebras are ordinary topological spaces (in the category of $C^\ast$-algebra to be precise), what are the noncommutative ones? How can we study them using the machinery that we are used to manipulate the topological spaces? Is there somewhere another category of algebras in which commutative algebras are differential functions on a differential manifold? If not (for the moment, it is no!), can we manipulate them with some kind of differential structures?

I hope to show you in the following that these questions make sense, and that some answers can be formulated. In section~\ref{castalgebrasfortopologists} we introduce $C^\ast$-algebras, and we make the precise statement about commutative $C^\ast$-algebras. In section~\ref{Ktheoryforbeginniers}, we show that one of the machineries developed on topological spaces can be used on their noncommutative counterparts, the $C^\ast$-algebras. In section~\ref{cyclichomologyforgeometers}, cyclic homology is shown to be a good candidate to fool us enough into thinking that we manipulate differential structures on algebras. Section~\ref{thenotmissinglinkthecherncharacter} is devoted to the Chern character, an object which can convince the more commutative geometer that noncommutative geometry does not only make sense, but also is one of the most beautiful developments in modern mathematics.

\section{$C^\ast$-algebras for topologists}
\label{castalgebrasfortopologists}

In this section, we will explore some aspects of the theory of $C^\ast$-algebras. The main result, we would like to explain, is the theorem by Gelfand and Neumark about commutative $C^\ast$-algebras.

\subsection{General definitions and results}

In order to be concise, only algebras over the field $\gC$ will be considered.

\begin{Definition}[Involutive, Banach and $C^\ast$ algebras]
An involutive algebra $\algA$ is an associative algebra equipped with map $a \mapsto a^\ast$ such that 
\begin{align*}
a^{\ast \ast} &= a & (a+b)^\ast &= a^\ast + b^\ast & (\lambda a)^\ast &= \overline{\lambda} a^\ast & (a b )^\ast &= b^\ast a^\ast
\end{align*}
for any $a,b \in \algA$ and $\lambda \in \gC$, and where $\overline{\lambda}$ denotes the ordinary conjugation on the complex numbers.

A Banach algebra $\algA$ is an associative algebra equipped with a norm $\Vert \cdot \Vert : \algA \rightarrow \gR^+$ such that the topological space $\algA$ is complete for this norm and such that
\begin{equation*}
\Vert a b \Vert \leq \Vert a \Vert \; \Vert b \Vert
\end{equation*}
If the algebra is unital, with unit denoted by $\bbbone$, then it is also required that $\| \bbbone \| = 1$. 

A $C^\ast$-algebra is an involutive and a Banach algebra $\algA$ such that the norm satisfies the $C^\ast$-condition
\begin{equation}
\label{eq-Castcondition}
\Vert a^\ast a \Vert = \Vert a \Vert^2
\end{equation}
\end{Definition}

A $C^\ast$-algebra is then a normed complete algebra equipped with an involution and a compatibility condition between the norm and the involution. One can show that this $C^\ast$-condition \eqref{eq-Castcondition} implies that $\Vert a \Vert \leq \Vert a^\ast \Vert \leq \Vert a^{\ast \ast} \Vert = \Vert a \Vert$. The adjoint is then an isometry in any $C^\ast$-algebra.

\begin{Definition}[self-adjoint, normal, unitary and positive elements]
An element $a$ in a $C^\ast$-algebra $\algA$ is self-adjoint if $a^\ast = a$, normal if $a^\ast a = a a^\ast$, unitary if $a^\ast a = a a^\ast = \bbbone$ when $\algA$ is unital, and positive if it is of the form $a = b^\ast b$ for some $b \in \algA$.
\end{Definition}
Self-adjoint and unitary elements are obviously normal, and positive elements are obviously self-adjoint.

\begin{Example}[The algebra of matrices]
\label{ex-algebraofmatrices}
Let $M_n(\gC)$ be the algebra of $n\times n$ matrices over $\gC$. This is an involutive algebra for the adjoint. This algebra is complete for the three equivalent norms
\begin{align*}
\| a \|_\text{max} &= \max \{ | a_{ij} | \ / \ i,j = 1, \dots, n \} & &\text{max norm}\\
\| a \| &= \sup \{ \| av \| \ / \ v \in \gC^n,\ \| v \| \leq 1 \} & &\text{operator norm}\\
\| a \|_{\sum} &= \sum_{i,j} | a_{ij} |  & &\text{sum norm}
\end{align*}
which are related by the inequalities
\begin{equation*}
\| a \|_\text{max} \leq \| a \| \leq \| a \|_{\sum} \leq n^2 \| a \|_\text{max}
\end{equation*}
The algebra $M_n(\gC)$ is then a Banach algebra for any of these norms. Only the operator norm defines on $M_n(\gC)$ a $C^\ast$-algebraic structure.
\end{Example}

\begin{Example}[The algebra of bounded linear operators]
Let $\ehH$ be a separable Hilbert space, and $\caB(\ehH) = \caB$ the algebra of bounded linear operators on $\ehH$. Equipped with the adjointness operation and the operator norm
\begin{equation*}
\| a \| = \sup \{ \| a u \| \ / \ u \in \ehH,\ \| u \| \leq 1 \}
\end{equation*}
this algebra is a $C^\ast$-algebra. In the finite dimensional case, one recovers $M_n(\gC)$.
\end{Example}

\begin{Example}[The algebra of compact operators]
Let $\ehH$ be an Hilbert space. A finite rank operator $a \in \caB$ is an operator such that $\dim \Ran a < \infty$. Let $\caB_F$ denote the subalgebra of finite rank operators in $\caB$. The algebra $\caK(\ehH) = \caK$ of compact operators is the closure of $\caB_F$ for the topology of the operator norm. The algebra $\caK$ is a $C^\ast$-algebra, which is not unital when $\ehH$ is infinite dimensional. In case $\ehH$ is finite dimensionnal, $\caK = M_n(\gC)$ for $n = \dim \ehH$.

For any integer $n \geq 1$, $M_n(\gC)$ is identified as a subalgebra of $\caB$, as the operators which act only on the first $n$ vectors of a fixed orthonormal basis of $\ehH$. Then one gets a direct system of $C^\ast$-algebras inside $\caB$, $i_n : M_n(\gC) \hookrightarrow M_{n+1}(\gC)$ with $i_n(a) = \spmatrix{a & 0 \\ 0 & 0}$. One has
\begin{equation*}
\caK = \varinjlim M_n(\gC)
\end{equation*}

Using this identification, it is easy to see that $\caK$ is an ideal in $\caB$. The quotient $C^\ast$-algebra $\caQ = \caB/\caK$ is the Calkin algebra.
\end{Example}

\begin{Example}[The algebra of continuous functions]
\label{ex-thealgebraofcontinousfunctions}
Let $X$ be a compact Hausdorff space. Denote by $C(X)$ the (commutative) algebra of continuous functions on $X$, for pointwise addition and multiplication of functions. Define the involution $f \mapsto \overline{f}$ and the sup norm
\begin{equation}
\label{eq-normesup}
\| f \|_\infty = \sup_{x \in X} |f(x)|
\end{equation}
With these definitions, $C(X)$ is a $C^\ast$-algebra.

If the topological space $X$ is only a locally compact Hausdorff space, one defines $C_0(X)$ to be the algebra of continous functions on $X$ vanishing at infinity : for any $\epsilon >0$, there exists a compact $K \subset X$ such that $f(x) <\epsilon$ for any $x \in X \backslash K$. Equipped with the same norm and involution as $C(X)$, this is a $C^\ast$-algebra.
\end{Example}

\begin{Example}[The tensor product]
One can perform a lot of operations on $C^\ast$-algebras, some of them being described in the following examples. 

Let us mention that there exist some well defined tensor products on $C^\ast$-algebras (see Chapter~11 in \cite{KadiRing:97a} or Appendix~T in \cite{Wegg:93}). In the following, we denote by $\otimeshat$ the spatial tensor product. One of its properties is that $C(X) \otimeshat C(Y) = C(X \times Y)$ for any compact spaces $X,Y$.
\end{Example}

\begin{Example}[The algebra $M_n(\algA)$] 
Let $\algA$ be a $C^\ast$-algebra. We denote by $M_n(\algA)$ the set of $n\times n$ matrices with entries in $\algA$. This is naturally an algebra. One can define the max norm and the sum norm on this algebra using $\| a_{ij} \|$ instead of $| a_{ij} |$ as in Example~\ref{ex-algebraofmatrices}. For the operator norm, the situation is more subtle. One has to take an injective representation $\rho : \algA \rightarrow \caB(\ehH)$, which induces an injective representation $\rho_n : M_n(\algA) \rightarrow \caB(\ehH^n)$. The operator norm of $a \in M_n(\algA)$ is defined as $\| a \| = \| \rho_n(a) \|$ where the last norm is on $\caB(\ehH^n)$. One can then show that this norm is independent of the choice of the injective representation $\rho$ and gives $M_n(\algA)$ a structure of $C^\ast$-algebra.

This construction corresponds also to define $M_n(\algA)$ as $M_n(\gC) \otimeshat \algA$.

The natural inclusion $M_n(\algA) \hookrightarrow M_{n+1}(\algA)$ defines a direct system of $C^\ast$-algebras. One can show that $\algA \otimeshat \caK = \varinjlim M_n(\algA)$. 
\end{Example}

\begin{Example}[The algebra $C_0(X,\algA)$]
\label{ex-algebraC0XA}
Let $X$ be a locally compact topological space and $\algA$ a $C^\ast$-algebra. The space $C_0(X,\algA)$ of continuous functions $a : X \rightarrow \algA$ vanishing at infinity, equipped with the involution induced by the involution on $\algA$ and the sup norm $\| a \|_\infty = \sup_{x \in X} \| a(x)\| $, is a $C^\ast$-algebra. Using the spatial tensor product, one has $C_0(X,\algA) = C_0(X) \otimeshat \algA$.

If $X$ is compact, we denote it by $C(X,\algA)$. If $\algA$ is unital and $X$ is compact, this algebra is unital.
\end{Example}

\begin{Example}[The convolution algebra]
The algebra $L^1(\gR)$ for the convolution product
\begin{equation*}
(f \ast g)(x) = \int_\gR f(x-y) g(y) dy
\end{equation*}
 with the norm
\begin{equation*}
\| f \|_1 = \int_\gR |f(x)| dx 
\end{equation*}
and equipped with the involution
\begin{equation*}
f^\ast(x) = \overline{f(-x)}
\end{equation*}
is a Banach algebra with involution but is not a $C^\ast$-algebra. 
\end{Example}

\begin{Definition}[Fréchet algebra]
A semi-norm of algebras $p$ on $\algA$ is a semi-norm on the vector space $\algA$ for which $p(a b) \leq p(a) p(b)$ for any $a,b \in \algA$. 

A Fréchet algebra is a topological algebra for the topology of a numerable set of algebra semi-norms, which is complete.
\end{Definition}

\begin{Example}[The Fréchet algebra $C^\infty(M)$]
Let $M$ be a $C^\infty$ finite dimension locally compact manifold. Then the algebra $C^\infty(M)$ of differentiable functions on $M$ is an involutive algebra but is not a Banach algebra for the sup norm because it is not complete. Nevertheless, this algebra can be equipped with a family of semi-norms $p_{K_r,N}$ to make it a Fréchet algebra. These semi-norms are defined as follows. For any $\alpha = (\alpha_1, \dots, \alpha_n)$ where $\alpha_r \in \gN$, let us use the compact notation $D^\alpha = \left( \frac{\partial\hfill}{\partial x^1} \right)^{\alpha_1} \cdots \left( \frac{\partial\hfill}{\partial x^n} \right)^{\alpha_n}$ with $| \alpha| = \alpha_1 + \cdots + \alpha_n$. Then, for any compact subspace $K \subset M$ and any integer $N \geq 0$, define $p_{K,N}(f) = \max \{ | (D^\alpha f)(x) | \ / \ x \in K, \ |\alpha| \leq N \}$. With a increasing numerable family of compact spaces $K_r \subset M$ such that $\bigcup_{r \geq 0} K_r = M$, one gets a numerable family of semi-norms $p_{K_r,N}$ for the topology of which $C^\infty(M)$ is complete.
\end{Example}

\begin{Example}[The irrational rotation algebra]
Let $\theta$ be an irrational number. On the Hilbert space $L^2(\gS^1)$, consider the two unitary operators
\begin{align*}
(Uf)(t) &= e^{2 \pi i t} f(t) 
&
(Vf)(t) &= f(t-\theta)
\end{align*}
where $f : \gS^1 \rightarrow \gC$ is considered as a periodic function in the variable $t \in \gR$. Then one has $U V = e^{2 \pi i \theta} V U \in \caB(L^2(\gS^1))$. The $C^\ast$-algebra $\tncA_\theta$ generated by $U$ and $V$ is called the irrational rotation algebra or the noncommutative torus. 

Let us consider the Schwartz space $\caS(\gZ^2)$ of sequences $(a_{m,n})_{m,n \in \gZ}$ of rapid decay: $(|m| + |n|)^q | a_{m,n} |$ is bounded for any $q \in \gN$. We define the algebra $\tncA^\infty_\theta$ as the set of elements in $\tncA_\theta$ which can be written as $\sum_{m,n \in \gZ} a_{m,n} U^m V^n$ for a sequence $(a_{m,n})_{m,n \in \gZ} \in \caS(\gZ^2)$. The family of semi-norms $p_q(a) = \sup_{m,n \in \gZ}\{ (1 + |m| + |n|)^q | a_{m,n} | \}$ gives to $\tncA^\infty_\theta$ a structure of Fréchet algebra.

This algebra admits two continuous non inner derivations $\delta_i$, $i=1,2$, defined by $\delta_1(U^m) = 2 \pi i m U^m$, $\delta_1(V^n) = V^n$, $\delta_2(U^m) = U^m$ and $\delta_2(V^n) = 2 \pi i n V^n$.

Using Fourier analysis, $\caS(\gZ^2)$ is isomorphic to the space $C^\infty(\gT^2)$ where $\gT^2$ is the two-torus. The algebra $\tncA^\infty_\theta$ is then the equivalent of smooth functions on the noncommutative torus $\tncA_\theta$. 
\end{Example}

Let us mention some general results about $C^\ast$-algebras:
\begin{Proposition}[Unitarisation]
Any $C^\ast$-algebra $\algA$ is contained in a unital $C^\ast$-algebra $\algA_+$ as a maximal ideal of codimension one.
\end{Proposition}

The construction of the unitarization $\algA_+$ is as follows: as a vector space, $\algA_+ = \algA + \gC$; as an algebra, $(a + \lambda)(b + \mu) = ab + \lambda b + \mu a + \lambda \mu$; as an involutive algebra, $(a + \lambda)^\ast = a^\ast + \overline{\lambda}$; as a normed algebra, $\| (a + \lambda) \| = \sup \{ \| a b + \lambda b \| \ / \ b \in \algA,\ \| b \| \leq 1 \}$.

\begin{Theorem}
For any $C^\ast$-algebra $\algA$, there exist a Hilbert space $\ehH$ and an injective representation $\algA \rightarrow \caB(\ehH)$. Then every $C^\ast$-algebra is a subalgebra of the bounded operators on a certain Hilbert space.
\end{Theorem}

The construction of this Hilbert space, which is not necessarily separable, is performed through the GNS construction. This theorem implies that any $C^\ast$-algebra can be concretely realized as an algebra of operators on a Hilbert space. Obviously, the converse is not true: there are many algebras of operators on Hilbert spaces which are not $C^\ast$-algebras.

\begin{Proposition}
Any morphism between two $C^\ast$-algebras is norm decreasing.

The norm on a $C^\ast$-algebra is unique. An isomorphism of $C^\ast$-algebras is an isometry.
\end{Proposition}

\subsection{The Gelfand transform}

Let $\algA$ be a unital Banach algebra. Let us use the notation $z = z \bbbone \in \algA$ for any $z \in \gC$.

\begin{Definition}[Resolvant, spectrum and spectral radius]
Let $a$ be an element in $\algA$.

The resolvant of $a$, denoted by $\rho(a)$, is the subspace of $\gC$:
\begin{equation*}
\rho(a) =\{ z \in \gC \ / \ (a-z)^{-1} \in \algA \}
\end{equation*}

The spectrum of $a$, denoted by $\sigma(a)$, is the complement of $\rho(a)$ in $\gC$: $\sigma(a) = \gC \backslash \rho(a)$. One can show that $\sigma(a)$ is a compact subspace of $\gC$ contained in the disk $\{ z \in \gC \ / \ |z| \leq \| a \| \}$.

The spectral radius of $a$ is defined as
\begin{equation*}
r(a) = \sup \{ |z| \ / \ z \in \sigma(a) \}
\end{equation*}

\end{Definition}

For any $a \in M_n(\gC)$, the spectrum of $a$ contains the set of eigenvalues of $a$, but can contain other values not associated to eigenvectors.

The spectrum of $a \in \algA$ depends on the algebra $\algA$. Nevertheless, we will see exceptions to that.

The spectral radius can be computed using the relation
\begin{equation*}
r(a) = \lim_{n \rightarrow \infty} \| a^n \|^{1/n}
\end{equation*}
which can look surprising at first: on the left, the radius is defined using only the algebraic structure of $\algA$ (is an element $(a-z)$ invertible?), on the right it is related to the norm\dots

Here are some interesting results about the spectrum of particular elements.
\begin{Proposition}
If $a$ is self-adjoint, then $\sigma(a) \subset \gR$. If $a$ is unitary, then $\sigma(a) \subset \gS^1$. If $a$ is positive, then $\sigma(a) \subset \gR_+$. 
\end{Proposition}

One can show the following:
\begin{Theorem}[Gelfand-Mazur]
Any unital Banach algebra in which every non zero element is invertible is isomorphic to $\gC$.
\end{Theorem}

\begin{Definition}[The spectrum of an algebra and the Gelfand transform]
Let $\algA$ be a Banach algebra. A (continuous) character on $\algA$ is a non zero continuous morphism of algebras $\chi : \algA \rightarrow \gC$. If $\algA$ is unital, we require $\chi(\bbbone) = 1$. 

The spectrum of $\algA$, denoted by $\Delta(\algA)$, is the set of characters of $\algA$. The spectrum $\Delta(\algA)$ is a topological space for the topology induced by the pointwise convergence $\chi_n \xrightarrow{n \rightarrow \infty} \chi \Leftrightarrow \forall a \in \algA, \ \chi_n(a) \xrightarrow{n \rightarrow \infty} \chi(a)$. 

There is a natural map $\algA \rightarrow C(\Delta(\algA))$ defined by $a \mapsto \widehat{a}$ where $\widehat{a}(\chi) = \chi(a)$. This is the Gelfand transform of $\algA$.
\end{Definition}

So, now that we have associated to elements in a Banach algebra, and to the algebra itself, some topological spaces, it is time to introduce some functional spaces on them! The next example gives us an insight to what will happen in the general case.

\begin{Example}[The spectrum of $C(X)$]
Let $\algA = C(X)$ as in Example~\ref{ex-thealgebraofcontinousfunctions}. For any function $f \in C(X)$, $\sigma(f)$ is the set of values of $f$: $\sigma(f)=f(X) \subset \gC$. Any point $x \in X$ defines a character $\chi_x \in \Delta(\algA)$ by $\chi_x(f) = f(x)$, so that $X \subset \Delta(\algA)$. The topologies on $X$ and $\Delta(\algA)$ makes this inclusion a continuous application.
\end{Example}

\begin{Example}[Unital commutative Banach algebra]
What happens when the Banach algebra is unital and commutative? In that case, one can show that the maximal ideals in $\algA$ are in a one-to-one correspondence with characters on $\algA$. Indeed, it is easy to associate to any character $\chi \in \Delta(\algA)$, the maximal ideal $\algI_\chi = \ker \chi$. In the other direction, for any maximal ideal $\algI$, one can show that every non zero element in the algebra $\algA/\algI$ is invertible, which means, by the Gelfand-Mazur's theorem, that  $\algA/\algI = \gC$. Associate to $\algI$ the projection $\algA \rightarrow \algA/\algI$. This is the desired character.

Then, one can show that $\Delta(\algA)$ is a compact Hausdorff space. The Gelfand transform connects two commutative unital Banach algebras $\algA \rightarrow C(\Delta(\algA))$ by a continuous morphism of algebras. What is now a pleasant surprise, is that the spectrum of $a$ in $\algA$ is exactly the spectrum of $\widehat{a}$ in $C(\Delta(\algA))$, which is the set of values of the function $\widehat{a}$ on $\Delta(\algA)$:
\begin{equation*}
\sigma(a) = \sigma(\widehat{a}) = \{ \widehat{a}(\chi) = \chi(a) \ / \ \chi \in \Delta(\algA) \} 
\end{equation*}

When the commutative Banach algebra is not unital, the Gelfand transform realizes a continuous morphism of algebras $\algA \rightarrow C_0(\Delta(\algA))$. One important result is that $\Delta(\algA)^+ = \Delta(\algA_+)$ where on the left $\Delta(\algA)^+$ is the one-point compactification of the topological space $\Delta(\algA)$ and on the right $\algA_+$ is the unitarization of $\algA$. 

\end{Example}

It is now possible to state the main theorem in this section:
\begin{Theorem}[Gelfand-Neumark]
For any commutative $C^\ast$-algebra $\algA$, the Gelfand transform is an isomorphism of $C^\ast$-algebras.
\end{Theorem}

In the unital case, one gets $\algA \simeq C(\Delta(\algA))$ and in the non unital case, $\algA \simeq C_0(\Delta(\algA))$ and $\algA_+ \simeq C(\Delta(\algA)^+)$.

In the language of categories, this theorem means that the category of locally compact Hausdorff spaces is equivalent to the category of commutative $C^\ast$-algebras.

\subsection{Functional calculus}

The demonstration of the Gelfand-Neumark theorem relies on some constructions largely known as functional calculus. These constructions are very important to understand the relations between commutative $C^\ast$-algebras and topological spaces. Their understanding opens the door to the comprehension of noncommutative geometry.

The first example we consider is the polynomial functional calculus. This gives us the general idea. Let $a \in \algA$, where $\algA$ is any unital associative algebra. To every polynomial function $p \in \gC[x]$ in the real variable $x$, we can associate $p(a) \in \algA$ as the element obtained by the replacement $x^n \mapsto a^n$ in $p$. In particular, for the polynomial $p(x)=x$ (resp. $p(x) = 1$), one gets $p(a)=a$ (resp. $p(a) = \bbbone$).

For algebras with supplementary structures, this can be generalized using other algebras of functions.

First, consider an involutive unital algebra $\algA$, and let $a \in \algA$ be a normal element. To every polynomial function $p \in \gC[z, \overline{z}]$ of the complex variable $z$ and its conjugate $\overline{z}$, we associate $p(a) \in \algA$ through the replacements $z^n \mapsto a^n$ and $\overline{z}^n \mapsto (a^\ast)^n$. Because $a$ is normal, $p(a)$ is a well defined element in $\algA$.

Let $\algA$ be now a unital Banach algebra. For any $\lambda \notin \sigma(a)$ one introduces the resolvant of $a$ at $\lambda$:
\begin{equation*}
R(a,\lambda) = \frac{1}{\lambda - a} \in \algA
\end{equation*}
Consider any holomorphic function $f : U \rightarrow \gC$, with $U$ an open subset of $\gC$ which strictly contains the compact subspace $\sigma(a)$, and $ \Gamma : [0,1] \rightarrow \gC$ a closed path in $U$ such that $\sigma(a)$ is strictly inside $\Gamma$. The usual Cauchy formula $f(z) = \frac{1}{2 \pi i} \int_\Gamma \frac{f(\lambda)}{\lambda - z} d\lambda \in \gC$ can be generalized in the form
\begin{equation*}
f(a) = \frac{1}{2 \pi i} \int_\Gamma \frac{f(\lambda)}{\lambda - a} d\lambda \in \algA
\end{equation*}
Indeed, $\lambda \mapsto R(a,\lambda)$ is a function which takes its values in a Banach space, and integration of $f(\lambda) R(a,\lambda)$ is meaningful along $\Gamma$. What can be shown is that this integration does not depend on the choice of the surrounding closed path $\Gamma$.

This relation defines what is called the holomorphic functional calculus on $\algA$. In case $f$ is a polynomial function (of the variable $z$, but not of the variable $\overline{z}$), $f(a)$ coincides with the polynomial functional calculus. 

Let us consider now a unital $C^\ast$-algebra $\algA$. In that case, one would like to mix the two previous functional calculi on involutive and Banach algebras.

In order to do that, consider a normal element $a \in \algA$. One can introduce $C^\ast(a)$, the smallest unital $C^\ast$-subalgebra of $\algA$ which contains $a$ and $a^\ast$ (and $\bbbone$ since it is unital). Because $\bbbone, a$ and $a^\ast$ commute among themselves, the $C^\ast$-algebra $C^\ast(a)$ is a commutative $C^\ast$-algebra. Let us summarize some facts about this algebra:
\begin{Proposition}
The spectrum of $a$ in $C^\ast(a)$ is the same as the spectrum of $a$ in $\algA$. It will be denoted by $\sigma(a)$.

The spectrum of the algebra $C^\ast(a)$ is the spectrum of the element $a$, $\Delta(C^\ast(a)) = \sigma(a)$, so that
\begin{equation*}
C^\ast(a) = C(\sigma(a))
\end{equation*}

The Gelfand transform maps $a$ into the continuous function $\widehat{a} : \sigma(a) \rightarrow \gC$ which is identity: $\sigma(a) \ni z \mapsto z \in \gC$.

The inverse of the Gelfand transform associates to any continuous function $f : \sigma(a) \rightarrow \gC$ a unique element $f(a) \in C^\ast(a) \subset \algA$ such that
\begin{align*}
\| f(a) \| &= \| f \|_\infty
&
\sigma(f(a)) &= f( \sigma(a)) \subset \gC 
\end{align*}
In particular, the norm of $f(a)$ in $C^\ast(a)$ is the norm of $f(a)$ in $\algA$.
\end{Proposition}

The association $f \mapsto f(a)$ in this Proposition is the continuous functional calculus associated to the normal element $a$. In case $f$ is a polynomial function in the variables $z$ and $\overline{z}$ (resp. $f$ is an holomorphic function), one recovers the polynomial functional calculus (resp. the holomorphic functional calculus). 

There are a lot of interesting normal elements in a $C^\ast$-algebras (self-adjoint, unitary, positive\dots) for which the continuous functional calculus is very convenient. The next example illustrates such a situation.

\begin{Example}[Absolute value in $\algA$]
One can associate to any element $a \in \algA$ its absolute value using the functional calculus associated to the normal (and positive) element $a^\ast a$. Consider the continuous function $\gR_+ \ni x \mapsto f(x)=|x|^{1/2}$ and define $|a| \in \algA^+$ by $|a| = f(a^\ast a)$.
\end{Example}

What do we learn from these constructions? The main result here is that it is not necessary to consider a commutative $C^\ast$-algebra in order to manipulate some topological spaces. Just consider some normal elements commuting among themselves, build upon them the smallest $C^\ast$-algebra they generate, and you have in hand a Hausdorff space!

The idea of noncommutative topology is to study $C^\ast$-algebras from the point of view that they are ``continuous functions on noncommutative spaces''. In order to do that, one needs some tools that are common to the topological situation and to the algebraic one. 

Such tools exist! One of them is $K$-theory.

\section{$K$-theory for beginners}
\label{Ktheoryforbeginniers}

The $K$-theory groups are defined through a universal construction, the Grothendieck group associated to an abelian semigroup.
\begin{Definition}[Grothendieck group of an abelian semigroup]
An abelian semigroup is a set $\grV$ equipped with an internal associative and abelian law $\boxplus : \grV \times \grV \rightarrow \grV$. A unit element is an element $0$ such that $v \boxplus 0 = v$ for any $v \in \grV$. Any abelian group is a semigroup.

The Grothendieck group associated to $\grV$ is the abelian group $(\Gr(\grV), +)$ which satisfies the following universal property. There exits a semigroup map $i : \grV \rightarrow \Gr(\grV)$ such that for any abelian group $(\grG,+)$ and any morphism of abelian semigroups $\phi : \grV \rightarrow \grG$, there exists a unique morphism of abelian groups $\widehat{\phi} : \Gr(\grV) \rightarrow \grG$ such that $\phi = \widehat{\phi} \circ i$. 
\end{Definition}

This means that the following diagram can be completed with $\widehat{\phi}$ to get a commutative diagram:
\begin{equation*}
\xymatrix{ 
{\Gr(\grV)} \ar[dr]^-{\widehat{\phi}} & {} \\
{\grV} \ar[u]^-{i} \ar[r]^-{\phi} & {\grG} 
}
\end{equation*}

It is convenient to have in mind one of the possible constructions of the Grothendieck group associated to $\grV$. On the set $\grV \times \grV$ consider the equivalence relation: $(v_1, v_2) \sim (v'_1, v'_2)$ if there exists $v \in \grV$ such that $v_1 \boxplus v'_2 \boxplus v = v'_1 \boxplus v_2 \boxplus v \in \grV$. Denote by $\langle v_1, v_2 \rangle$ an equivalence class in $\grV \times \grV$ for this relation and let $\Gr(\grV) = (\grV \times \grV)/\sim$. The group structure on $\Gr(\grV)$ is defined by $\langle v_1, v_2 \rangle + \langle v'_1, v'_2 \rangle = \langle v_1 \boxplus v'_1, v_2  \boxplus v'_2 \rangle$, the unit is $\langle v, v \rangle$ for any $v \in \grV$, the inverse of $\langle v_1, v_2 \rangle$ is $\langle v_2, v_1 \rangle$. The morphism of abelian semigroups $i : \grV \rightarrow \Gr(\grV)$ is $v \mapsto \langle v \boxplus v', v' \rangle$ (independent of the choice of $v'$). Notice that $\langle v_1 + v, v_2 + v \rangle = \langle v_1, v_2 \rangle$ in $\Gr(\grV)$.

Then one can show that $\Gr(\grV)$ is indeed the Grothendieck group associated to $\grV$ and that
\begin{equation*}
\Gr(\grV) = \{ i(v) - i(v') \ / \ v,v' \in \grV\}
\end{equation*}
A useful relation is that $i(v + w) - i(v' + w) = i(v) - i(v') \in \Gr(\grV)$ for any $w \in \grV$.

\begin{Example}[The semigroup $\gN$]
\label{ex-semigroupN}
The set of natural numbers defines an abelian semigroup $(\gN, +)$. Its Grothen\-dieck's group is $(\gZ,+)$. In this situation, the morphism $i : \gN \rightarrow \gZ$ is injective. This is not always the case. Another particular property is that any relation $n_1 + n = n_2 + n$ in $\gN$ can be simplified into $n_1 = n_2$. This is the simplification property, which is not satisfied by all abelian semigroups.
\end{Example}

\begin{Example}[The semigroup $\gN \cup \{\infty \}$]
\label{ex-semigroupNinfty}
Consider the set $\gN \cup \{\infty \}$ with the ordinary additive law for two elements in $\gN$ and the new law $\infty + n = \infty + \infty = \infty$. Then its Grothendieck group is $\algzero$. Indeed, all couples $(n,m)$, $(n, \infty)$, $(\infty, m)$ and $(\infty, \infty)$ are equivalent.
\end{Example}

\subsection{The topological $K$-theory}

It is useful to recall some facts and constructions about vector bundles over topological spaces. We will restrict ourselves to locally trivial complex vector bundles over Hausdorff spaces.

\begin{Definition}
Let $\pi : E \rightarrow X$ and $\pi' : E' \rightarrow X$ two vector bundles over $X$, of rank $n$ and $n'$. Then one defines the Whitney sum $E \oplus E' \rightarrow X$ of rank $n+n'$ by $E \oplus E' = \cup_{x \in X} (E_x \oplus E'_x) = \{ (e,e') \in E \times E'\ / \ \pi(e) = \pi'(e')\} \subset E \times E'$ and the tensor product $E \otimes E' = \cup_{x \in X} (E_x \otimes E'_x)$ of rank $n n'$.

We denote by $\underline{\gC}^n = X \times \gC^n \rightarrow X$ the trivial vector bundles of rank $n$.

For any continuous map $f :Y \rightarrow X$ and any vector bundle $E \rightarrow X$, we define the pull-back $f^\ast E \rightarrow Y$ as $f^\ast E = \{ (y,e) \in Y \times E \ / \ f(y)=\pi(e) \} \subset Y \times E$.
\end{Definition}

When $i : Y \hookrightarrow X$ is an inclusion, the pull-back $i^\ast E = E_{|Y}$ is just the restriction of $E$ to $Y \subset X$. When $f_0, f_1 : Y \rightarrow X$ are homotopic, the two vector bundles $f^\ast_0 E$ and $f^\ast_1 E$ are isomorphic. 

We will use the following very important result in the theory of vector bundles:
\begin{Theorem}[Serre-Swan]
\label{thm-serreswan}
Let $X$ be a locally compact topological space. For any vector bundle $E \rightarrow X$ there exist an integer $N$ and a second vector bundle $E' \rightarrow X$ such that $E \oplus E' \simeq \underline{\gC}^N$.
\end{Theorem}

We introduce $\grV(X)$, the set of isomorphic classes of vector bundles over $X$. Let use the notation $[E]$ for the isomorphic class of $E$. The set $\grV(X)$ is an abelian semigroup for the law induced by the Whitney sum: $[E] + [E'] = [E \oplus E']$.

For any continuous map $f :Y \rightarrow X$, the pull-back construction defines a morphism of abelian semigroups $f^\ast : \grV(X) \rightarrow \grV(Y)$, which depends only on the homotopic class of $f$.

\begin{Definition}[$K^0(X)$ for $X$ compact]
\label{def-K0pourXcompact}
For any compact topological space $X$, we define $K^0(X)$ as the Grothendieck group of $\grV(X)$.
\end{Definition}

\begin{Remark}[Representatives in $K^0(X)$]
From the construction of the Grothendieck group, any element in $K^0(X)$ can be realized as a formal difference $[E]-[F]$ of two isomorphic classes in $\grV(X)$. Adding the same vector bundle to $E$ and $F$ does not change this element in the Grothendieck group. So, one can always find a representative of the form $[E] - [\underline{\gC}^n]$ for some integer $n$.
\end{Remark}

For any continuous map $f :Y \rightarrow X$, the morphism $f^\ast : \grV(X) \rightarrow \grV(Y)$ induces a morphism of abelian groups $f^\indast : K^0(X) \rightarrow K^0(Y)$.

\begin{Example}[$K^0(\espast) = \gZ$]
Let $\espast$ denote the space reduced to a point. In that case any vector bundle $E \rightarrow \espast$ is just a finite dimensional vector space. It is well known that the isomorphic classes of finite dimensional vector spaces are classified by their dimension, so that $\grV(\espast) = \gN$, with the abelian semigroup structure of Example~\ref{ex-semigroupN}. Then one gets $K^0(\espast) = \gZ$. Obviously this result is true for any contractible topological space.
\end{Example}

Let $x_0 \in X$ be a fixed point. Denote by $i : \espast=\{x_0\} \rightarrow X$ the inclusion, and $p : X \rightarrow \espast$ the projection. Then $p\circ i$ is $\Id_{\espast}$. These maps define two morphisms 
\begin{align*}
p^\indast : \gZ = K^0(\espast) &\rightarrow K^0(X)  
&
i^\indast : K^0(X) &\rightarrow K^0(\espast) = \gZ
\end{align*}
Because $i^\indast p^\indast = \Id_{\gZ} : \gZ \rightarrow \gZ$, $p^\indast$  is injective. 

\begin{Definition}[Reduced $K$-group for pointed compact spaces]
We define the reduced $K$-group of the pointed compact topological space $X$ by
\begin{equation*}
\widetilde{K}^0(X) = \Ker( i^\indast : K^0(X) \rightarrow \gZ ) = K^0(X) / p^\indast\gZ
\end{equation*}
\end{Definition}

The injective morphism $p^\indast$ splits the short exact sequence of abelian groups
\begin{equation*}
\xymatrix@1{{\algzero} \ar[r] & {\widetilde{K}^0(X)} \ar[r]^-{} & {K^0(X)} \ar[r]^-{i^\indast} & {K^0(\espast) = \gZ} \ar[r] & {\algzero}}
\end{equation*}
so that $K^0(X) = \widetilde{K}^0(X) \oplus \gZ$. The reduced $K$-theory, like the reduced singular homology, is the natural $K$-theory of pointed compact spaces.

\begin{Remark}[Interpretation of $\widetilde{K}^0(X)$]
An element in $\widetilde{K}^0(X)$ is a formal difference $[E]-[F]$ where now $E$ and $F$ have the same rank because they must coincide over $x_0$ in order to be in the kernel of $i^\indast$. It is possible to choose $F = \underline{\gC}^N$ with $N = \rank E$.
\end{Remark}

\begin{Definition}[$K^0(X)$ for any $X$]
Let $X$ be a locally compact topological space $X$, not necessarily compact. Denote by $X^+$ its one-point compactification. Then one defines $K^0(X) = \widetilde{K}^0(X^+)$.
\end{Definition}

Let us make some comments about this construction.

\begin{Remark}
\label{rmk-definitionofreducedKtheory}
In this situation, the natural fixed point in $X^+$ is the point at infinity, so that $i : \espast \rightarrow X^+$ sends $\espast$ into $\infty$. The compactification adds a point to $X$ and the reduced $K$-group construction removes the contribution from this point.

In case $X$ is compact, it is then easy to verify that $\widetilde{K}^0(X^+)$ identifies with the $K$-group as in Definition~\ref{def-K0pourXcompact}.
\end{Remark}

\begin{Remark}[Interpretation of $K^0(X)$]
In $\widetilde{K}^0(X^+)$, an element is a formal difference $[E] - [F]$ with $\rank E = \rank F$. Because this element is in the kernel of $i^\indast$, one has $E_{|\infty} = F_{|\infty}$. So these two vector bundles are also isomorphic in a neighborhood of $\infty \in X^+$. By definition of the one-point compactification, such a neighborhood is the complement of a compact in $X$, so that $E$ and $F$ can be considered as vector bundles over $X$ which coincide outside some compact $K \subset X$. 

One can go a step further. Because $E$ and $F$ coincide outside some compact $K$, one can add to them a third vector bundle such that outside $K$ the sums are isomorphic to a trivial vector bundle. Adding such a vector bundle does not change the element $[E] - [F] \in K^0(X)$. Therefore, any element in $K^0(X)$ can be represented by a formal difference $[E] - [F]$ where $E$ and $F$ are not only isomorphic outside some compact, but also trivial.

Owing to this interpretation, this definition of $K^0(X) = \widetilde{K}^0(X^+)$ is also called, in the literature, the $K$-theory with compact support (the non trivial part of the vector bundles is inside a compact). For instance, it is denoted by $K_{\text{cpt}}$ in \cite{LawsMich:89}.
\end{Remark}

The indice $0$ in the definition of the $K$-group suggests that others $K$-groups can be defined. This is indeed the case, but we will see that there are not so many!

\begin{Definition}[Higher orders $K$-groups]
Let $X$ be a locally compact topological space $X$. For any $n \geq 1$, we define $K^{-n}(X) = K^0(X \times \gR^n)$. 
\end{Definition}

The corresponding reduced $K$-group is $\widetilde{K}^{-n}(X) = \widetilde{K}^0(X \wedge \gS^n)$, where we recall that for two pointed compact spaces $(X,x_0)$ and $(Y, y_0)$, their wedge product is $X\wedge Y = X \times Y /(\{x_0\}\times Y \cup X \times\{y_0\})$. For any $n$, one can show that $K^{-n}(X) = \widetilde{K}^{-n}(X^+)$.

\begin{Proposition}[Long exact sequences]
\label{propo-longexactsequencesKtheory}
Let $X$ be a locally compact space and $Y \subset X$ a closed subspace. Then there exist boundary maps $\delta : K^{-n}(Y) \rightarrow K^{-n+1}(X\backslash Y)$ and a long exact sequence
\begin{multline*}
\xymatrix@1{{\cdots} \ar[r]^-{\delta} & {K^{-n}(X\backslash Y)} \ar[r] & {K^{-n}(X)} \ar[r] & {K^{-n}(Y)} \ar[r]^-{\delta} & {K^{-n+1}(X\backslash Y)} \ar[r] & {\cdots}}\\
\xymatrix@1{{\cdots} \ar[r]^-{\delta} & {K^0(X\backslash Y)} \ar[r] & {K^0(X)} \ar[r] & {K^0(Y)}}
\end{multline*}

In reduced $K$-theory, for pointed compact spaces $Y \subset X$, one has the corresponding long exact sequence
\begin{multline*}
\xymatrix@1{{\cdots} \ar[r]^-{\delta} & {\widetilde{K}^{-n}(X/Y)} \ar[r] & {\widetilde{K}^{-n}(X)} \ar[r] & {\widetilde{K}^{-n}(Y)} \ar[r]^-{\delta} & {\widetilde{K}^{-n+1}(X/Y)} \ar[r] & {\cdots}}\\
\xymatrix@1{{\cdots} \ar[r]^-{\delta} & {\widetilde{K}^0(X/Y)} \ar[r] & {\widetilde{K}^0(X)} \ar[r] & {\widetilde{K}^0(Y)}}
\end{multline*}
\end{Proposition}

\begin{Proposition}[The ring structure]
The tensor product of vector bundles induces a ring structure on $K^0(X)$ and $\widetilde{K}^0(X)$.

The external tensor product induces graded ring structures on $K^\grast(X) = \bigoplus_{n \geq 0} K^{-n}(X)$ and on $\widetilde{K}^\grast(X) = \bigoplus_{n \geq 0} \widetilde{K}^{-n}(X)$ which extend the ring structures on $K^0(X)$ and $\widetilde{K}^0(X)$.
\end{Proposition}

\begin{Example}[The $2$-sphere]
\label{ex-the2sphere}
Any vector bundle on the $2$-sphere is characterized by its clutching function on the equator. This is a continuous map $\gS^1 \rightarrow U(n)$ for a vector bundle of rank $n$. In order to consider all the possible ranks at the same time, the maps to consider are $\gS^1 \rightarrow U(\infty) = \varinjlim U(n)$. Studying these functions, in particular their homotopic equivalence classes, gives the following result. Let $H$ denote the tautological vector bundle of rank $1$ over $\gCP^1 = \gS^2$. Then one has $(H \otimes H) \oplus \underline{\gC} \simeq H \oplus H$ and as rings $K^0(\gS^2) \simeq \gZ[H]/\langle(H-\underline{\gC})^2\rangle$ where $\langle(H-\underline{\gC})^2\rangle$ is the ideal in $\gZ[H]$ generated by $(H-\underline{\gC})^2$, so that $K^0(\gS^2) \simeq \gZ \cdot \underline{\gC} \oplus \gZ \cdot (H-\underline{\gC}) \simeq \gZ \oplus \gZ$. Because $H - \underline{\gC} \in \Ker( i^\indast : K^0(\gS^2) \rightarrow \gZ )$, one has $\widetilde{K}^0(\gS^2) = \gZ \cdot (H-\underline{\gC}) \simeq \gZ$ with a null product. 
\end{Example}

\begin{Example}[The ring $K^\grast(\espast)$]
The ring structure of $K^\grast(\espast)$ is easy to describe. One can show that $K^{-2}(\espast) = K^0(\gR^2) = \widetilde{K}^0(\gS^2) = \gZ$. Denote by $\xi$ the generator of $K^{-2}(\espast)$. Then, one can show that $K^\grast(\espast) = \gZ[\xi]$. As $\xi$ is of degree $-2$, one has $K^{-2n}(\espast) = \gZ$ and $K^{-(2n+1)}(\espast) = {\algzero}$.
\end{Example}

Here is now the main result in $K$-theory:
\begin{Theorem}[Bott periodicity]
For any locally compact space $X$, one has a natural isomorphism
\begin{equation*}
K^0(X \times \gR^2) = K^{-2}(X) \simeq K^{0}(X)
\end{equation*}

For any pointed compact space $X$, one has a natural isomorphism
\begin{equation*}
\widetilde{K}^0(X \wedge \gS^2) = \widetilde{K}^{-2}(X) \simeq \widetilde{K}^0(X)
\end{equation*}
\end{Theorem}

In reduced $K$-theory, for two pointed compact spaces $X, Y$, there is a natural (graded) product
\begin{equation*}
\widetilde{K}^\grast(X) \otimes \widetilde{K}^\grast(Y) \rightarrow \widetilde{K}^\grast(X \wedge Y)
\end{equation*}
Using $Y=\gS^2$, this product gives us an isomorphism
\begin{equation*}
\widetilde{K}^0(X) \otimes \widetilde{K}^0(\gS^2) \overset{\simeq}{\rightarrow} \widetilde{K}^0(X \wedge \gS^2)
\end{equation*}
which is exactly the Bott periodicity. Indeed, we saw in Example~\ref{ex-the2sphere} that $\widetilde{K}^0(\gS^2)$ is generated by $H - \underline{\gC}$ (with $(H-\underline{\gC})^2=0$). The Bott periodicity is the isomorphism
\begin{align*}
\beta : \widetilde{K}^0(X) &\overset{\simeq}{\rightarrow} \widetilde{K}^0(X \wedge \gS^2) = \widetilde{K}^{-2}(X)\\
 a & \mapsto (H-\underline{\gC})\cdot a
\end{align*}

\begin{Example}[$K$-theories of spheres]
One has $\gS^n \wedge \gS^m \simeq \gS^{n+m}$, so that $\widetilde{K}^0(\gS^{2n}) = \widetilde{K}^0(\gS^{2n-2} \wedge \gS^2) = \widetilde{K}^0(\gS^2) = \gZ$. For odd degrees, one only needs to know $\widetilde{K}^0(\gS^{1})$. Using standard arguments from topology of fiber bundles (see \cite{Stee:51} for instance), there are no non trivial (complex) vector bundles over $\gS^1$, so that $\grV(\gS^1) = \gN$ and then $K^0(\gS^{1}) = \gZ$ and $\widetilde{K}^0(\gS^{1}) = {\algzero}$. This shows that $\widetilde{K}^0(\gS^{2n+1}) = \widetilde{K}^0(\gS^{1}) = {\algzero}$. Notice that because $\gR^+ = \gS^{1}$ (one-point compactification), one has $K^0(\gR) = {\algzero}$.
\end{Example}

\begin{Proposition}[Six term exact sequences in $K$-theory]
The Bott periodicity reduces the long exact sequences of Proposition~\ref{propo-longexactsequencesKtheory} into two six term exact sequences
\begin{equation}
\label{suiteexacte6termestopologique}
\xymatrix{ 
{K^0(X\backslash Y)} \ar[r]  & {K^0(X)} \ar[r]    & {K^0(Y)} \ar[d]^-{\delta} \\
{K^{-1}(Y)} \ar[u]^-{\delta} & {K^{-1}(X)} \ar[l] & {K^{-1}(X\backslash Y)} \ar[l] 
}
\end{equation}
for locally compact spaces $Y \subset X$ with $Y$ closed, and
\begin{equation*}
\xymatrix{ 
{\widetilde{K}^0(X/ Y)} \ar[r]  & {\widetilde{K}^0(X)} \ar[r]    & {\widetilde{K}^0(Y)} \ar[d]^-{\delta} \\
{\widetilde{K}^{-1}(Y)} \ar[u]^-{\delta} & {\widetilde{K}^{-1}(X)} \ar[l] & {\widetilde{K}^{-1}(X/ Y)} \ar[l] 
}
\end{equation*}
for pointed compact spaces $Y \subset X$ with $Y$ closed.
\end{Proposition}

\begin{Example}[$K$-groups for some topological spaces] Here is a table of some known $K$-groups for ordinary topological spaces.

\centerline{
\begin{tabular}{rcc}
Topological space & $K^0$ & $K^{-1}$ \\ \hline
$\espast$, compact contractible Hausdorff space & $\gZ$ & $\algzero$ \\
$]0,1]$ & $\algzero$ & $\algzero$ \\
$\gR$, $]0,1[$ & $\algzero$ & $\gZ$ \\
$\gR^{2n}$, $n\geq 1$ & $\gZ$ & $\algzero$ \\
$\gR^{2n+1}$, $n \geq 0$ & $\algzero$ & $\gZ$ \\
$\gS^{2n}$, $n\geq 1$ & $\gZ \oplus \gZ$ & $\algzero$ \\
$\gS^{2n+1}$, $n \geq 0$ & $\gZ$ & $\gZ$ \\
$\gT^n$ & $\gZ^{2^{n-1}}$ & $\gZ^{2^{n-1}}$
\end{tabular}
}
\end{Example}

\begin{Remark}[Real topological $K$-theory]
We have introduced the topological $K$-theory using the complex vector bundles over topological spaces. It is possible to define a real topological $K$-theory in exactly the same way using real vector bundles. The theory is different. For instance, there are non trivial real vector bundles over $\gS^1$ (think at the Moebius trip), but there are no non trivial complex vector bundles. In real $K$-theory the Bott periodicity is of period 8, and the six term exact sequence is replaced by a 24 terms exact sequence.
\end{Remark}

\begin{Remark}[The origin of Bott periodicity]
The first paper mentioning Bott periodicity, \cite{Bott:59}, was concerned with the homotopy of classical groups, in particular $U(n)$ and $O(n)$. What Bott discovered is that if one denotes by $U(\infty) = \varinjlim U(n)$ and $O(\infty) = \varinjlim O(n)$ the inductive limits for the natural inclusions $U(n) \hookrightarrow U(n+1)$ and $O(n) \hookrightarrow O(n+1)$, then 
\begin{align*}
\pi_k(U(\infty)) &= \pi_{k+2}(U(\infty)) 
&
\pi_k(O(\infty)) &= \pi_{k+8}(O(\infty)) 
\end{align*}
In fact, for $n$ large enough, $\pi_k(U(n)) = \pi_k(U(n+1))$ for $n > k/2$, so that this periodicity expresses itself before infinity. The period $2$ for the complex case $U(\infty)$ (resp. $8$ for the real case $O(\infty)$) is related to the period $2$ for $K$-theory (resp. real $K$-theory). See \cite{Karo:05} for a review and references.
\end{Remark}

\subsection{$K$-theory for $C^\ast$-algebras}

Topological $K$-theory is defined using some explicit geometrical constructions on vector bundles over a compact topological space $X$. These constructions, except the tensor product of vector bundles, can be described using the $C^\ast$-algebra 
$C(X)$.

Indeed, it is a well known fact that continuous sections of a vector bundle $E \rightarrow X$ is a $C(X)$-module. Recall the following definitions about modules. From now on, every modules are left modules.
\begin{Definition}[Finite projective modules]
Let $\algA$ be a unital associative algebra. 

$\modM$ is a free $\algA$-module if it admits a free basis. 

$\modM$ is a projective $\algA$-module if there exists a $\algA$-module $\modN$ such that $\modM \oplus \modN$ is a free module.

$\modM$ is a finite projective $\algA$-module if there exist a $\algA$-module $\modN$ and an integer $N$ such that $\modM \oplus \modN \simeq \algA^N$.
\end{Definition}

Theorem~\ref{thm-serreswan} can then be written in the following algebraic form:
\begin{Theorem}[Serre-Swan, algebraic version]
\label{thm-serreswanalgebraicversion}
The functor ``continuous sections'' realizes an equivalence of categories between the category of vector bundles over a compact topological space $X$ and the category of finite projective modules over $C(X)$.
\end{Theorem}

Any finite projective module is characterized by the morphism of $\algA$-modules $p: \algA^N \rightarrow  \algA^N$ which projects onto $\modM$. This morphism is representable as a projection $p \in M_N(\algA)$, $p^2=p$, $p^\ast = p$, such that $\modM = \algA^N p$. In particular, any vector bundle over $X$ is  given by a projection $p \in M_N(C(X)) = C(X, M_n(\gC))$ (see Example~\ref{ex-algebraC0XA}).

We have defined the topological $K$-theory via the isomorphic classes of vector bundles. In the algebraic language, isomorphic classes correspond to some equivalence classes on projections. Let us define some possible equivalence relations on projections in $C^\ast$-algebras. 

Let us denote by $\caP(\algA) = \{ p \in \algA \ / \ p^2 = p^\ast = p \}$ the set of projections in a unital $C^\ast$-algebra $\algA$.

\begin{Definition}[Equivalences of projections]
A partial isometry is an element $v \in \algA$ such that $v^\ast v \in \caP(\algA)$. In that case, one can show that $v v^\ast \in \caP(\algA)$. An isometry is an element $v \in \algA$ such that $v^\ast v = \bbbone$. Unitaries are in particular isometries.

Two projections $p,q \in \caP(\algA)$ are orthogonal if $pq=qp = 0 \in \algA$. This means that they project on direct summands of $\algA$. In this situation $p \oplus q \in \caP(\algA)$ is well defined.

There are three notions of equivalence for two projections $p,q \in \caP(\algA)$:
\begin{description}
\item[homotopic equivalence: ] $p \sim_h q$ if there exists a continuous path of projections in $\algA$ connecting $p$ and $q$.

\item[unitary equivalence: ] $p \sim_u q$ if there exists a unitary element $u \in \algA$ such that $u^\ast p u = q$.

\item[Murray-von~Neumann equivalence:] $p \sim_\text{M. v.N.} q$ if there exists a partial isometry $v \in \algA$ such that $v^\ast v = p$ and $v v^\ast = q$.
\end{description}
\end{Definition} 

One can show that 
\begin{equation*}
p \sim_h q \Longrightarrow p \sim_u q \Longrightarrow p \sim_\text{M. v.N.} q
\end{equation*}

Define $\caP_n(\algA) \subset M_n(\algA)$, the set of projections in $M_n(\algA)$. The natural inclusions $i_n : M_n(\algA) \hookrightarrow M_{n+1}(\algA)$ with $i_n(a) = \spmatrix{a & 0 \\ 0 & 0}$ permits one to define $M_\infty(\algA) = \bigcup_{n \geq 1} M_n(\algA)$ and $\caP_\infty(\algA) = \bigcup_{n \geq 1} \caP_n(\algA)$. The three equivalence relations defined above are well defined on $\caP_\infty(\algA)$.

\begin{Proposition}[Stabilisation of the equivalence relations]
In $\caP_\infty(\algA)$, the three equivalence relations coincide. 
\end{Proposition}

We will denote this relation by $\sim$.

\begin{Definition}[$K_0(\algA)$ for unital $C^\ast$-algebra]
Let $\grV(\algA)$ denote the set of equivalence classes in $\caP_\infty(\algA)$ for the relation $\sim$. This is an abelian semigroup for the addition $p \oplus q = \spmatrix{p & 0 \\ 0 & q} \in \caP_\infty(\algA)$.

The group $K_0(\algA)$ is the Grothendieck group associated to $(\grV(\algA), \oplus)$.
\end{Definition}

\begin{Example}[$K_0(M_n(\gC))$]
Let us look at the algebra $\algA = \gC$. In that case, a projection $p \in \caP_\infty(\gC)$ is represented by a projection $p \in M_N(\gC)$ for a sufficiently large $N$. Such a projection defines the vector space $\Ran p \subset \gC^N$ of dimension $\rank p$. It is easy to see that the equivalence relation $\sim$ detects only this dimension, so that $\grV(\gC) = \gN$ and $K_0(\gC) = \gZ$.

Let us consider now $\algA = M_n(\gC)$. One has $M_N(M_n(\gC)) = M_{Nn}(\gC)$, so that $\caP_\infty(M_n(\gC)) = \caP_\infty(\gC)$, with the same equivalence relation. Then one has $K_0(M_n(\gC)) = \gZ$. This result is an example of Morita invariance of the $K$-theory.
\end{Example}

More generally, by the same argument, one can show:
\begin{Proposition}[Morita invariance of the $K$-theory]
\begin{equation*}
K_0(M_n(\algA)) = K_0(\algA)
\end{equation*}
\end{Proposition}

\begin{Example}[$K_0(C(X))$]
Let $X$ be a compact topological space. Then by Theorem~\ref{thm-serreswanalgebraicversion} one has
\begin{equation*}
K_0(C(X)) = K^0(X)
\end{equation*}
\end{Example}

Any morphism of $C^\ast$-algebras $\phi : \algA \rightarrow \algB$ gives rise to a natural map $\phi : \caP_\infty(\algA) \rightarrow \caP_\infty(\algB)$ compatible with the relation $\sim$ on both sides. This induces a morphism of semigroups $\grV(\algA) \rightarrow \grV(\algB)$ and a morphism of abelian groups $\phi_\indast : K_0(\algA) \rightarrow K_0(\algB)$.

When the algebra $\algA$ is not unital, consider its unitarization $\algA_+$. Then one has the short exact sequence of $C^\ast$-algebras
\begin{equation*}
\xymatrix@1{{\algzero} \ar[r] & {\algA} \ar[r]^-{i} & {\algA_+} \ar[r]^-{\pi} & {\gC} \ar[r] & {\algzero}}
\end{equation*}

\begin{Definition}[$K_0(\algA)$ for non unital $C^\ast$-algebra]
For a non unital algebra $\algA$, one defines
\begin{equation*}
K_0(\algA) = \Ker( \pi_\indast : K_0(\algA_+) \rightarrow K_0(\gC) = \gZ )
\end{equation*}
\end{Definition}

\begin{Remark}
Exactly as in Remark~\ref{rmk-definitionofreducedKtheory}, this construction adds a point (the unity) and removes its contribution afterwards. In case $\algA$ is unital, one can show that the two definitions coincide. More generally, as abelian groups, one has $K_0(\algA_+) = K_0(\algA) \oplus \gZ$.
\end{Remark}

\begin{Remark}[Interpretation of $K_0(\algA)$]
An element in $K_0(\algA)$ is a difference $[p] - [q]$ for some projections $p, q \in \caP_n(\algA_+)$, for large enough $n$, such that $[\pi(p)] - [\pi(q)] = 0$. In fact, it is possible to choose $p$ and $q$ such that $p - q \in M_n(\algA) \subset M_n(\algA_+)$ ($p-q$ is not a projection in this relation!). Adding a common projection, one can always represent an element in $K_0(\algA)$ as $[p] - [\bbbone_n]$, where $\bbbone_n \in M_n(\algA_+)$ is the unit matrix.
\end{Remark}

\begin{Example}[$K_0(\caB)$]
For any integer $n$ and infinite dimensional separable Hilbert space $\ehH$, one has $M_n(\caB) \simeq \caB$ (because $\ehH^n \simeq \ehH$ here), so we only need to consider projections in $\caB$. Two projections in $\caB$ are equivalent precisely when their ranges are isomorphic. So that only the dimension (possibly infinite) is an invariant, and one gets $\grV(\caB) = \gN \cup \{ \infty \}$, which produces $K_0(\caB) = \algzero$ (see Example~\ref{ex-semigroupNinfty}). 
\end{Example}

\begin{Definition}[Higher orders $K$-groups]
\label{def-higherordersKgroups}
Let $\algA$ be a $C^\ast$-algebra. The suspension of $\algA$ is the $C^\ast$-algebra $S\algA = C_0(\gR, \algA)$ (see Example~\ref{ex-algebraC0XA}).

For any $n \geq 1$, we define $K_n(\algA) = K_0(S^n \algA)$ where $S^n \algA = S(S^{n-1}\algA)$ is the $n$-th suspension of $\algA$.
\end{Definition}

This definition leads to the following useful result:
\begin{Proposition}[Long exact sequences]
For any short exact sequence of $C^\ast$-algebras
\begin{equation*}
\xymatrix@1{{\algzero} \ar[r] & {\algI} \ar[r] & {\algA} \ar[r] & {\algA/\algI} \ar[r] & {\algzero}}
\end{equation*}
there exist boundary maps $\delta : K_{n}(\algA/\algI) \rightarrow K_{n-1}(\algI)$ and a long exact sequence
\begin{multline*}
\xymatrix@1{{\cdots} \ar[r]^-{\delta} & {K_n(\algI)} \ar[r] & {K_n(\algA)} \ar[r] & {K_n(\algA/\algI)} \ar[r]^-{\delta} & {K_{n-1}(\algI)} \ar[r] & {\cdots}}\\
\xymatrix@1{{\cdots} \ar[r]^-{\delta} & {K_0(\algI)} \ar[r] & {K_0(\algA)} \ar[r] & {K_0(\algA/\algI)}}
\end{multline*}
\end{Proposition}

\begin{Remark}[Other definition of $K_1(\algA)$]
\label{rmk-otherdefinitionofK1A}
Let us introduce the following groups 
\begin{itemize}
\item $GL_n(\algA_+)$, invertible elements in $M_n(\algA_+)$
\item $GL_n^+(\algA) = \{ a \in GL_n(\algA_+) \ / \ \pi(a) = \bbbone_n \}$ where $\pi : \algA_+ \rightarrow \gC$ is the projection associated to the unitarization
\item $\caU_n(\algA_+)$, unitaries in $M_n(\algA_+)$
\item $\caU_n^+(\algA) = \{ u \in \caU_n(\algA_+) \ / \ \pi(u) = \bbbone_n \}$
\end{itemize}
These groups define some direct systems for the natural inclusion $g \mapsto \spmatrix{g & 0 \\ 0 & \bbbone}$. Denote by $GL_\infty(\algA_+)$, $GL_\infty^+(\algA)$, $\caU_\infty(\algA_+)$ and $\caU_\infty^+(\algA)$ their respective inductive limits.

One can show that for any $C^\ast$-algebra $\algA$, one has
\begin{align*}
K_1(\algA) &= GL_\infty(\algA_+)/GL_\infty(\algA_+)_0 = GL_\infty^+(\algA)/GL_\infty^+(\algA)_0 \\
 & = \caU_\infty(\algA_+)/\caU_\infty(\algA_+)_0 = \caU_\infty^+(\algA)/\caU_\infty^+(\algA)_0 
\end{align*}
where the index $0$ means the connected component of the unit element.
\end{Remark}

\begin{Proposition}[Continuity for direct systems]
Let $(\algA_i, \alpha_i)$ be a direct system of $C^\ast$-algebras. Then for any $n$ one has $K_n( \varinjlim A_i) = \varinjlim K_n( A_i)$.
\end{Proposition}

\begin{Example}[$K_0(\caK)$]
The algebra of compact operators is the direct limit $\caK = \varinjlim M_n(\gC)$. As $K_0(M_n(\gC)) = \gZ$ is a stationary system, one has $K_0(\caK) = \gZ$. Explicitly, the isomorphism is realized as the trace $[p] \mapsto \tr(p)$.
\end{Example}

More generally we have:
\begin{Proposition}[Morita invariance of $K$-theory]
For any $C^\ast$-algebra $\algA$, and any $n$, one has $K_n(\algA \otimeshat \caK) = K_n(\algA)$.
\end{Proposition}

Here is the version of Bott periodicity for $K$-theory of $C^\ast$-algebras:
\begin{Theorem}[Bott periodicity]
For any $C^\ast$-algebra $\algA$, one has
\begin{equation*}
K_0(S^2\algA) = K_2(\algA) \simeq K_0(\algA)
\end{equation*}
\end{Theorem}

\begin{Proposition}[Six term exact sequence]
\label{prop-sixtermsexactsequencesK}
The Bott periodicity theorem reduces the long exact sequence associated to any short exact sequence of $C^\ast$-algebras to a six term exact sequence
\begin{equation*}
\xymatrix{ 
{K_0(\algI)} \ar[r]  & {K_0(\algA)} \ar[r]  & {K_0(\algA/\algI)} \ar[d]^-{\delta} \\
{K_{1}(\algA/\algI)} \ar[u]^-{\delta} & {K_{1}(\algA)} \ar[l] & {K_{1}(\algI)} \ar[l] 
}
\end{equation*}
\end{Proposition}

\begin{Remark}[$K$-groups via homotopy groups]
\label{rmk-Kggroupsviahomotopygroups}
We have seen in Remark~\ref{rmk-otherdefinitionofK1A} that the $K_1$-group can be defined using the $0$-th homotopy group as $K_1(\algA) = \pi_0( \caU_\infty(\algA_+))$. It is possible to show that more generally 
\begin{equation*}
K_n(\algA) = \pi_{n-1}( \caU_\infty(\algA_+))
\end{equation*}
Bott periodicity is then directly equivalent to
\begin{equation*}
\pi_{n+2}( \caU_\infty(\algA_+)) \simeq \pi_{n}( \caU_\infty(\algA_+))
\end{equation*}
Because $\caU_n(\algA_+)$ and $GL_n(\algA_+)$ have the same topology (one is the retraction of the other), these relations make sense with $GL_\infty(\algA_+)$.
\end{Remark}

\begin{Example}[$K$-groups for some $C^\ast$-algebras] Here is a table of some known $K$-groups for ordinary $C^\ast$-algebras.

\centerline{
\begin{tabular}{rcc}
Algebra & $K_0$ & $K_1$ \\ \hline
$\gC$, $M_n(\gC)$, $\caK(\ehH)$ (compacts op.) & $\gZ$ & $\algzero$ \\
$\caB(\ehH)$ (bounded op.) & $\algzero$ & $\algzero$ \\
$\caQ(\ehH)$ (Calkin's alg.) & $\algzero$ & $\gZ$ \\
$\caT$ (Tœplitz' alg.) & $\gZ$ & $\algzero$ \\
$\caO_n$, $n \geq 2$ (Cuntz' alg.) & $\gZ_{n-1}$ & $\algzero$ \\
$\tncA_\theta$, $\theta$ irrationnal & $\gZ^2 \simeq \theta \gZ + \gZ$ & $\gZ^2$ \\
$C^\ast(\gF_n)$ ($\gF_n$ free group with $n$ generators) & $\gZ$ & $\gZ^n$ \\
$M_n(\algA)$, $\algA \otimeshat \caK$ (stabilisation) & $K_0(\algA)$ & $K_1(\algA)$ \\
$\algA_+$ (unitarization) & $K_0(\algA) \oplus \gZ$ & $K_1(\algA)$ \\
$S\algA = C_0(]0,1[, \algA)$ (suspension) & $K_1(\algA)$ & $K_0(\algA)$ \\
$C\algA = C_0(]0,1], \algA)$ (cone) & $\algzero$ & $\algzero$ 
\end{tabular}
}
\end{Example}

\begin{Remark}[$K$-theory computed on dense subalgebras]
\label{rmk-Ktheorycomputedondensesubalgebras}
For a lot of examples, one can compute the $K$-groups of a $C^\ast$-algebra $\algA$ using a dense subalgebra $\algB$. For instance, for any compact finite dimensional manifold $M$, the $K$-theory of $C(M)$ (continuous functions) is the same as the $K$-theory of the Fréchet algebra $C^\infty(M)$. The same situation occurs for the irrational rotation algebra: $K_n(\tncA_\theta) = K_n(\tncA^\infty_\theta)$.

In the geometric situation, it is possible to understand this result. Smooth structures are sufficiently dense in continuous structures: any continuous vector bundle can be deformed into a smooth one\dots

Here is a description of some more general situations. Let $\algA$ be $C^\ast$-algebra (or a Banach algebra) and $\algA^\infty \subset \algA$ a dense subalgebra (but not necessarily a $C^\ast$-subalgebra). The exponent $\infty$ does not mean that we consider ``differentiable'' functions, even if in practice this can happen: think about $\tncA^\infty_\theta \subset \tncA_\theta$ as a typical example. Let $\algA_+$ and $\algA^\infty_+$ their unitarizations. Suppose that $\algA^\infty_+$ is stable under holomorphic functional calculus, which means that for any $a \in \algA^\infty_+$  and any holomorphic function $f$ in a neighborhood of the spectrum of $a$, $f(a) \in \algA^\infty_+$. 

Using the topologies induced on $\algA^\infty_+$ and $GL_n(\algA^\infty_+)$ by the topologies on $\algA_+$ and $GL_n(\algA_+)$, it is possible to define $K$-groups by using the relations in Remark~\ref{rmk-Kggroupsviahomotopygroups}. Then one has the density theorem: the inclusion $i : \algA^\infty \rightarrow \algA$ induces isomorphisms
\begin{equation*}
i_\indast : K_n(\algA^\infty) \overset{\simeq}{\rightarrow} K_n(\algA)
\end{equation*}
for any $n \geq 0$. 
\end{Remark}

\begin{Remark}[$K$-homology]
\label{rmk-Khomology}
As for many other ordinary homologies, there exists a dual version of the $K$-theory of $C^\ast$-algebras, named $K$-homology, which we outline here.

A Fredholm module over the $C^\ast$-algebra $\algA$ is a triplet $(\ehH, \rho, F)$ where $\ehH$ is a Hilbert space, $\rho$ is  an involutive representation of $\algA$ in $\caB(\ehH)$, and $F$ is an operator on $\ehH$  such that for any $a \in \algA$, $(F^2 - 1) \rho(a)$, $(F - F^\ast) \rho(a)$ and $[F, \rho(a)]$ are in $\caK$. Such a Fredholm module is called odd.

A $\gZ_2$-graded Fredholm module is a Fredholm module $(\ehH, \rho, F)$ such that $\ehH = \ehH^+ \oplus \ehH^-$ and $\rho(a)$ is of even parity in this decomposition, and $F$ is of odd parity: $\rho(a) = \spmatrix{\rho^+(a) & 0 \\ 0 & \rho^-(a)}$ and $F = 
\spmatrix{0 & U^+ \\ U^- & 0}$. With these notations, $U^\pm : \ehH^\mp \rightarrow \ehH^\pm$ are essentially adjoint (adjoint modulo compact operators). Such a Fredholm module is called even.  

In the even case, one has a natural grading map $\gamma : \ehH \rightarrow \ehH$ defined by $\gamma = \spmatrix{1 & 0 \\ 0 & -1}$ on the decomposition $\ehH = \ehH^+ \oplus \ehH^-$. It satisfies $\gamma = \gamma^\ast$, $\gamma^2 = 1$, $\gamma \rho(a) = \rho(a) \gamma$ and $\gamma F = -F \gamma$.

Two Fredholm modules $(\ehH, \rho, F)$ and $(\ehH', \rho', F')$ are unitary equivalent if there exists a unitary map $U : \ehH' \rightarrow \ehH$ such that $\rho' = U^\ast \rho U$ and $F' = U^\ast F U$. This defines an equivalence relation $\sim_U$ of Fredholm modules.

A homotopy of Fredholm modules is a familly $t \mapsto (\ehH, \rho, F_t)$ with $[0,1] \ni t \mapsto F_t$ continuous for the operator norm in $\ehH$. Two Fredholm modules are homotopic equivalent if they are connected by a homotopy of Fredholm modules. This defines an equivalence relation $\sim_h$.

The direct sum of two Fredholm modules $(\ehH, \rho, F)$ and $(\ehH', \rho', F')$, denoted by $(\ehH, \rho, F) \oplus (\ehH', \rho', F')$, is defined by $\left( \ehH \oplus \ehH', \spmatrix{\rho & 0 \\ 0 & \rho'},\spmatrix{F & 0 \\ 0 & F'}\right)$.

The $K$-homology group $K^0(\algA)$ of $\algA$ is the Grothendieck group of the abelian semigroup of equivalence classes of even Fredholm modules for $\sim_U$ and $\sim_h$. The unit for the addition is the class of the Fredholm module $(0,0,0)$, the inverse of the class of $(\ehH, \rho, F)$ is the class of $(\ehH, \rho, -F)$.

The $K$-homology group $K^1(\algA)$ is defined in the same manner using odd Fredholm modules.

A degenerated Fredholm module is a Fredholm module for which $(F -F^\ast) \rho(a) = 0$, $(F^2 -1)\rho(a) = 0$ and $[F, \rho(a)] = 0$ for any $a \in \algA$. The equivalence class of such a Fredholm module is zero.

In each equivalence class, there is a representative for which $F^\ast = F$ (self-adjoint Fredholm module) and $F^2 = 1$ (involutive Fredholm module).

For $\algA = \gC$, the representation $\rho$ defines a projection $p = \rho(1)$ on $\ehH$, and one can show that modulo compact operators, one has $(\ehH, \rho, F) = (p\ehH, \rho, pFp) \oplus ((1-p)\ehH, \rho, (1-p)F(1-p))$. The representation for the second Fredholm module is zero, so that its class is zero. $pFp$ is an ordinary Fredholm operator on $\ehH$, and its index induces an isomorphism $\ind : K^0(\gC) \xrightarrow{\simeq} \gZ$.

For any $C^\ast$-algebra $\algA$, let $p$ be a projection in $\caP_n(\algA)$, and $(\ehH, \rho, F)$ a Fredholm module. Then, in the Hilbert space $\rho(p)(\ehH \otimes \gC^n)$, the operator $\rho(p)(F\otimes \bbbone_n)\rho(p)$ is a Fredholm operator, and its index defines a pairing between $K_0(\algA)$ and $K^0(\algA)$: $\langle [p], [(\ehH, \rho, F)] \rangle = \ind \rho(p)(F\otimes \bbbone_n)\rho(p) \in \gZ$.

For more developments in $K$-homology, see \cite{Blac:98} and \cite{HigsRoe:04}.
\end{Remark}

\subsection{Algebraic $K$-theory}

Until now, $K$-theory has been defined using topological structures, either at the level of a space or at the level of an algebra (remember that they are the manifestations of the same topological structure in the commutative case).

Nevertheless the group $K_0(\algA)$ can be defined in the pure algebraic context. Indeed, to any ring $\algA$, which we take unital from now on, one can associate its category of finite projective modules. 

\begin{Definition}[$K^\text{alg}_0(\algA)$ for unital ring $\algA$]
The group $K^\text{alg}_0(\algA)$ is the Grothendieck group associated to the semigroup of isomorphic classes of finite projective modules on $\algA$, on which the additive law is induced by the direct sum of modules.
\end{Definition}

As in the topological case, every finite projective $\algA$-module $\modM$ is characterized by a (non unique) projector $p \in M_m(\algA)$. Be aware of the different terminologies that are used. ``Projection'' is reserved to the $C^\ast$-algebra context, because in that case $p$ satisfies $p^2=p$ and $p^\ast = p$. ``Projector'' is more general, in that case $p$ satisfies only $p^2=p$.

The equivalence relation we use on these projectors is the following:
\begin{Definition}[Equivalence relation on projectors]
Two projectors $p \in M_m(\algA)$ and $q \in M_n(\algA)$ are equivalent if there exist an integer $r \geq m,n$ and an invertible $u \in GL_r(\algA)$ such that $p, q \in M_r(\algA)$ are conjugated by $u$: $p = u^{-1} q u$. We donote by $\sim$ this equivalence relation.
\end{Definition}

If $p \sim q$, then they define isomorphic finite projective modules.

In case $\algA$ is a $C^\ast$-algebra, one can show that in the equivalence class of any projector $p$ one can find a projection. This means that the two semigroups which define the $K$-theories are the same :
\begin{equation*}
K^\text{alg}_0(\algA) = K_0(\algA) \text{ (as a $C^\ast$-algebra)}
\end{equation*}

For higher order groups, the situation is no more equivalent. The Definition~\ref{def-higherordersKgroups} (or their equivalent ones given in Remark~\ref{rmk-Kggroupsviahomotopygroups}) uses extensively the topological structure of the algebra, either to define continuous functions $\gR \rightarrow \algA$ or to compute the homotopy groups of the spaces $\caU_\infty(\algA_+)$ in Remark~\ref{rmk-Kggroupsviahomotopygroups}.

Nevertheless, one can define $K^\text{alg}_1(\algA)$ as follows:
\begin{Definition}[$K^\text{alg}_1(\algA)$ for unital ring $\algA$]
One defines 
\begin{equation*}
K^\text{alg}_1(\algA) = GL_\infty(\algA) / [ GL_\infty(\algA), GL_\infty(\algA)] = GL_\infty(\algA)_\text{ab}
\end{equation*}
\end{Definition}

Let $\algA$ be a $C^\ast$-algebra. A well known fact about invertibles is that if $u, v \in GL_\infty(\algA)$ then $uv$ and $vu$ are homotopic. So that there is a natural morphism of groups
\begin{equation*}
K^\text{alg}_1(\algA) \rightarrow K_1(\algA) \text{ (as a $C^\ast$-algebra)}
\end{equation*}
which factors out by the homotopic relation.

For every $n \geq 2$, there is a definition which surprisingly uses some topological objects: $K^\text{alg}_n(\algA)$ is the $\pi_n$ group of a topological space associated to the classifying space $BGL_\infty(\algA)$ where $GL_\infty(\algA)$ is considered as a discrete group.

In algebraic $K$-theory, there is no Bott periodicity, but there are some other beautiful and powerful results which are beyond the scope of this introduction: in the following, we will only make use of $K^\text{alg}_0(\algA)$ and $K^\text{alg}_1(\algA)$. For a review, see \cite{Karo:03} or \cite{Rose:94}.

\section{Cyclic homology for (differential) geometers}
\label{cyclichomologyforgeometers}

Now we have in hand some tools to characterize noncommutative topological spaces. But topology is not everything in life. Differential geometry has to be considered also! In this section, we will explore other concepts that look very much like differential forms.

\subsection{Differential calculi}

Differential forms on a differentiable manifold define a differential graded commutative algebra. This concept can be generalised:
\begin{Definition}[Differential calculus on an algebra]
Let $\algA$ be an associative algebra. A differential calculus on $\algA$ is a graded differential algebra $(\Omega^\grast, d)$ such that $\Omega^0 = \algA$.
\end{Definition}

Remember the definition of a graded differential algebra: $\Omega^\grast$ is a graded algebra on which the differential satisfies $d (\omega \eta) = (d \omega) \eta + (-1)^{|\omega|} \omega (d \eta)$ for any $\omega, \eta \in \Omega^\grast$ and where $|\omega|$ is the degree of $\omega$.

In this definition, one does not suppose this graded algebra to be a graded commutative algebra.

\begin{Example}[de~Rham differential calculus]
Let $M$ be a finite dimensional differential manifold. The graded differential algebra $(\Omega^\grast(M), d)$ of differential forms is a differential calculus on $C^\infty(M)$.
\end{Example}

There are many possibilities to define a differential calculus on an algebra. One can summarize the questing after noncommutative differential geometry to the search of some reasonable definition of such a differential calculus on any algebra. Many propositions have been made, depending on the context: associative algebra without any additional structure, topological or involutive algebras, quantum groups\dots 

Some examples will be given after we introduce three main examples which are the universal differential calculi.

\begin{Example}[Universal differential calculus for associative algebra]
This differential calculus is defined to be the free graded differential algebra generated by $\algA$ as elements in degree $0$. It is denoted by $(\Omega^\grast(\algA), d)$.

Because it is freely generated, it has the following universal property: for any differential calculus $(\Omega^\grast, d)$ on $\algA$, there exists a unique morphism of differential calculi $\phi : \Omega^\grast(\algA) \rightarrow \Omega^\grast$ (of degree $0$) such that $\phi(a) = a$ for any $a\in \algA = \Omega^0(\algA) = \Omega^0$.

This implies that if $(\Omega^\grast, d)$ is generated (possibly with relations) by $\algA = \Omega^0$ then it is a quotient of $(\Omega^\grast(\algA), d)$ by a differential two-side ideal.

Concretely, any element in $\Omega^n(\algA)$ is a sum of terms either of the form $adb_1\dots db_n$ or of the form $db_1\dots db_n$. This property gives us an identification of left $\algA$-modules
\begin{equation*}
\Omega^n(\algA) = \algA_+ \otimes \algA^{\otimes n}
\end{equation*}
by the morphism $adb_1\dots db_n \mapsto (0+a)\otimes b_1 \otimes \dots \otimes b_n$ and $db_1\dots db_n \mapsto (1+0)\otimes b_1 \otimes \dots \otimes b_n$, where $(0+a)$ and $(1+0)$ are elements in $\algA_+ = \gC \oplus \algA$. Be aware of the fact that this identification is not an identification of graded differential algebras, neither of bimodules. 
\end{Example}

In the differential calculus $(\Omega^\grast(\algA), d)$, if $\algA$ is unital, $d\bbbone$ is not zero, because it is identified with $1 \otimes \bbbone \in \algA_+ \otimes \algA$.  It is the aim of the following example to show that in the more restrictive situation where $\algA$ is unital, one can promote the unit of $\algA$ to a unit of the differential calculus.

\begin{Example}[Universal differential calculus for associative unital algebra]
Let $\algA$ be an associative unital algebra. The differential calculus $(\Omega^\grast_U(\algA), d_U)$ is defined to be the free unital graded differential algebra generated by $\algA$ in degree $0$.

Because this algebra is required to have an unit, this unit is necessarily the unit in $\algA = \Omega^0_U(\algA)$. Then the derivative law for $d_U$ gives $d_U \bbbone = 0$. This differential calculus admits a universal property as the previous one does: for any unital differential calculus $(\Omega^\grast, d)$ on $\algA$, there exists a unique morphism of unital differential calculi $\phi : \Omega^\grast_U(\algA) \rightarrow \Omega^\grast$ (of degree $0$) such that $\phi(a) = a$ for any $a\in \algA = \Omega^0_U(\algA) = \Omega^0$. 

Because $(\Omega^\grast_U(\algA), d_U)$ is a differential calculus generated by $\algA$, it is a quotient of $(\Omega^\grast(\algA), d)$. This quotient reveals itself in the concrete identification of $\Omega^\grast(\algA)$: any element in $\Omega_U^n(\algA)$ is a sum of terms of the form $ad_Ub_1 \cdots d_U b_n$. Here $a$ can be $\bbbone$, and in this case $\bbbone d_Ub_1 \cdots d_U b_n = d_Ub_1 \cdots d_U b_n$. If one of the $b_k$ is proportional to $\bbbone$, one has $ad_Ub_1 \cdots d_U b_n = 0$. This leads to the identification of left modules
\begin{equation*}
\Omega_U^n(\algA) = \algA \otimes \overline{\algA}^{\otimes n}
\end{equation*}
by the map $ad_Ub_1\dots d_Ub_n \mapsto a\otimes \overline{b_1} \otimes \dots \otimes \overline{b_n}$, where $\overline{b}$ is the projection of $b\in \algA$ onto the vector space $\overline{\algA} = \algA/\gC\bbbone$.
\end{Example}

In these two examples, even if the algebra $\algA$ is commutative, the graded algebras are not graded commutative. For commutative algebras, it is possible to construct a differential calculus with a graded commutative algebra.

\begin{Example}[Kähler differential calculus for commutative unital algebra]
\label{ex-Kahlerdifferentialcalculusforcommutativeunitalalgebra}
Let $\algA$ be an associative commutative unital algebra over a field $\gK$. The Kähler differential calculus $(\Omega^\grast_{\algA|\gK}, d_K)$ is defined to be the free unital graded commutative differential algebra generated by $\algA$ in degree $0$.

One can show that the algebra $\Omega^\grast_{\algA|\gK}$ is an exterior algebra over $\algA$: $\Omega^\grast_{\algA|\gK} = \exter^\grast_\algA \Omega^1_{\algA|\gK}$. Moreover, let $\algI \subset \algA\otimes \algA$ be the kernel of the product map $\mu : \algA\otimes \algA \rightarrow \algA$. Consider $\algA\otimes \algA$ as an algebra (commutative) and introduce $\algI^2$, generated by the products of elements in $\algI$. Then one has the explicit construction $\Omega^1_{\algA|\gK} = \algI/\algI^2$.

We denote by $\pi_\grast : \Omega^\grast(\algA) \rightarrow \Omega^\grast_{\algA|\gC}$ the universal projection, given explicitly by 
\begin{align*}
\pi_n(a_0 da_0 \cdots da_n) &= a_0 d_K a_1 \wedge \cdots \wedge d_K a_n
&
\pi_n(da_0 \cdots da_n) &= d_K a_1 \wedge \cdots \wedge d_K a_n
\end{align*}

The cohomology of this differential algebra is denoted by $H_\text{dR}^\grast(\algA)$, and is called the de~Rham cohomologie of the commutative unital algebra $\algA$. This terminology comes from the fact that this differential calculus looks very much like the de~Rham differential calculus (see Example~\ref{ex-polynomialalgebra-differentialcalculus}).
\end{Example}

\begin{Example}[Polynomial algebra]
\label{ex-polynomialalgebra-differentialcalculus}
Let $\evV$ be a finite dimensional vector space over $\gC$. Consider the commutative algebra $\symes \evV$ of polynomials on $\evV$. Then one has the identification $\symes \evV \otimes \exter^n \evV = \Omega^n_{\symes \evV|\gC}$ by the map $a\otimes v \mapsto ad_K v$ in degree $1$. This differential calculus is the ``restriction'' of the de~Rham differential calculus of $C^\infty$ functions to the subalgebra of polynomial functions.
\end{Example}

\begin{Example}[Spectral triplet]
Let $\algA$ be an involutive unital associative algebra. A spectral triplet on $\algA$ is a triplet $(\algA, \ehH, D)$ where $\ehH$ is a Hilbert space on which an involutive representation $\rho$ of $\algA$ is given, and $D$ is a self-adjoint operator on $\ehH$ (not necessarily bounded), whose resolvant is compact, and such that $[D,\rho(a)]$ is bounded for any $a \in \algA$. The operator $D$ is called a Dirac operator. 

The map $\pi : \Omega^\grast_U(\algA) \rightarrow \caB(\ehH)$ defined by $\pi(a_0 d_U a_1 \cdots d_U a_n) = a_0 [D,a_1] \cdots [D,a_n]$ is an involutive representation of $\Omega^\grast_U(\algA)$ on $\ehH$.

Define $J_0 = \bigoplus_{n\geq 0} ( \Ker \pi \cap \Omega^n_U(\algA))$. One can show that $J = J_0 + d_U J_0$ is a differential two-sided ideal in $\Omega^\grast_U(\algA)$. The differential calculus defined by the spectral triplet $(\algA, \ehH, D)$ is the graded differential algebra $\Omega^\grast_D(\algA) = \Omega^\grast_U(\algA)/J$.

This construction is inspired by the definition of Fredholm modules, which are the building blocks of $K$-homology (see Remark~\ref{rmk-Khomology}). 

See \cite{Conn:94} and \cite{GracVariFigu:01} for more details and examples.
\end{Example}

\begin{Example}[Derivations based differential calculus for associative algebra]
Let $\algA$ be an associative algebra. The space of derivations on $\algA$, $\der(\algA) = \{ X : \algA \rightarrow \algA\ / \ X \text{ linear map and } X(ab) = (Xa) b + a (Xb) \}$, is a Lie algebra and a module over the center $\caZ(\algA)$ of $\algA$.

Let $\underline{\Omega}^n_\der(\algA)$ be the set of $\caZ(\algA)$-multilinear antisymmetric maps $\der(\algA)^n \rightarrow \algA$. Define on $\underline{\Omega}^\grast_\der(\algA) = \bigoplus_{n \geq 0} \underline{\Omega}^n_\der(\algA)$ the product
\begin{equation*}
(\omega\eta)(X_1,\dots, X_{p+q}) = \frac{1}{p!q!} \sum_{\sigma\in \kS_{p+q}} (-1)^{\sign(\sigma)} \omega(X_{\sigma(1)}, \dots, X_{\sigma(p)}) \eta(X_{\sigma(p+1)}, \dots, X_{\sigma(p+q)})
\end{equation*}
for any $X_i \in \der(\algA)$, any $\omega \in \underline{\Omega}^p_\der(\algA)$ and any $\eta \in \underline{\Omega}^q_\der(\algA)$. Introduce on this graded algebra the differential $\hat{d} : \underline{\Omega}^n_\der(\algA) \rightarrow \underline{\Omega}^{n+1}_\der(\algA)$:
\begin{multline*}
\hat{d}\omega(X_1, \ldots , X_{n+1}) =  \sum_{i=1}^{n+1} (-1)^{i+1} X_i \omega( X_1, \ldots \omi{i} \ldots, X_{n+1}) \\
 + \sum_{1\leq i < j \leq n+1} (-1)^{i+j} \omega( [X_i, X_j], \ldots \omi{i} \ldots \omi{j} \ldots , X_{n+1})
\end{multline*}
Then $(\underline{\Omega}^\grast_\der(\algA), \hat{d})$ is a differential calculus on $\algA$.

This differential calculus is not \textsl{a priori} generated by $\algA$ in degree $0$. The differential calculus generated by $\algA$ in $(\underline{\Omega}^\grast_\der(\algA), \hat{d})$ is denoted by $(\Omega^\grast_\der(\algA), \hat{d})$.

For $\algA = C^\infty(M)$, the Lie algebra $\der(\algA)$ is the Lie algebra of vector fields on the manifold $M$, and this differential calculus (the two coincide here) is the de~Rham differential calculus.

For $\algA=M_n(\gC)$, the Lie algebra  $\der(\algA)$ identifies with $\ksl_n(\gC)$, and the differential calculus identifies with the Lie complex $M_n(\gC) \otimes \exter^\grast \ksl_n(\gC)^\ast$ for the adjoint representation of $\ksl_n(\gC)$ on $M_n(\gC)$.

See \cite{DuVi:88}, \cite{DuViKernMado:90a}, \cite{DuViKernMado:90b}, \cite{Mass:14}, \cite{Mass:15}, \cite{Mass:25} for more details, examples and applications.
\end{Example}

\subsection{Hochschild homology}

The Hochschild homology will not be presented here in its full generality. We refer to \cite{Pier:82}, \cite{Loda:98} and \cite{GersScha:88} (for instance) to get further developments. What will be presented here is the relation between Hochschild homology with values in the algebra itself and the differential calculi introduced above. These constructions are necessary to introduce and understand cyclic homology.

Let $\algA$ be an associative algebra, not necessarily unital. As usual we denote by $\algA_+ = \gC \oplus \algA$ its unitarization.

The Hochschild homology we are interested in is defined using the following bicomplex $CC_{\grast,\grast}^{(2)}(\algA)$ with only two non zero columns:

\newlength{\largeurdiagramme}
\setlength{\largeurdiagramme}%
{\widthof{$\xymatrix{ 
{\algA^{\otimes n+1}} & {\algA^{\otimes n+1}} \ar[l]_-{1-t}}\quad\quad$}}

\medskip
\noindent\begin{minipage}[c]{\largeurdiagramme}
$\xymatrix{ 
{\vdots} \ar[d]_-{b} & {\vdots} \ar[d]_-{-b'} \\
{\algA^{\otimes n+1}} \ar[d]_-{b} & {\algA^{\otimes n+1}} \ar[l]_-{1-t} \ar[d]_-{-b'} \\
{\algA^{\otimes n}} \ar[d]_-{b} & {\algA^{\otimes n}} \ar[l]_-{1-t} \ar[d]_-{-b'} \\
{\vdots} \ar[d]_-{b} & {\vdots} \ar[d]_-{-b'} \\
{\algA} & {\algA} \ar[l]_-{1-t}
}$
\end{minipage}%
\begin{minipage}[c]{\textwidth-\largeurdiagramme}
$t : \algA^{\otimes n} \rightarrow \algA^{\otimes n}$ is the cyclic operator:
\begin{equation*}
t(a_1\otimes \cdots \otimes a_n) = (-1)^{n+1}a_n\otimes a_1\otimes\cdots \otimes a_{n-1}
\end{equation*}

$b : \algA^{\otimes n+1} \rightarrow \algA^{\otimes n}$ is the Hochschild boundary for the Hochschild complex with values in $\algA$:
\begin{multline*}
b(a_0\otimes \dots \otimes a_n) = \sum_{p=0}^{n-1} (-1)^p a_0\otimes \dots \otimes a_p a_{p+1} \otimes \dots \otimes a_n \\ 
 + (-1)^n a_n a_0 \otimes a_1 \otimes \dots \otimes a_{n-1} 
\end{multline*}

$b' : \algA^{\otimes n+1} \rightarrow \algA^{\otimes n}$ is the first part of the Hochschild boundary $b$:
\begin{equation*}
b'(a_0\otimes \dots \otimes a_n) = \sum_{p=0}^{n-1} (-1)^p a_0\otimes \dots \otimes a_p a_{p+1} \otimes \dots \otimes a_n
\end{equation*}
\end{minipage}

\medskip
One can show the relations
\begin{align*}
b^2=b'^2 &=0
&
b(1-t) &= (1-t)b'
\end{align*}

The total complex of this bicomplex is given in degree $n$ by $CC_{n}^{(2)}(\algA) = \algA^{\otimes n+1} \oplus \algA^{\otimes n}$, with the total differential 
\begin{equation*}
b_H= \begin{pmatrix}
b & 1-t \\
0 & -b'
\end{pmatrix}
\end{equation*}
in matrix form.

Now, notice that $CC_{n}^{(2)}(\algA) = \algA^{\otimes n+1} \oplus \algA^{\otimes n} = \algA_+ \otimes \algA^{\otimes n} = \Omega^n(\algA)$ in degree $n \geq 1$ and $CC_{0}^{(2)}(\algA) = \algA$. In this identification, the differential $b_H$ takes the very simple expression
\begin{equation*}
b_H(\omega da) = (-1)^{n} [\omega, a]
\end{equation*}

\begin{Definition}[Hochschild homology with values in the algebra]
Let $\algA$ be an associative algebra. The Hochschild homology $HH_\grast(\algA)$ is the homology of the total complex of the bicomplex $CC_{\grast,\grast}^{(2)}(\algA)$ defined above, \textsl{i.e.} the homology of the complex $(\Omega^\grast(\algA), b_H)$.
\end{Definition}

This second complex takes the form
\begin{equation*}
\xymatrix@1{ {\Omega^0(\algA)} & {\cdots} \ar[l]_-{b_H} & {\Omega^n(\algA)} \ar[l]_-{b_H} & {\Omega^{n+1}(\algA)} \ar[l]_-{b_H} & {\cdots} \ar[l]_-{b_H} }
\end{equation*}
Notice that $b_H$ is of degree $-1$, but the differential, which has not appeared in this construction, is of degree $1$.

\begin{Remark}[The unital case]
When the algebra is unital, the second column (the one with $b'$) is exact: it admits the homotopy 
\begin{equation}
\label{eq-shomotopyhochschildunital}
s(a_1\otimes \cdots \otimes a_n) = \bbbone \otimes a_1\otimes \cdots \otimes a_n 
\end{equation}
Using standard spectral sequence arguments on this bicomplex, the homology of the total complex is then the homology of the first column. In this case, one recovers the definition of the Hochschild complex which is usually given in textbooks:
\begin{equation}
\label{eq-usualhochschildcomplex}
\xymatrix@1{ {\algA} & {\cdots} \ar[l]_-{b} & {\algA^{\otimes n}} \ar[l]_-{b} & {\algA^{\otimes n+1}} \ar[l]_-{b} & {\cdots} \ar[l]_-{b} }
\end{equation}

One can even go a step further. It is possible to consider a quotient of this complex, called the normalized complex, and to show, by standard arguments coming from the theory of simplicial modules, that its homology is the same as the homology of the previous complex. 

This normalized complex is defined by removing any contributions coming from elements proportional to the unit in the last $n$ factors in $\algA^{\otimes n+1}$. The first factor is not affected because it is in fact the $\algA$-bimodule in which the Hochschild homology takes its values. The normalized complex is then defined on the spaces $\algA \otimes \overline{\algA}^{\otimes n}$ (where as before $\overline{\algA} = \algA/\gC\bbbone$) on which it is easy to check that the differential $b$ is well-defined. But now, one has the identification $\Omega_U^n(\algA) = \algA \otimes \overline{\algA}^{\otimes n}$, so that in the unital case, the Hochschild homology can be computed from the complex
\begin{equation}
\label{eq-hochschildcomplexuniversalunitaldifferentialcalculus}
\xymatrix@1{ {\Omega_U^0(\algA)} & {\cdots} \ar[l]_-{b} & {\Omega_U^n(\algA)} \ar[l]_-{b} & {\Omega_U^{n+1}(\algA)} \ar[l]_-{b} & {\cdots} \ar[l]_-{b} }
\end{equation}

\end{Remark}

\begin{Definition}[The trace map]
\label{def-tracemap}
The trace map $\tr : CC_{n}^{(2)}(M_n(\algA)) \rightarrow CC_{n}^{(2)}(\algA)$ is the morphism of complexes defined by
\begin{equation*}
\tr(\alpha_0 \otimes \cdots \otimes \alpha_n) = \sum_{(i_0, \dots , i_n)} a_{0,i_0 i_1} \otimes \dots \otimes a_{n,i_n i_0}
\end{equation*}
where $\alpha_r = (a_{r,i j})_{i,j} \in M_n(\algA)$.
\end{Definition}

\begin{Proposition}[Morita invariance of Hochschild homology]
For any unital algebra $\algA$ and any integer $n$, the trace map induces an isomorphism
\begin{equation*}
HH_\grast(M_n(\algA)) \simeq HH_\grast(\algA)
\end{equation*}
\end{Proposition}

In fact, Morita invariance of the Hochschild homology of unital algebras is stronger than the one presented here. It is invariant for the Morita equivalence defined which we now define:
\begin{Definition}[Morita equivalence of algebras]
One says that two algebras $\algA$ and $\algB$ are Morita equivalent if there exist an $\algA$-$\algB$-module $\modM$ and a $\algB$-$\algA$-module $\modN$ such that $\algA \simeq \modM \otimes_{\algB} \modN$ and $\algB \simeq \modN \otimes_{\algA} \modM$ as bimodules over $\algA$ and $\algB$ respectively.
\end{Definition}

For instance, $\algA$ is Morita equivalent to $\algB = M_n(\algA)$ using $\modM = \algA^n$ written as a row and $\modN = \algA^n$ written as a column.

Morita invariance of the Hochschild homology can be extended to the class of $H$-unital algebras, which contains the unital algebras. 

\begin{Definition}[$H$-unital algebras]
\label{def-Hunitalalgebras}
A $H$-unital algebra is an algebra $\algA$ for which the complex $(\algA^{\otimes \grast}, b')$ has trivial homology.
\end{Definition}

\begin{Example}[The algebra $\gC$]
\label{ex-thealgebraCHochschildhomology}
In the case $\algA=\gC$, one has
\begin{align*}
HH_0(\gC) &= \gC & HH_n(\gC) &= \algzero \text{ for $n \geq 1$}
\end{align*}
\end{Example}

\begin{Example}[Tensor algebra]
\label{ex-tensoralgebra}
Let $\evV$ be a finite dimensional vector space and $\algA = \caT \evV$ the tensor algebra over $\evV$. Denote by $t$ the cyclic permutation acting on $\algA$ in each degree (the $t$ defining the bicomplex). Then
\begin{align*}
HH_0(\algA) &= \oplus_{m \geq 0} \left( \evV^{\otimes m} / \Ran (1 - t) \right)
& \text{co-invariants under the action of $t$}\\
HH_1(\algA) &= \oplus_{m \geq 1} \left( \evV^{\otimes m} \right)^t
& \text{invariants under the action of $t$} \\
HH_n(\algA) &= \algzero & \text{for $n \geq 2$}
\end{align*}
\end{Example}

\begin{Example}[Relation with Lie algebra homology]
\label{ex-relationwithLiealgebrahomology}
Any associative algebra $\algA$ gives rise to a Lie algebra $\algA_\lie$ where the vector space is $\algA$ and the Lie bracket is the commutator: $[a,b] = ab - ba$. In the following, $\algA$ is supposed to be unital.

The permutation group $\kS_n$ acts on $\algA^{\otimes n}$ by $\sigma(a_1 \otimes \cdots \otimes a_n) = a_{\sigma^{-1}(1)} \otimes \cdots \otimes a_{\sigma^{-1}(n)}$. Let us define $\epsilon_n = \sum_{\sigma \in \kS_n} (-1)^{\sign(\sigma)} \sigma : \algA^{\otimes n} \rightarrow \algA^{\otimes n}$ the total antisymmetrisation. It induces a natural morphism
\begin{align*}
\epsilon_n : \exter^n \algA & \rightarrow \algA^{\otimes n} \\
a_1 \wedge \cdots \wedge a_n & \mapsto \epsilon_n(a_1 \otimes \cdots \otimes a_n)
\end{align*}
which can be shown to commute with the boundary $\partial$ of the Lie algebra complex $\exter^\grast \algA_\lie$ and the boundary $b$ of the Hochschild complex, so that the morphism of differential complexes $\epsilon_\grast : (\exter^\grast \algA_\lie, \partial) \rightarrow (\algA^{\otimes \grast}, b)$ induces a morphism in homologies
\begin{equation*}
\epsilon_\indast : H_\grast(\exter^\grast \algA_\lie, \partial) \rightarrow HH_\grast(\algA)
\end{equation*}

If one consider the universal enveloping algebra $\Ug$ of a finite dimensional Lie algebra $\kg$, one can show that $HH_\grast(\Ug) \simeq H_\grast(\kg; \Ug)$ where on the right it is the ordinary Lie algebra homology defined with the complex $(\exter^\grast \kg, \partial)$.
\end{Example}

\begin{Example}[The commutative case]
\label{ex-thecommutativecaseHochschildhomology}
Let us suppose that $\algA$ is a commutative unital algebra. Consider the constructions of Example~\ref{ex-relationwithLiealgebrahomology}. Here the Lie structure on $\algA_\lie$ is trivial, so that $\partial = 0$, and the morphism $\epsilon_\indast$ is in fact a morphism $\epsilon_\indast : \exter^\grast \algA \rightarrow HH_\grast(\algA)$.

One can show that there is a natural map
\begin{align*}
\exter^n \algA &\rightarrow \Omega^n_{\algA|\gC}\\
a_1 \wedge \cdots \wedge a_n & \mapsto da_1 \wedge \cdots \wedge da_n
\end{align*}
through which $\epsilon_\indast$ factors. One then get a natural map (also denoted by $\epsilon_\indast$):
\begin{align*}
\epsilon_\indast : \Omega^\grast_{\algA|\gC} &\rightarrow HH_\grast(\algA)\\
da_1 \wedge \cdots \wedge da_n & \mapsto [\epsilon_n(a_1 \otimes \cdots \otimes a_n)]
\end{align*}
where on the right hand side the brackets mean the homology class.
\end{Example}

\begin{Example}[Polynomial algebra]
For a finite dimensional vector space $\evV$ and the commutative unital algebra $\symes \evV$ of polynomials on $\evV$, one has
\begin{equation*}
HH_\grast(\symes \evV) = \symes \evV \otimes \exter^\grast \evV = \Omega^\grast_{\symes \evV|\gC}
\end{equation*}

\end{Example}

This is a particular situation of a more general theorem for which we need the following definition:
\begin{Definition}[Smooth algebras]
A commutative algebra $\algA$ is a smooth algebra if for any algebra $\algB$ and any ideal $\algI$ in $\algB$ such that $\algI^2=0$, the map $\Hom_\gC(\algA, \algB) \rightarrow \Hom_\gC(\algA, \algB/\algI)$ is surjective. This means that every morphism of algebras $\algA \rightarrow \algB/\algI$ can be lifted to a morphism of algebras $\algA \rightarrow \algB$.
\end{Definition}

Then one has the following result:
\begin{Theorem}[Hochschild-Kostant-Rosenberg]
\label{thm-HochschildKostantRosenberg}
For any unital smooth commutative algebra $\algA$, the map $\epsilon_\indast : \Omega^\grast_{\algA|\gC} \rightarrow HH_\grast(\algA)$ of Example~\ref{ex-thecommutativecaseHochschildhomology} is an isomorphism of graded commutative algebras:
\begin{equation*}
HH_\grast(\algA) \simeq \Omega^\grast_{\algA|\gC}
\end{equation*}
\end{Theorem}

The natural map which identifies a differential forms in $\Omega^n(\algA)$ to a differential form in $\Omega^n_{\algA|\gC}$ is explicitly given by $a_0 da_0 \cdots d a_n \mapsto \frac{1}{n!} a_0 d_K a_0 \wedge \cdots \wedge d_K a_n$. Notice the extra factor $\frac{1}{n!}$ compared to the universal projection $\pi_n$ of Example~\ref{ex-Kahlerdifferentialcalculusforcommutativeunitalalgebra}. This factor is required to get a further identification of the differential on the Kähler differential calculus with the $B$ operator in cyclic homology (see \cite{Loda:98}) and to get a morphism of graded commutative algebras. 

\begin{Remark}[Extension to topological algebras]
One can generalize the definition of the Hochschild homology given above to take into account some topological structure on the algebra $\algA$. In order to do that, one defines the spaces $\algA^{\otimes n}$ using a tensor product adapted to the topological structure on the algebra. The Hochschild homology one obtains in this way is called the continuous Hochschild homology.

For Fréchet algebras, such a continuous homology is well defined and leads to the next two very interesting examples.
\end{Remark}

\begin{Example}[The Fréchet algebra $C^\infty(M)$]
\label{ex-thefrechetalgebraCinfty(M)}
Let $M$ be a $C^\infty$ finite dimensional locally compact manifold. Then Connes computed its continuous Hochschild homology in \cite{Conn:85} and found the following result which generalizes the Hochschild-Kostant-Rosenberg theorem:
\begin{equation*}
HH_\grast^\text{Cont}(C^\infty(M)) = \Omega^\grast_\gC(M) \text{ (complexified de~Rham forms)}
\end{equation*}
For reasons that will be explained later, this isomorphism between vector spaces, which we denote by $\phi$, is explicitly given in terms of universal forms by 
\begin{align*}
\Omega^{2k}(C^\infty(M)) & \rightarrow \Omega^{2k}_\gC(M)
&
\Omega^{2k+1}(C^\infty(M)) & \rightarrow \Omega^{2k+1}_\gC(M)\\
\omega &\mapsto \left(\frac{i}{2 \pi}\right)^k \frac{1}{(2k)!} \pi_{2k}(\omega)
&
\omega &\mapsto \left(\frac{i}{2 \pi}\right)^{k+1} \frac{1}{(2k+1)!} \pi_{2k+1}(\omega)
\end{align*}
where $\pi_\grast : \Omega^{\grast}(C^\infty(M)) \rightarrow \Omega^\grast_\gC(M)$ is the universal map defined in Example~\ref{ex-Kahlerdifferentialcalculusforcommutativeunitalalgebra}.
\end{Example}

\begin{Example}[The irrational rotation algebra]
\label{ex-theirrationalrottionalgebraHochschimdhomology}
The continuous Hochschild homology of the Fréchet algebra $\tncA^\infty_\theta$ ($\theta$ irrational) has been computed by Connes in \cite{Conn:85}. Let $\lambda = \exp(2 i \pi \theta)$.

If $\lambda$ satisfies some diophantine condition (there exists an integer $k$ such that $|1 - \lambda^n|^{-1}$ is $O(n^k)$), then
\begin{align*}
HH^\text{Cont}_0(\tncA^\infty_\theta) &= \gC &
HH^\text{Cont}_1(\tncA^\infty_\theta) &= \gC^2 
\end{align*}
For any $\lambda$:
\begin{align*}
HH^\text{Cont}_2(\tncA^\infty_\theta) &= \gC &
HH^\text{Cont}_n(\tncA^\infty_\theta) &= \algzero \text{ for $n \geq 3$} 
\end{align*}
If $\lambda$ does not satisfy some diophantine condition, $HH^\text{Cont}_0(\tncA^\infty_\theta)$ and $HH^\text{Cont}_1(\tncA^\infty_\theta)$ are infinite dimensional.
\end{Example}

\begin{Definition}[Hochschild cohomology]
\label{def-Hochschildcohomology}
Recall that the dual $\algA^\ast$ of an algebra $\algA$ is a bimodule on $\algA$ for the definition $(a\phi b)(c) = \phi(bca)$ for any $\phi \in \algA$ and $a,b,c \in \algA$. The Hochschild complex $(C^\grast(\algA), \delta)$ for the Hochschild cohomology with values in the bimodule $\algA^\ast$ is defined by 
\begin{equation*}
C^n(\algA) = \Hom(\algA^{\otimes n}, \algA^\ast) = \Hom(\algA^{\otimes n+1}, \gC)
\end{equation*}
and by
\begin{multline*}
\delta\phi(a_0 \otimes a_1 \otimes \cdots \otimes a_{n+1}) = \sum_{p=0}^n (-1)^p \phi(a_0 \otimes a_1 \otimes \cdots \otimes a_p a_{p+1} \otimes \cdots \otimes a_{n+1})\\
+ (-1)^{n+1} \phi( a_{n+1} a_0 \otimes a_1 \otimes \cdots \otimes a_n)
\end{multline*}
By construction, $\delta$ is the adjoint to $b$ in homology.

The cohomology of this complex is denoted by $HH^\grast(\algA)$.
\end{Definition}

\subsection{Cyclic homology}

Cyclic homology is defined using a bicomplex $CC_{\grast, \grast}(\algA)$ constructed using the bicomplex $CC_{\grast,\grast}^{(2)}(\algA)$ of the Hochschild homology. In order to do that, we need a new operator.

$N : \algA^{\otimes n} \rightarrow \algA^{\otimes n}$ is the norm operator defined by 
\begin{equation*}
N = 1 +t + \cdots + t^n
\end{equation*}
Then one has the relations
\begin{align*}
(1-t)N = N(1-t) &= 0
&
b'N &= Nb
\end{align*}
The bicomplex $CC_{\grast, \grast}(\algA)$ is a repetition of the bicomplex $CC_{\grast,\grast}^{(2)}(\algA)$ infinitely on the right, using $N$ to connect them. In terms of the algebra $\algA$, one has
\begin{equation*}
\xymatrix{ 
{\vdots} \ar[d]_-{b} & {\vdots} \ar[d]_-{-b'} & {\vdots} \ar[d]_-{b} & {\vdots} \ar[d]_-{-b'} & \\
{\algA^{\otimes n+1}} \ar[d]_-{b} & {\algA^{\otimes n+1}} \ar[l]_-{1-t} \ar[d]_-{-b'} &{\algA^{\otimes n+1}} \ar[l]_-{N} \ar[d]_-{b} & {\algA^{\otimes n+1}} \ar[l]_-{1-t} \ar[d]_-{-b'} & {\cdots} \ar[l]_-{N} \\
{\algA^{\otimes n}} \ar[d]_-{b} & {\algA^{\otimes n}} \ar[l]_-{1-t} \ar[d]_-{-b'} &{\algA^{\otimes n}} \ar[l]_-{N} \ar[d]_-{b} & {\algA^{\otimes n}} \ar[l]_-{1-t} \ar[d]_-{-b'} & {\cdots} \ar[l]_-{N} \\
{\vdots} \ar[d]_-{b} & {\vdots} \ar[d]_-{-b'} &{\vdots} \ar[d]_-{b} & {\vdots} \ar[d]_-{-b'} & \\
{\algA} & {\algA} \ar[l]_-{1-t} &{\algA} \ar[l]_-{N} & {\algA} \ar[l]_-{1-t} & {\cdots} \ar[l]_-{N} 
}
\end{equation*}

\begin{Definition}[Cyclic homology]
Let $\algA$ be an associative algebra. The cyclic homology $HC_\grast(\algA)$ of $\algA$ is the homology of the total complex of the bicomplex $CC_{\grast,\grast}(\algA)$ defined above.
\end{Definition}

Any morphism of algebras $\varphi : \algA \rightarrow \algB$ induces a natural map of bicomplexes $CC_{\grast,\grast}(\algA) \rightarrow CC_{\grast,\grast}(\algB)$, so that one gets an induced map in cyclic homology $\varphi_\indast : HC_\grast(\algA) \rightarrow HC_\grast(\algB)$.

\begin{Remark}[The Connes complex]
In \cite{Conn:85}, Connes introduced cyclic cohomology, a dual version of cyclic homology. The way he introduced it did not rely on a bicomplex, but on a subcomplex of the Hochschild complex for cohomology. Some details of this construction are given in Example~\ref{ex-firstversionofcycliccohomology}. In a dual version, one can define the Connes complex to compute cyclic homology as a quotient of the Hochschild complex for homology for a unital algebra.

To the bicomplex defined above, add a column on the left whose spaces are the cokernels of the morphisms $(1-t) : \algA^{\otimes n+1} \rightarrow \algA^{\otimes n+1}$, which we denote by $C_n^\lambda(\algA) = \algA^{\otimes n+1}/\Ran (1-t)$. One can then check that the operator $b$ is a well-defined operator on $C_\grast^\lambda(\algA) = \oplus_{n\geq 0} C_n^\lambda(\algA)$ which satisfies $b^2=0$. Denote by $H_\grast^\lambda(\algA)$ the homology of this complex. The total complex $TCC_\grast(\algA)$ of $CC_{\grast,\grast}(\algA)$ projects onto the complex $C_\grast^\lambda(\algA)$, sending the column $p=0$ onto $C_\grast^\lambda(\algA)$ and the other columns onto $\algzero$. One then gets a morphism in homology
\begin{equation*}
HC_\grast(\algA) \rightarrow H_\grast^\lambda(\algA)
\end{equation*}
When the field over which the algebra is defined contains $\gQ$, this is an isomorphism. To show that, one introduces an explicit homotopy for the horizontal operators which shows that the horizontal homology of $CC_{\grast,\grast}(\algA)$ is trivial. By standard arguments on bicomplexes, this proves the assertion.
\end{Remark}

\begin{Remark}[The horizontal homology of $CC_{\grast,\grast}(\algA)$]
One can show that for any algebra $\algA$, the homology of any row of $CC_{\grast,\grast}(\algA)$ is the group homology $H_\grast(\kC_{n+1}; \algA^{\otimes n+1})$ of the cyclic group $\kC_{n+1}$ with values in the $\kC_{n+1}$-module $\algA^{\otimes n+1}$ (for the action induced by $t$).
\end{Remark}

We have seen that the total complex of $CC_{\grast,\grast}^{(2)}(\algA)$ can be written in terms of the universal differential calculus $\Omega^\grast(\algA)$ with the boundary operator $b_H$. We can do something similar here. Every grouping of two columns isomorphic to $CC_{\grast,\grast}^{(2)}(\algA)$ can be ``compressed'' as we did for the Hochschild bicomplex. The operators $b,b'$ and $(1-t)$ are then replaced by the unique operator $b_H : \Omega^n(\algA) \rightarrow \Omega^{n-1}(\algA)$. The operator $N$ is replaced by a new operator $B : \Omega^n(\algA) \rightarrow \Omega^{n+1}(\algA)$, which takes the matrix form
\begin{equation*}
B= \begin{pmatrix}
0 & 0 \\
N & 0
\end{pmatrix}
\end{equation*}
in the decomposition $\Omega^n(\algA) = \algA^{\otimes n+1} \oplus \algA^{\otimes n}$. In order to have a pleasant diagram representing the new bicomplex, lift vertically each column on the right in proportion to its degree in the horizontal direction. We then get the following (triangular) bicomplex
\begin{equation*}
\xymatrix{ 
{\vdots} \ar[d]_-{b_H} & {\vdots} \ar[d]_-{b_H} & & {\vdots} \ar[d]_-{b_H} & {\vdots} \ar[d]_-{b_H} \\
{\Omega^{n+1}(\algA)} \ar[d]_-{b_H} & {\Omega^{n}(\algA)} \ar[l]_-{B} \ar[d]_-{b_H} & {\cdots} \ar[l]_-{B} & {\Omega^{1}(\algA)} \ar[d]_-{b_H} \ar[l]_-{B} & {\Omega^{0}(\algA)} \ar[l]_-{B} \\
{\Omega^{n}(\algA)} \ar[d]_-{b_H} & {\Omega^{n-1}(\algA)} \ar[l]_-{B} \ar[d]_-{b_H} & {\cdots} \ar[l]_-{B} & {\Omega^{0}(\algA)} \ar[l]_-{B} & \\
{\vdots} \ar[d]_-{b_H} & {\vdots} \ar[d]_-{b_H} & {\vdots} \ar[d]_-{b_H} & & \\
{\Omega^{2}(\algA)} \ar[d]_-{b_H} & {\Omega^{1}(\algA)} \ar[d]_-{b_H} \ar[l]_-{B} & {\Omega^{0}(\algA)} \ar[l]_-{B} & & \\
{\Omega^{1}(\algA)} \ar[d]_-{b_H} & {\Omega^{0}(\algA)} \ar[l]_-{B} & & & \\
{\Omega^{0}(\algA)} & & & & 
}
\end{equation*}
The total homology of this bicomplex is again the cyclic homology of $\algA$.

\begin{Definition}[Mixed bicomplex]
A mixed bicomplex is a $\gN$-graded complex $M_\grast = \bigoplus_{n \geq 0} M_n$ equipped with a differential $b_M$ of degree $-1$ and a differential $B_M$ of degree $+1$ such that 
\begin{equation*}
b_M B_M + B_M b_M = 0
\end{equation*}

The homology of the complex $(M, b_M)$ is called the Hochschild homology of the mixed bicomplex, and it is denoted by $HH_\grast M = H_\grast(M, b_M)$.

We associate to such a mixed bicomplex the $\gN$-graded complex $\widetilde{M}_\grast$ defined by 
\begin{equation*}
\widetilde{M}_n = \bigoplus_{p \geq 0} M_{n-2p}
\end{equation*}
on which we introduce the differential operator $B'_M - b_M$ where $B'_M : \widetilde{M}_n \rightarrow \widetilde{M}_{n-1}$ is $B_M$ on $M_{n-2p}$ such that $0< 2p \leq n$ and $0$ on $M_n$. The cyclic homology of the mixed bicomplex is the homology of this differential complex: $HC_\grast M = H_\grast(\widetilde{M}, B'_M - b_M)$

Two mixed bicomplexes $(M_\grast, b_M, B_M)$ and $(N_\grast, b_N, B_N)$ are said to be $b$-quasi-isomorphic if there exists a morphism of mixed bicomplexes $\varphi :(M_\grast, b_M, B_M) \rightarrow (N_\grast, b_N, B_N)$ ($\varphi$ is of degree $0$ and commutes with the $b$'s and $B$'s) which induces an isomorphism in Hochschild homology.
\end{Definition}

\begin{Proposition}
\label{prop-bquasiisomorphismHC}
Two $b$-quasi-isomorphic mixed bicomplexes have the same cyclic homology.
\end{Proposition}

\begin{Example}[The mixed bicomplex $(\Omega^\grast(\algA), b_H, B)$]
\label{ex-mixedbicomplexeOmegaAbHB}
The motivation for the above definition is the example of the $\gN$-graded module $\Omega^\grast(\algA)$ with two differentials $b_H$ and $B$. The definitions of Hochschild and cyclic homology reproduce the ones we introduced before.
\end{Example}

\begin{Example}[The mixed bicomplex $(HH_\grast(\algA), 0, B_\indast)$]
\label{ex-mixedbicomplexHHA0B}
Because the differential $B$ commutes with the differential $b_H$, it defines a morphism $B_\indast$ on the Hochschild homology $HH_\grast(\algA)$. With this induced morphism, the triplet $(HH_\grast(\algA), 0, B_\indast)$ is a mixed bicomplex, whose Hochschild homology is the Hochschild homology of $\algA$. Indeed, taking Hochschild homology maps $b_H$ to the zero operator.

Now, using standard argument on the spectral sequence constructed on the filtration by vertical degree, one can see that this mixed complex computes the cyclic homology of $\algA$.
\end{Example}

\begin{Remark}[Some ideas to compute cyclic homology]
\label{rmk-awaytocomputecyclichomology}
Example~\ref{ex-mixedbicomplexHHA0B} teaches us that in order to compute cyclic homology, one can first compute Hochschild homology, then look at the operator $B_\indast$ induced by $B$, and compute the $B_\indast$-homology. Many examples of concrete computations of cyclic homology are performed this way. Obviously, this supposes that Hochschild homology is computable!

Another approach is to consider the simplest possible differential complex which computes the Hochschild homology of the algebra, and then guess a $B$ operator on it in order to build a mixed bicomplex $b$-quasi-isomorphic to one of the standard mixed bicomplexes given here. Because Hochschild homology can be defined through projective resolutions, such simple differential complexes are usually possible to find.
\end{Remark}

\begin{Example}[Mixed bicomplexes for the unital case]
Let us suppose now that the algebra $\algA$ is unital. Then we know that the Hochschild homology can be computed with the complex \eqref{eq-usualhochschildcomplex}. Using the ideas of Remark~\ref{rmk-awaytocomputecyclichomology}, one can find a $B$ operator on this complex in order to make it into a mixed bicomplex. 

This operator is defined by $B= (1-t) s N : \algA^{\otimes n} \rightarrow \algA^{\otimes (n+1)}$ where $t$ and $N$ have been introduced before, and $s: \algA^{\otimes n} \rightarrow \algA^{\otimes (n+1)}$ is the homotopy \eqref{eq-shomotopyhochschildunital}. Explicitly, one has
\begin{multline*}
B(a_0 \otimes a_1 \otimes \cdots \otimes a_n) = \sum_{p=0}^{n-1} \big[ (-1)^{np} \bbbone \otimes a_p \otimes \cdots \otimes a_n \otimes a_0 \otimes \cdots \otimes a_{p-1} \\
 - (-1)^{n(p-1)} a_{p-1}\otimes \bbbone \otimes a_p \otimes \cdots \otimes a_n \otimes a_0 \otimes \cdots \otimes a_{p-2} \big]
\end{multline*}
In low degrees, these expressions take the following forms
\begin{align*}
B(a_0) &= \bbbone \otimes a_0 + a_0 \otimes \bbbone \\
B(a_0 \otimes a_1) &= (\bbbone \otimes a_0 \otimes a_1 - \bbbone \otimes a_1 \otimes a_0) + (a_0 \otimes \bbbone \otimes a_1 - a_1 \otimes \bbbone \otimes a_0)
\end{align*}

This mixed bicomplex $(\algA^{\otimes (\grast+1)}, b, B)$ is represented by the diagram
\begin{equation*}
\xymatrix{ 
{\vdots} \ar[d]_-{b} & {\vdots} \ar[d]_-{b} & {\vdots} \ar[d]_-{b} \\
{\algA^{\otimes 3}} \ar[d]_-{b} & {\algA^{\otimes 2}} \ar[d]_-{b} \ar[l]_-{B} & {\algA} \ar[l]_-{B} \\
{\algA^{\otimes 2}} \ar[d]_-{b} & {\algA} \ar[l]_-{B} & \\
{\algA} & & 
}
\end{equation*} 

Now, one can perform this procedure with the complex \eqref{eq-hochschildcomplexuniversalunitaldifferentialcalculus}. Then one obtains the same operator $B$ and the mixed bicomplex $(\Omega_U^\grast(\algA), b, B)$ which is $b$-quasi-isomorphic to the mixed bicomplex of Example~\ref{ex-mixedbicomplexeOmegaAbHB} through the natural projection $\Omega^\grast(\algA) \rightarrow \Omega_U^\grast(\algA)$.
\end{Example}

\begin{Example}[The commutative case]
\label{ex-thecommutativecasecyclichomology}
Let us consider the notations and results of Example~\ref{ex-thecommutativecaseHochschildhomology}, where the algebra is over the field $\gC$. One can introduce the mixed bicomplex $(\Omega^{\grast}_{\algA|\gC}, 0, d_K)$ based on the Kähler differential calculus, which takes the diagrammatic form
\begin{equation*}
\xymatrix{ 
{\vdots} \ar[d]_-{0} & {\vdots} \ar[d]_-{0} & {\vdots} \ar[d]_-{0} \\
{\Omega^{2}_{\algA|\gC}} \ar[d]_-{0} & {\Omega^{1}_{\algA|\gC}} \ar[d]_-{0} \ar[l]_-{d_K} & {\Omega^{0}_{\algA|\gC}} \ar[l]_-{d_K} \\
{\Omega^{1}_{\algA|\gC}} \ar[d]_-{0} & {\Omega^{0}_{\algA|\gC}} \ar[l]_-{d_K} & \\
{\Omega^{0}_{\algA|\gC}} & & 
}
\end{equation*}
One can show that there is a natural morphism of mixed bicomplexes $(\Omega_U^\grast(\algA), b, B) \rightarrow (\Omega^{\grast}_{\algA|\gC}, 0, d_K)$, so that there is a natural map
\begin{equation*}
HC_n(\algA) \rightarrow \Omega^{n}_{\algA|\gC}/d_K\Omega^{n-1}_{\algA|\gC} \oplus H_\text{dR}^{n-2}(\algA) \oplus H_\text{dR}^{n-4}(\algA) \oplus \cdots
\end{equation*}
the last term being $H_\text{dR}^{0}(\algA)$ or $H_\text{dR}^{1}(\algA)$ depending on the parity of $n$.

Using Theorem~\ref{thm-HochschildKostantRosenberg}, for any smooth unital commutative algebra $\algA$, one has the isomorphism
\begin{equation*}
HC_{\grast}(\algA) \simeq \Omega^{\grast}_{\algA|\gC}/d_K\Omega^{\grast-1}_{\algA|\gC} \oplus H_\text{dR}^{\grast-2}(\algA) \oplus H_\text{dR}^{\grast-4}(\algA) \oplus \cdots
\end{equation*}
In particular, this is the case for the polynomial algebra $\algA = \symes \evV$ over a finite dimensional vector space $\evV$.
\end{Example}

There are a lot of structural properties on the cyclic homology groups which help a lot to compute them. We refer to \cite{Loda:98} to explore them. Let us just mention the following result:
\begin{Proposition}[Connes long exact sequence]
There are morphisms $I$ and $S$ which induce the following long exact sequence
\begin{equation*}
\xymatrix@1{ {\cdots} \ar[r] & {HH_n(\algA)} \ar[r]^-{I} & {HC_n(\algA)} \ar[r]^-{S} & {HC_{n-2}(\algA)} \ar[r]^-{B} & {HH_{n-1}(\algA)} \ar[r]^-{I} & {\cdots} }
\end{equation*}
\end{Proposition}

In low degrees, one gets
\begin{equation*}
\xymatrix@1{ {\cdots} \ar[r] & {HH_2(\algA)} \ar[r]^-{I} & {HC_2(\algA)} \ar[r]^-{S} & {HC_{0}(\algA)} \ar[r]^-{B} & {HH_{1}(\algA)} \ar[r]^-{I} & {HC_1(\algA)} \ar[r]^-{S} & {\algzero} }
\end{equation*}
and the isomorphism
\begin{equation*}
\xymatrix@1{ {\algzero} \ar[r]^-{B} & {HH_0(\algA)} \ar[r]^-{I} & {HC_0(\algA)} \ar[r]^-{S} & {\algzero} }
\end{equation*}

This long exact sequence is a direct consequence of the fact that the bicomplex $CC^{(2)}_{\grast, \grast}(\algA)$ is included as pairs of columns in the bicomplex $CC_{\grast, \grast}(\algA)$. This inclusion gives rise to the short exact sequence of bicomplexes
\begin{equation*}
\xymatrix@1{ {\algzero} \ar[r] & {CC^{(2)}_{\grast, \grast}(\algA)} \ar[r]^-{I} & {CC_{\grast, \grast}(\algA)} \ar[r]^-{S} & {CC_{\grast-2, \grast}(\algA)} \ar[r] & {\algzero} }
\end{equation*}
which defines $I$ and $S$. In homology this short exact sequence produces Connes long exact sequence. The morphism $S$ is called the periodic morphism.

\begin{Example}[$HC_\grast(\gC)$]
Using the results of Example~\ref{ex-thealgebraCHochschildhomology} and the mixed bicomplex of Example~\ref{ex-mixedbicomplexHHA0B}, one easily gets
\begin{align*}
HC_{2n}(\gC) &= \gC & HC_{2n+1}(\gC) &= \algzero
\end{align*}
Using Connes long exact sequence, one has an isomorphism $S : HC_n(\gC) \rightarrow HC_{n-2}(\gC)$. Denote by $u_n \in HC_{2n}(\gC) = \gC$ the canonical generator. Then, one can show that there is an isomorphism of coalgebras $HC_\grast(\gC) \xrightarrow{\simeq} \gC[u]$ explicitly given by $u_n \mapsto u^n$, where the coproduct on $\gC[u]$ is $\Delta(u^n) = \sum_{p+q=n} u^p \otimes u^q$.

For any algebra $\algA$, $HC_\grast(\algA)$ is a comodule over the coalgebra $HC_\grast(\gC)$:
\begin{align*}
HC_\grast(\algA) & \rightarrow HC_\grast(\algA) \otimes \gC[u]\\
x & \mapsto \sum_{p \geq 0} S^p(x) \otimes u^p
\end{align*}
where $S^p$ is the $p$-th iteration of $S$. This is a concrete interpretation of the periodic morphism $S$.
\end{Example}

Let us now define the periodic and negative cyclic homologies. In order to do that, let us introduce the bicomplex $CC_{\grast, \grast}^\text{per}(\algA)$, infinite in the two horizontal directions:
\begin{equation*}
\xymatrix{ 
{} & {\vdots} \ar[d]_-{b} & {\vdots} \ar[d]_-{-b'} & {\vdots} \ar[d]_-{b} & {\vdots} \ar[d]_-{-b'} & \\
{\cdots} & {\algA^{\otimes n+1}} \ar[d]_-{b} \ar[l]_-{N} & {\algA^{\otimes n+1}} \ar[l]_-{1-t} \ar[d]_-{-b'} &{\algA^{\otimes n+1}} \ar[l]_-{N} \ar[d]_-{b} & {\algA^{\otimes n+1}} \ar[l]_-{1-t} \ar[d]_-{-b'} & {\cdots} \ar[l]_-{N} \\
{\cdots} & {\algA^{\otimes n}} \ar[d]_-{b} \ar[l]_-{N} & {\algA^{\otimes n}} \ar[l]_-{1-t} \ar[d]_-{-b'} &{\algA^{\otimes n}} \ar[l]_-{N} \ar[d]_-{b} & {\algA^{\otimes n}} \ar[l]_-{1-t} \ar[d]_-{-b'} & {\cdots} \ar[l]_-{N} \\
& {\vdots} \ar[d]_-{b} & {\vdots} \ar[d]_-{-b'} &{\vdots} \ar[d]_-{b} & {\vdots} \ar[d]_-{-b'} & \\
{\cdots} & {\algA} \ar[l]_-{N} & {\algA} \ar[l]_-{1-t} &{\algA} \ar[l]_-{N} & {\algA} \ar[l]_-{1-t} & {\cdots} \ar[l]_-{N} \\
{p=} & {-2} & {-1} & {0} & {1} & {\cdots}
}
\end{equation*}
The bicomplex $CC_{\grast, \grast}(\algA)$ is naturally included in $CC_{\grast, \grast}^\text{per}(\algA)$ as the sub-bicomplex for which $p\geq 0$. Denote by $CC^-_{\grast, \grast}(\algA)$ the sub-bicomplex defined by $p\leq 1$.

\begin{Definition}[Periodic and negative cyclic homology]
The periodic cyclic homology $HP_\grast(\algA)$ of $\algA$ is the homology of the total complex (for the product) defined from $CC_{\grast, \grast}^\text{per}(\algA)$ by
\begin{equation*}
TCC_{n}^\text{per}(\algA) = \prod_{p+q=n} CC_{p, q}^\text{per}(\algA)
\end{equation*}
for any $n \in \gZ$.

The negative cyclic homology $HC^-_\grast(\algA)$ of $\algA$ is the homology of the total complex (for the product) defined from $CC^-_{\grast, \grast}(\algA)$ by
\begin{equation*}
TCC_{n}^-(\algA) = \prod_{\substack{p+q=n \\ (p \leq 1)}} CC_{p, q}^-(\algA)
\end{equation*}
\end{Definition}

Let us recall that in this situation, as the two bicomplexes we consider are infinite in the left direction, direct sum and product do not coincide. An element in the direct sum contains only a finite number of non zero elements in the spaces $CC_{p, q}^\text{per}(\algA)$ for $p+q=n$, but an element in the product can be non zero in all of these spaces. If we were using direct sums to define their total complexes, then it would be possible to show that the associated homologies were trivial if the base field contains $\gQ$. 

Using an adaptation of the procedure described for the cyclic homology, one can define the cyclic periodic homology of a mixed bicomplex, as well as its cyclic negative homology. Then one has:

\begin{Proposition}
\label{prop-bquasiisomorphismHPHCneg}
Two $b$-quasi-isomorphic mixed bicomplexes have the same cyclic periodic homology and the same cyclic negative homology.
\end{Proposition}

The natural inclusion $I : CC^-_{\grast, \grast}(\algA) \rightarrow CC_{\grast, \grast}^\text{per}(\algA)$ and the natural projection $p : CC_{\grast, \grast}^\text{per}(\algA) \rightarrow CC_{\grast, \grast}(\algA)$ induce morphisms in homology
\begin{align*}
I : HC^-_n(\algA) & \rightarrow HP_n(\algA)
&
p : HP_n(\algA) & \rightarrow HC_n(\algA)
\end{align*}

\begin{Proposition}[$2$-periodicity of $HP_\grast(\algA)$]
The periodic map $S$ defined on the periodic bicomplex by translating on the left through two columns is an isomorphism. It induces the natural $2$-periodicity:
\begin{equation*}
HP_n(\algA) \simeq HP_{n-2}(\algA)
\end{equation*}
which means that $HP_\grast(\algA)$ is $\gZ_2$-graded. 
\end{Proposition}

From now on, we will use the notation $HP_\nu(\algA)$, with $\nu=0,1$.  

We saw a $\gZ_2$-graded situation earlier for complex $K$-theory. Here is another similitude proved in \cite{CuntQuil:97}:
\begin{Proposition}[Six term exact sequence]
\label{prop-sixtermsexactsequencesHP}
For any short exact sequence of associative algebras $\xymatrix@1{{\algzero} \ar[r] & {\algI} \ar[r] & {\algA} \ar[r] & {\algA/\algI} \ar[r] & {\algzero}}$, one has the six term exact sequence
\begin{equation*}
\xymatrix{ 
{HP_0(\algI)} \ar[r]  & {HP_0(\algA)} \ar[r]  & {HP_0(\algA/\algI)} \ar[d]^-{\delta} \\
{HP_{1}(\algA/\algI)} \ar[u]^-{\delta} & {HP_{1}(\algA)} \ar[l] & {HP_{1}(\algI)} \ar[l] 
}
\end{equation*}
\end{Proposition}

\begin{Proposition}[Morita invariance]
For any integer $n \geq 1$, the trace map defined in Definition~\ref{def-tracemap} induces an isomorphism $\tr_\indast : HP_\nu(M_n(\algA)) \xrightarrow{\simeq} HP_\nu(\algA)$.
\end{Proposition}

Notice that we did not mention such a result for cyclic homology, because it is not true! There is a Morita invariance for cyclic homology on $H$-unital algebras (see Definition~\ref{def-Hunitalalgebras}), but not on all algebras.

\begin{Example}[$HP_\nu(\gC)$ and $HC_\grast^-(\gC)$]
\label{ex-periodiccyclicC}
One has 
\begin{align*}
HP_0(\gC) &= \gC & HP_1(\gC) &= \algzero 
\end{align*}

There is an isomorphism of algebras $HC_\grast^-(\gC) \simeq \gC[v]$ for a generator $v \in HC_{-2}^-(\gC)$. The product by $v$ corresponds to the operation $S : HC_{n}^-(\gC) \rightarrow HC_{n-2}^-(\gC)$.
\end{Example}

\begin{Example}[Tensor algebra]
Let us use the notations of Example~\ref{ex-tensoralgebra}. The inclusion $\gC \rightarrow \caT \evV$ induces an isomorphism in periodic cyclic homology:
\begin{align*}
HP_0(\caT \evV) &= \gC & HP_1(\caT \evV) &= \algzero 
\end{align*}
\end{Example}

\begin{Example}[The Laurent polynomials]
\label{ex-thelaurentpolynomials}
Let $\algA = \gC[z,z^{-1}]$ be the Laurent polynomials for the variable $z$. Then one has
\begin{align*}
HP_0(\gC[z,z^{-1}]) &= \gC & HP_1(\gC[z,z^{-1}]) &= \gC 
\end{align*}

\end{Example}

\begin{Example}[Unital smooth commutative algebras]
\label{ex-smoothcommutativeunitalalgebras}
For any unital smooth commutative algebra $\algA$, one has
\begin{align*}
HP_0(\algA) &= H_\text{dR}^\text{even}(\algA) = \prod_{p \geq 0} H^{2p}_\text{dR}(\algA)
&
HP_1(\algA) &= H_\text{dR}^\text{odd}(\algA) = \prod_{p \geq 0} H^{2p+1}_\text{dR}(\algA)
\end{align*}
If we compare this result with the cyclic homology groups given at the end of Example~\ref{ex-thecommutativecasecyclichomology}, one sees that periodic cyclic homology is not ill with the edge effects on the left of the cyclic bicomplex which produce the contributions $\Omega^{\grast}_{\algA|\gC}/d_K\Omega^{\grast-1}_{\algA|\gC}$. 
\end{Example}

\begin{Remark}[Extension to topological algebras]
As for Hochschild homology, one can generalize the definitions given above to take into account some topological structure on the algebra $\algA$, by replacing the tensor products with topological tensor products. We then obtain ``continuous'' versions of these cyclic homologies.

In \cite{Cunt:97}, Cuntz proved a six term exact sequence as in Proposition~\ref{prop-sixtermsexactsequencesHP} for a restricted class of topological algebras, called $m$-algebras (see also \cite{CuntSkanTsyg:04}).
\end{Remark}

\begin{Example}[Continuous cyclic homology of Banach algebras]
\label{ex-continuouscyclichomologyofbanachalgebras}
On Banach algebras, the continuous cyclic homologies are not interesting. For instance, for commutative $C^\ast$-algebras, one gets
\begin{align*}
HP_{0}^{\text{cont}}(C(X)) &= \{\text{bounded measures on $X$}\} & 
HP_{1}^{\text{cont}}(C(X)) &= \algzero
\end{align*}
\end{Example}

This example shows that cyclic homology is not a very powerful theory for noncommutative \emph{topological} spaces. The following result confirms this fact. We need a
\begin{Definition}[Diffeotopic morphisms]
Let $\algA$ and $\algB$ be two associative algebras. Two morphisms of algebras $\varphi_0, \varphi_1 : \algA \rightarrow \algB$ are said to be diffeotopic if there exists a morphism of algebras $\varphi : \algA \rightarrow \algB \otimes C^\infty([0,1])$ such that $\varphi_t$ coincides with $\varphi_0$ (resp. $\varphi_1$) when evaluated at $t=0$ (resp. at $t=1$) in the target algebra. Notice that the tensor product $\algB \otimes C^\infty([0,1])$ is purely algebraic. 
\end{Definition}

\begin{Proposition}[Diffeotopic invariance]
If $\varphi_0$ and $\varphi_1$ are diffeotopic then they induce the same morphism $HP_\nu(\algA) \rightarrow HP_\nu(\algB)$.\end{Proposition}

There is no general homotopic invariance result on periodic cyclic homology.

If you need a result more to convince you that cyclic homology is well adapted to differential structures, here is the main result, obtained by Connes in \cite{Conn:85}, which is a generalisation of Example~\ref{ex-smoothcommutativeunitalalgebras}:

\begin{Example}[Continuous periodic cyclic homology of $C^\infty(M)$]
\label{ex-continuousperiodiccyclichomologyofCinftyM}
Let $M$ be a $C^\infty$ finite dimensional locally compact manifold. Then one has
\begin{align*}
HP_{0}^{\text{cont}}(C^\infty(M)) &= H_\text{dR}^\text{even}(M) & 
HP_{1}^{\text{cont}}(C^\infty(M)) &= H_\text{dR}^\text{odd}(M)
\end{align*}
\end{Example}

\begin{Remark}[Comparing cyclic homology with $K$-theory]
\label{rmk-cyclichomologyandKtheory}
In the next section, we will establish a very strong relation between $K$-theory and periodic cyclic homology. $\gZ_2$-graduation, Morita invariance and the six term exact sequence give us obvious similarities between these two theories. 

But, we would like to make it clear that the two theories are very different on an essential point. We notice in Remark~\ref{rmk-Ktheorycomputedondensesubalgebras} that $K$-theory can be computed using some dense subalgebra (stable by holomorphic functional calculus). The situation is clearly not the same for periodic cyclic homology: compare Example~\ref{ex-continuouscyclichomologyofbanachalgebras} with Example~\ref{ex-continuousperiodiccyclichomologyofCinftyM}.

$K$-theory is an homology theory for noncommutative topological spaces. Periodic cyclic homology is an homology theory for algebras concealing some differentiable properties.
\end{Remark}

\begin{Example}[First version of cyclic cohomology]
\label{ex-firstversionofcycliccohomology}
We define here the cyclic cohomology using the differential complex $(C^\grast_\lambda(\algA), \delta)$, which is the first version of cyclic cohomology exposed in \cite{Conn:85}. 

By definition, 
\begin{align*}
C^n_\lambda(\algA) &= \{ \phi \in \Hom(\algA^{\otimes n+1}, \gC)\ / \ \phi(a_1 \otimes \cdots \otimes a_n \otimes  a_0) = (-1)^{n} \phi(a_0 \otimes a_1 \otimes \cdots \otimes a_n) \}\\
 &\subset C^n(\algA)
\end{align*}
and $\delta$ is the Hochschild boundary operator for the complex $C^\grast(\algA)$ restricted to this subspace (see Definition~\ref{def-Hochschildcohomology}). Indeed, the main remark made by Connes to define its cyclic complex as a subcomplex of a Hochschild complex was that the cyclic condition defining the $\phi$'s in $C^n(\algA)$ which are elements of $C^n_\lambda(\algA)$ is compatible with the boundary $\delta$. 

The cohomology of the complex $(C^\grast_\lambda(\algA), \delta)$ is the cyclic cohomology $HC^\grast(\algA)$ of $\algA$. 

For $n=0$, a cycle $\phi$ is a trace on $\algA$, because $(\delta\phi)(a_0 \otimes a_1) = \phi(a_0 a_1) - \phi(a_1 a_0) = 0$. The cyclic complex of Connes is explicitly defined to generalize this property to higher degrees. So, cyclic cohomology is a theory of generalized traces.

On this cohomology, the inclusion $C^\grast_\lambda(\algA) \hookrightarrow C^\grast(\algA)$ at the level of complexes induces a map $I : HC^\grast(\algA) \rightarrow HH^\grast(\algA)$ and Connes long exact sequence
\begin{equation*}
\xymatrix@1{ {\cdots} \ar[r] & {HH^n(\algA)} \ar[r]^-{B} & {HC^{n-1}(\algA)} \ar[r]^-{S} & {HC^{n+1}(\algA)} \ar[r]^-{I} & {HH^{n+1}(\algA)} \ar[r]^-{B} & {\cdots} }
\end{equation*}
In this long exact sequence, the two maps $B$ and $S$ are not so easy to define as in the previous construction for cyclic homology.

Nevertheless, one can show that the periodic map $S : HC^{n}(\algA) \rightarrow HC^{n+2}(\algA)$ can be used to define periodic cyclic cohomology as 
\begin{align*}
HP^{0}(\algA) &= \varinjlim (HC^{2n+1}(\algA), S)
&
HP^{1}(\algA) &= \varinjlim (HC^{2n}(\algA), S)
\end{align*}
which explains the name ``periodic'' for the cohomology group and the map $S$.
\end{Example}

One defines the dual bicomplex $CC^{\grast, \grast}(\algA)$ of $CC_{\grast, \grast}(\algA)$ replacing in each bidegree $(p,n)$ the space $\algA^{\otimes n+1}$ by the space $\Hom(\algA^{\otimes n}, \algA^\ast) = \Hom(\algA^{\otimes n+1}, \gC)$, and adjoining the four maps $b$, $b'$, $t$ and $N$. Recall that the adjoint of $b$ is the $\delta$ map introduced in Definition~\ref{def-Hochschildcohomology}. 

In the same manner, one defines the bicomplex $CC^{\grast, \grast}_\text{per}(\algA)$ from the bicomplex $CC_{\grast, \grast}^\text{per}(\algA)$.

\begin{Definition}[Cyclic cohomology]
The cyclic cohomology $HC^\grast(\algA)$ of $\algA$ is the cohomology of the total complex of the bicomplex $CC^{\grast, \grast}(\algA)$.

The cyclic periodic cohomology $HP^\grast(\algA)$ of $\algA$ is the cohomology of the total complex of the bicomplex $CC^{\grast, \grast}_\text{per}(\algA)$. Here, the total complex is constructed using direct sums.
\end{Definition}

As for periodic cyclic homology, $HP^\grast(\algA)$ is $\gZ_2$-graded.

\begin{Remark}[Entire cyclic cohomology]
In the definition of $HP^\grast(\algA)$, one uses the direct sum to construct the total complex. This is the dual version of the direct product used for periodic cyclic homology. Indeed, one can show that the direct product would produce a trivial cohomology. Using direct sum in periodic cyclic cohomology permits one to define a natural pairing with periodic cyclic homology: cochains in periodic cyclic cohomology have finite support, so that only a finite number of terms are non zero when evaluated on a (infinite) chain in cyclic periodic homology.

Let $\algA$ be a Banach algebra. Then one defines a norm on $\Hom(\algA^{\otimes n+1}, \gC)$ by $\Vert \phi_n \Vert = \sup \{ | \phi(a_0 \otimes \cdots \otimes a_n) | \ / \ \Vert a_i \Vert \leq 1 \}$.

Denote by $TCC^\grast_{\prod}(\algA)$ the total complex of $CC^{\grast, \grast}_\text{per}(\algA)$ obtained using direct product. Each element in $TCC^p_{\prod}(\algA)$ is an infinite sequence $(\phi_{2n})$ or $(\phi_{2n+1})$ according to parity of $p$. One defines a subcomplex $ECC^\grast(\algA)$ of $TCC^\grast_{\prod}(\algA)$ imposing a growing condition on such an infinite sequence: the radius of convergence of the series $\sum_{n \geq 0} \Vert \phi_{2n} \Vert z^n / n!$ (resp. $\sum_{n \geq 0} \Vert \phi_{2n+1} \Vert z^n / n!$) is infinity. 

The entire cyclic cohomology $HE^\grast(\algA)$ is defined as the cohomology of the complex $ECC^\grast(\algA)$. One can show that $HE^\grast(\algA)$ is $\gZ_2$-graded, as is the periodic cyclic cohomology. 

Any cochain defining an element in $HP^\grast(\algA)$ has finite support, so that there is an natural map $HP^\grast(\algA) \rightarrow HE^\grast(\algA)$. This map is an isomorphism in some cases, for instance $\algA=\gC$, but not in general. See \cite{Khal:94} for examples of such isomorphisms.
\end{Remark}

\begin{Example}[The irrational rotation algebra]
For irrational $\theta$, one has
\begin{align*}
HP_{0}^{\text{cont}}(\tncA^\infty_\theta) & = \gC^2 
& 
HP_{1}^{\text{cont}}(\tncA^\infty_\theta) & = HH_1(\tncA^\infty_\theta)/ \Ran B_\indast = \gC^2
\end{align*}
There is no need to mention any diophantine condition here (see Example~\ref{ex-theirrationalrottionalgebraHochschimdhomology}).

In periodic cyclic cohomology, one of the two classes in $HP^{\nu}_{\text{cont}}(\tncA^\infty_\theta) = \gC^2$ is $S \tau$ where $\tau$ is the unique normalised trace on $\tncA^\infty_\theta$, $\tau(\sum_{m,n \in \gZ} a_{m,n} U^m V^n) = a_{0,0}$, and the second one is expressed in terms of the continuous derivations $\delta_1$ and $\delta_2$:
\begin{equation*}
\varphi(a_0\otimes a_1 \otimes a_2) = \frac{1}{2 i \pi} \tau[a_0 (\delta_1(a_1) \delta_2(a_2) - \delta_2(a_1) \delta_1(a_2))]
\end{equation*}
\end{Example}

\begin{Remark}[Pairing with $K$-theories]
A Fredholm module $(\ehH, \rho, F)$ over the $C^\ast$-algebra $\algA$ is called $p$-summable if $[F,a] \in \caL^p(\ehH)$ for any $a \in \algA$ (we write $a$ for $\rho(a)$ from now on). The space $\caL^p(\ehH) = \{ T \in \caK \ / \ \sum_{n=0}^\infty \mu_n(T)^p < \infty \}$ with $\mu_n(T)$ the $n$-th eigenvalue of $|T| = (T^\ast T)^{1/2}$, is the Schatten class. It is a two-sided ideal in $\caB$, and a Banach space for the norm $\Vert T \Vert_p = (\sum_{n=0}^\infty \mu_n(T)^p)^{1/p} = \tr(|T|^p)^{1/p}$. For any $S \in \caL^p(\ehH)$ and $T \in \caL^q(\ehH)$, one has $ST \in \caL^r(\ehH)$ for $\frac{1}{r} = \frac{1}{p} + \frac{1}{q}$ and $\Vert ST \Vert_r \leq \Vert S \Vert_p \Vert T \Vert_q$.  

Let $(\ehH, \rho, F)$ be a $p$-summable Fredholm module (odd or even) and denote by $\gamma$ its grading map if it is even. For any $n \geq 0$, and $a_i \in \algA$, the expressions
\begin{align*}
\varphi_{2n+1} &= \tr(a_0 [F, a_1] \cdots [F, a_{2n+1}]) \text{ in the odd case}\\
\varphi_{2n} &= \tr(\gamma a_0 [F, a_1] \cdots [F, a_{2n}]) \text{ in the even case}\\
\end{align*}
make sense and define an odd or an even cocyle in $HC^\grast(\algA)$, which depends only on the $K$-homology class of the Fredholm module. This defines a pairing $HP_\nu(\algA) \times K^\nu(\algA) \rightarrow \gC$ for $\nu = 0,1$.

In the next section, the Chern character will realize a pairing $HP^\nu(\algA) \times K_\nu(\algA) \rightarrow \gC$.
\end{Remark}

\section{The not-missing link: the Chern character}
\label{thenotmissinglinkthecherncharacter}

The Chern character is a special characteristic class defined first in the topological context. It was used to related the $K$-theory of a topological space to its cohomology. When Connes introduced cyclic homology, he saw immediately that a purely algebraic generalisation was possible, which connects the $K$-theory for algebras and the periodic cyclic homology. Now, the Chern character is extensively studied, because it helps interpret a lot of previous results in different areas of mathematics, which where not so well understood.

\subsection{The Chern character in ordinary differential geometry}

Let us recall some basic facts about characteristic classes for vector bundles.

Let $G$ be a topological group. Then one has:
\begin{Proposition}[Classifying space $BG$]
There exists a $G$-principal fibre bundle $EG \rightarrow BG$ such that for any $G$-principal fibre bundle $P$ over a topological space $X$, there exists a continuous map $f_P : P \rightarrow BG$ such that $P = f_P^\ast EG$ (the pull-back fibre bundle). $BG$ is called the classifying space of the topological group $G$ and $f_P$ the classifying map of $P$.
\end{Proposition}

Recall that the pull-back $P=f^\ast Q$ of a fibre bundle $Q \rightarrow Y$ through a continuous map $f : X \rightarrow Y$ is defined by $P_x = Q_{f(x)}$ for any $x \in X$. If $g : X \rightarrow Y$ is homotopic to $f$ then $f^\ast Q$ and $g^\ast Q$ are isomorphic.

One can show that $EG$ is a contractible space, so that its homology is not very interesting. The important object in this proposition is $BG$:
\begin{Proposition}[Classification of $G$-principal fibre bundles]
The space of isomorphic classes of $G$-principal fibre bundles over $X$ is $[X; BG]$, the space of homotopic classes of continuous maps $X \rightarrow BG$.
\end{Proposition}

This space is not easy to compute, so that this classification remains just an identification without any practical utility in general. This leads us to consider other objects to try to classify $G$-principal fibre bundles, in terms of cohomology classes: 
\begin{Definition}[Characteristic classes]
A characteristic class of $P$ in a cohomology group $H^\grast(X; \grA)$ with coefficient in the abelian group $\grA$, is the pull-back by $f_P$ of any cohomology class $c \in H^\grast(BG; \grA)$.
\end{Definition}

Characteristic classes depend upon the coefficient group $\grA$, which is often essential to make some concrete interpretations of certain characteristic classes.

If one is interested in vector bundles instead of principal bundles, the previous construction can be performed with the associated principal bundle. Any vector bundle is the pull-back through the classifying map $f_P$ of a canonical vector bundle over $BG$. So that for any vector bundle $E \xrightarrow{\pi} X$ with structure group $G$, one can introduce its characteristic classes as pull-back of classes in $H^\grast(BG; \grA)$. We will use the notation $c(E)$ for the pull-back of $c \in H^\grast(BG; \grA)$ in $H^\grast(X; \grA)$.

\begin{Proposition}[Functoriality of characteristic classes]
Let $\varphi : X \rightarrow Y$ be a continuous map, and $E \rightarrow Y$ a vector bundle on $Y$. Then for any characteristic class $c$ one has $c(\varphi^\ast E) = \varphi^\indast c(E)$ where $\varphi^\indast : H^\grast(Y) \rightarrow H^\ast(X)$ is the ring morphism induced in cohomology.
\end{Proposition}

\begin{Example}[Discrete groups]
In the case of a discrete group $G$, one can show that $BG=K(G,1)$ is the Eilenberg-MacLane space of type $(G,1)$, so that $H^\grast(BG; \gZ)$ is the ordinary cohomology of groups $H^\grast(G)$.
\end{Example}

It is possible to construct explicitly the classifying spaces $BG$ for a large class of groups. Here are some examples.

\begin{Example}[Some usual classifying spaces]
\begin{equation*}
\begin{array}{r|cccccc}
G  & \gZ   & \gZ^n & \gZ_2       & U(1) = \gS^1 &  U(n)               & O(n)               \\ 
EG & \gR   & \gR^n & \gS^\infty  &              &                     &                    \\  
BG & \gS^1 & \gT^n & \gRP^\infty & \gCP^\infty  &  \gG(n, \gC^\infty) & \gG(n, \gR^\infty)                 
\end{array}
\end{equation*}
$\gS^\infty$ is the sphere in $\gR^\infty$, $\gRP^\infty = \varinjlim \gRP^{n}$, $\gCP^\infty = \varinjlim \gCP^{n}$, $\gG(n, \gC^\infty) = \varinjlim \gG(n, \gC^p)$ where $\gG(n, \gC^p)$ is the complex Grassmanian manifold\dots
\end{Example}

\begin{Example}[Cohomology groups of some classifying spaces]
\label{cohomologygroupsofsomeclassifyingspaces}
Here are some examples of cohomology groups of some classifying spaces. We denote by $\grA[a_1, \dots, a_p]$ the graded commutative ring generated over the abelian groups $\grA$ by the $p$ elements $a_i$ (whose degrees will be given):
\begin{align*}
H^\grast( BU(n); \gZ) &= \gZ[c_1, c_2, \dots, c_n]
&
H^\grast( BSU(n); \gZ) &= \gZ[c_2, \dots, c_n]
\end{align*}
where $\deg c_k = 2k$. The class $c_k$ is the $k$-th Chern class. The class $c = 1 + c_1 + c_2 + \dots + c_n$ is the total Chern class. It satisfies $c(E \oplus E') = c(E) c(E')$ for any vector bundles $E$ and $E'$.
\begin{equation*}
H^\grast( BO(n); \gZ) = \gZ[p_1, p_2, \dots, p_{[n/2]}]
\end{equation*}
where $\deg p_k = 4k$ and $[n/2]$ is the integer part of $n/2$. The class $p_k$ is the $k$-th Pontrjagin class. The class $p = 1 + p_1 + p_2 + \dots + p_{[n/2]}$ is the total Pontrjagin class which satisfies $p(E \oplus E') = p(E) p(E')$.
\begin{align*}
H^\grast( BSO(2m+1); \gZ) &= \gZ[p_1, p_2, \dots, p_m]
&
H^\grast( BSO(2m); \gZ) &= \gZ[p_1, p_2, \dots, p_{m-1}, e]
\end{align*}
where $\deg p_k = 4k$ and $\deg e = 2m$. The class $e$ is called the Euler class, it satisfies $e(E \oplus E') = e(E) e(E')$. 
\begin{align*}
H^\grast( BO(n); \gZ_2) &= \gZ_2[w_1, \dots, w_n]
&
H^\grast( BSO(n); \gZ_2) &= \gZ_2[w_2, \dots, w_n]
\end{align*}
where $\deg w_k = k$ is the $k$-th Stiefel-Whitney class. The class $w = 1 + w_1 + w_2 + \dots + w_n$ is the total Stiefel-Whitney class which satisfies $w(E \oplus E') = w(E) w(E')$.
\end{Example}

\begin{Example}[Interpretation of $w_1$ and $w_2$]
Let $M$ be a locally compact finite dimensional manifold. $M$ is orientable if and only if the first Stiefel-Whitney class $w_1(TM)$ of its tangent space $TM$ is zero. If $M$ is orientable, it admits a spin structure if and only if the second Stiefel-Whitney class $w_2(TM)$ is zero.
\end{Example}

\begin{Example}[Classification of complex line vector bundles]
The first Chern class of $c_1(L) \in H^2(X; \gZ)$ of a complex line vector bundle $L \rightarrow X$ is a total invariant in the space of isomorphic classes of line vector bundles over $X$.
\end{Example}

\begin{Example}[Compact connected Lie groups]
\label{ex-compactconnectedliegroups}
For any compact connected Lie group $G$, one has
\begin{align*}
H^{2 n}(BG; \gR) &= \caP^n_I(\kg)
&
H^{2 n+1}(BG; \gR) &= \algzero
\end{align*}
where $\kg$ is the Lie algebra of $G$ and $\caP^\grast_I(\kg)$ is the graded algebra of invariant polynomials on $\kg$.

For the compact Lie groups in Example~\ref{cohomologygroupsofsomeclassifyingspaces}, these invariant polynomials are generated by the formulas:
\begin{equation*}
\det( \lambda \bbbone + \frac{i}{2 \pi} X ) = \lambda^n + c_1(X) \lambda^{n-1} + c_2(X) \lambda^{n-2} + \dots + c_n(X)
\end{equation*}
for any $X \in \ku(n)$;
\begin{equation*}
\det( \lambda \bbbone - \frac{1}{2 \pi} X ) = \lambda^n + p_1(X) \lambda^{n-2} + p_2(X) \lambda^{n-4} + \dots + p_m(X) \lambda^{n - 2m}
\end{equation*}
for any $X \in \ko(n)$;
\begin{equation*}
e(X) = {\frac{ (-1)^m}{ 2^{2m} \pi^m m !} } \sum_{i_1, \dots, i_{2m} } \epsilon_{i_1 i_2 \dots i_{2m-1} i_{2m} } X_{i_1 i_2} \dots X_{ i_{2m-1} i_{2m} }
\end{equation*}
for any $X \in \kso(2m)$, where $\epsilon_{i_1 i_2 \dots i_{2m-1} i_{2m} }$ is completely antisymmetric with $\epsilon_{1 2 \dots 2m} = 1$. The quantity $\Pf(X)=(2 \pi)^m h(X)$ is called the Pfaffian of $X$. It is a square root of the determinant. The Euler class is then associated to a very particular invariant polynomial.
\end{Example}

\begin{Example}[Characteristic classes through connections]
\label{ex-characteristicclassesthroughconnections}
It is possible to construct characteristic classes directly using invariant polynomials in $\caP^\grast_I(\kg)$. In order to do that, consider a differentiable principal fibre bundle $P\rightarrow M$ over a differential manifold with structure group $G$. Let us denote by $\omega \in \Omega^1(P)\otimes \kg$ a connection on $P$ and $\Omega$ its curvature. Recall that $\omega$ is a covariant object for the action $\widetilde{R}_g$ of $G$ on $P$ by right multiplication and the adjoint action $Ad$ on $\kg$ : $\widetilde{R}_g^\ast \omega = Ad_{g^{-1}} \omega$ for any $g \in G$. Its curvature is also covariant, $\widetilde{R}_g^\ast \Omega = Ad_{g^{-1}} \Omega$, and satisfies the Bianchi identity $d\Omega + [\omega, \Omega] =0$.

Let $(U, \phi)$ be a local trivialisation of $P$, where $U$ is an open subset of $M$ and $\phi : U \times G \rightarrow P_{|U}$ is a diffeomorphism which intertwines the actions of $G$ on $P$ and $G$. Define by $s_U(x) = \phi(x,e)$ the section which trivializes $P_{|U}$ and by $A^U = s_U^\ast \omega$ and $F^U = s_U^\ast \Omega$ the local connection $1$-form and the local curvature $2$-form. If $(V, \psi)$ is a second trivialization of $P$, with $U \cap V \neq \ensvide$, then one has the relations
\begin{align*}
A^V & = g_{UV}^{-1} A^U g_{UV} + g_{UV}^{-1} d g_{UV}
&
F^V & = g_{UV}^{-1} F^U g_{UV}
\end{align*}
where $g_{UV} : U \cap V \rightarrow G$ is the transition function between the two trivializations. The local forms $F^U$ satisfy to a Bianchi identity.

Let us consider $p$ an invariant polynomial on $\kg$, of degree $k$. Then one can define $p(F^U, \dots, F^U)$ as a local $2k$-form on $U$ by evaluating $p$ on the values of $F^U$ in $\kg$. Because $p$ is invariant, one has $p(F^U, \dots, F^U) = p(F^V, \dots, F^V)$ so that it defines a global $2k$-form on $M$. Using the Bianchi identity, one can then show that its differential is zero. We then have associated to $p$ a cohomology class in $H^{2k}(M; \gR)$, which can be shown to be independent of the choice of the connection $\omega$. This map $\caP^k_I(\kg) \rightarrow H^{2k}(M; \gR)$ is the Chern-Weil map.

This class is exactly the characteristic class given by the invariant polynomial $p$ in the identification $H^{2 n}(BG; \gR) = \caP^n_I(\kg)$ in Example~\ref{ex-compactconnectedliegroups}.

One does not really need to express the connection $1$-form and its curvature $2$-form locally on an open set of the base space $M$. Indeed, $p(\Omega, \dots, \Omega)$ makes sense as a $2k$-form on $P$. Using the properties of the curvature $2$-form $\Omega$ and the invariance of the polynomial $p$, one can show that it is a basic form for the action of $G$ on $P$, and as such, it identifies with a $2k$-form on the base space $M$. 
\end{Example}

\begin{Proposition}[Decomposition principle]
Let $E_1, \dots, E_p \rightarrow X$ be $p$ complex vector bundles. Then there exist a manifold $\gF$ and a continuous map $\sigma : \gF \rightarrow X$ such that the pull-backs $\sigma^\ast E_i \rightarrow \gF$ are all decomposed as direct sum of complex line vector bundles, and such that the map induced in cohomology $\sigma^\indast : H^\grast(X) \rightarrow H^\grast(\gF)$ is injective.
\end{Proposition}

Why decompose a vector bundle in a direct sum of line vector bundles? The answer is in the following construction. 

Let $R(c(E_1), \dots, c(E_p))$ be a polynomial relation in $H^\grast(X)$ between the Chern classes of the vector bundles $E_i$. We would like to establish the relation $R(c(E_1), \dots, c(E_p)) = 0$ for any vector bundles over $X$, and for any $X$. Using the decomposition map $\sigma : \gF \rightarrow X$ and the functoriality of the Chern classes (and the fact that the relation is a polynomial relation) we have 
\begin{equation*}
\sigma^\indast(R(c(E_1), \dots, c(E_p))) = R(c(\sigma^\ast E_1), \dots, c(\sigma^\ast E_p))
\end{equation*}
Now, let us assume that for any base space $Y$ and any vector bundles $F_i$ over $Y$ which are direct sum of line vector bundles, the relation $R(c(F_1), \dots, c(F_p)) = 0$ can be established. Then, for any $E_i$ over $X$, the $F_i = \sigma^\ast E_i$ over $Y = \gF$ are direct sums of line vector bundles, so that the relation is true for them. The right hand side of the relation is then zero, which implies by injectivity of $\sigma^\indast : H^\grast(X) \rightarrow H^\grast(\gF)$ that the relation is also zero for the $E_i$'s. 

So, in order to establish an abstract relation between the Chern classes, it is sufficient to show it for \emph{any} vector bundle decomposed as a direct sum of line vector bundles over \emph{any} space.

\begin{Example}[Chern classes and elementary symmetric polynomials]
\label{chernclassesandelementarysymmetricpolynomials}
Let us apply the relation $c(E \oplus E') = c(E) c(E')$, where $c$ is the total Chern class, to a direct sum of line vector bundles $E = \ell_1 \oplus \cdots \oplus \ell_n$. Then $c(E) = c(\ell_1) \cdots c(\ell_n)$. For a line vector bundle, one has $c(\ell) = 1 + c_1(\ell)$. Denote by $x_i = c_1(\ell_i)$ the first Chern classes of these line vector bundles. Then one has
\begin{equation*}
c(E) = \prod_{i=1}^n (1+ x_i) = \sum_{j=0}^n \sigma_j(x_1, \dots, x_n)
\end{equation*}
where the functions $\sigma_j$ are the elementary symmetric polynomials of total degree $j$. They are explicitly given in terms of the $n$ (commuting) variables $X_i$ by
\begin{align*}
\sigma_0(X_1, \dots, X_n) &= 1
&
\sigma_1(X_1, \dots, X_n) &= \sum_{1\leq i \leq n} X_i
&
\sigma_2(X_1, \dots, X_n) &= \sum_{1\leq i < j \leq n} X_i X_j\\
& 
&\cdots
&
& 
\sigma_n(X_1, \dots, X_n) &= \prod_{1\leq i \leq n} X_i
\end{align*}
Any symmetric polynomial (resp. any formal symmetric series) in the $n$ variables $X_i$ can be expressed as a polynomial (resp. a formal series) in these elementary symmetric polynomials: 
\begin{align*}
\gC[X_1, \dots, X_n]^{\kS_n} &= \{ p \in \gC[X_1, \dots, X_n]\ / \ p(X_1, \dots, X_n) = p(X_{\sigma^{-1}(1)}, \dots, X_{\sigma^{-1}(n)}) \}\\
 &= \gC[\sigma_1, \dots, \sigma_n]
\end{align*}
The previous computation shows us that the Chern classes can be written as $c_j(E) = \sigma_j(x_1, \dots, x_n)$ when $E$ is decomposed. If $E$ is not decomposed, then use $\sigma^\ast E$ over $\gF$.
\end{Example}

\begin{Example}[Characteristic class associated to a symmetric polynomial]
The previous Example gives us another application of the decomposition principle, which is to construct a new characteristic class in terms of the Chern classes, but writing it down explicitly only in terms of the first Chern classes and a symmetric polynomial. Indeed, let $p(X_1, \dots, X_n)$ be a symmetric polynomials. Then it is a polynomial of the form $R(\sigma_1, \dots, \sigma_n)$. For any vector bundle $E \rightarrow X$ decomposed as $E = \ell_1 \oplus \cdots \oplus \ell_n$, define the characteristic class $c_p(E)= p(x_1, \dots, x_n)$ where $x_i = c_1(\ell_i)$. Then $c_p(E) = R(\sigma_1(x_1, \dots, x_n), \dots, \sigma_n(x_1, \dots, x_n)) = R(c_1(E), \dots, c_n(E))$. Now, if $E$ is not decomposed as a direct sum of line vector bundles, the last relation can be used to define, without ambiguities, the class $c_p(E)$, thanks to the decomposition principle and the functoriality of the Chern classes.

This construction can be generalised to any invariant formal series in $n$ variables.
\end{Example}

\begin{Definition}[The Chern character]
Let $E$ be a vector bundle over $X$. The Chern character $\ch(E)$ of $E$ is defined to be the characteristic class associated to the formal series 
\begin{equation*}
p(x_1, \dots, x_n) = e^{x_1} + \cdots + e^{x_n} = n + \sum_{i=1}^n x_i + \frac{1}{2} \sum_{i=1}^n (x_i)^2 + \cdots
\end{equation*}
Notice that the coefficient group for the cohomology of this class is necessarily $\gQ$, because the defining expression for $\ch(E)$ makes use of rational numbers. 
\end{Definition}

\begin{Example}[The invariant polynomial of the Chern character]
\label{ex-theinvariantpolynomialoftheCherncharacter}
We saw in Example~\ref{ex-characteristicclassesthroughconnections} that characteristic classes can be defined using a connection on the vector bundle and an invariant polynomial. The Chern character is a particular characteristic class, and its invariant polynomial (in fact an invariant formal series) is $P(X) = \tr \exp(\frac{i}{2 \pi}X)$, so that $\ch(E) = \tr \circ \exp\left(\frac{i F}{2\pi}\right)$ for any local curvature $2$-form of a connection on $E$.

As a form on the principal fibre bundle, this expression is 
\begin{equation*}
\ch(\omega) = \tr \circ \exp\left(\frac{i \Omega}{2\pi}\right) = \sum_{k=0}^\infty \frac{1}{k!} \left(\frac{i}{2\pi}\right)^{k} \tr(\Omega^k)
\end{equation*}
\end{Example}

\begin{Proposition}[Product and additive properties of $\ch$]
Using the decomposition principle, one can show that for any vector bundles $E$ and $E'$:
\begin{align*}
\ch(E \oplus E') &= \ch(E) + \ch(E')
&
\ch(E \otimes E') &= \ch(E) \ch(E')
\end{align*}
\end{Proposition}

\begin{Theorem}[The Chern character as an isomorphism]
\label{thm-thecherncharacterasanisomorphism}
The Chern Character defines a natural morphism of rings $\ch : K^0(X) \rightarrow H^{\text{even}}(X; \gQ)$ which induces an isomorphism
\begin{equation*}
\ch : K^0(X)\otimes_{\gZ} \gQ \xrightarrow{\simeq} H^{\text{even}}(X; \gQ)
\end{equation*}
for locally compact finite dimensional manifolds $X$. In that case, the Chern character can be extended to a isomorphism $\ch : K^{1}(X)\otimes_{\gZ} \gQ \xrightarrow{\simeq} H^{\text{odd}}(X; \gQ)$.
\end{Theorem}

\begin{Example}[The Chern character $K^{-1}(M) \rightarrow H^{\text{odd}}(M; \gQ)$]
\label{ex-thecherncharacterK1(m)rightarrowHodd(MQ)}
It is possible to give an expression of the Chern character in odd degrees using connections. Let $\omega_0$ and $\omega_1$ be connections on the principal fibre bundle $P$. Then $\omega_t = \omega_0 + t(\omega_1 - \omega_0)$ is also a connection for any $t \in [0,1]$. We denote by $\Omega_t$ its curvature. One can show that the Chern-Simons form
\begin{equation*}
\cs(\omega_0, \omega_1) = \int_0^1 dt \tr\left((\omega_1 - \omega_0) \exp\left(\frac{i \Omega_t}{2\pi} \right)\right)
\end{equation*}
satisfies $d \cs(\omega_0, \omega_1) = \ch(\omega_1) - \ch(\omega_0)$ where $\ch(\omega)$ is given as in Example~\ref{ex-theinvariantpolynomialoftheCherncharacter}.

Let $g : M \rightarrow U(n)$ be a smooth map. Consider the trivial fibre bundle $P = M \times U(n)$, with the two connections $\omega_0 = 0$ and $\omega_1 = g^{-1} dg$. Then one defines
\begin{equation*}
\ch(g) = \cs(0, g^{-1} dg) = \sum_{k=0}^\infty (-1)^k \frac{k!}{(2k+1)!} \left(\frac{i}{2\pi}\right)^{k+1} \tr( (g^{-1} dg)^{2k+1})
\end{equation*}
This defines a map from the class of $g$ in $K^{-1}(M)$ into $H^{\text{odd}}(M; \gQ)$.
\end{Example}

\subsection{Characteristic classes and Chern character in noncommutative geometry}

It is possible to construct some characteristic classes, and in particular the Chern character, using the algebraic setting of modules and differential calculi. The construction of these classes are based upon some generalisation of the construction of the Chern classes in terms of the curvature of some connection. In order to do that, one need to define the so-called noncommutative connections.

Let $(\Omega^\grast, d)$ be a differential calculus on an associative unital algebra $\algA$, and let $\modM$ be a finite projective left module over $\algA = \Omega^0$. 

\begin{Definition}[Noncommutative connection]
A noncommutative connection on $\modM$ for the differential calculus $(\Omega^\grast, d)$ is linear map $\nabla : \modM \rightarrow \Omega^1 \otimes_\algA \modM$ such that $\nabla (am) = da \otimes m + a (\nabla m)$ for any $m \in \modM$ and $a \in \algA$.
\end{Definition}

Let us introduce $\widetilde{\modM} = \Omega^\grast \otimes_\algA \modM$ as a graded left module over $\Omega^\grast$, and $\End^\grast_\Omega(\widetilde{\modM})$ the graded algebra of $\Omega^\grast$-linear endomorphisms on $\widetilde{\modM}$ (any $T \in \End^{k}_\Omega(\widetilde{\modM})$ satisfies $T(\eta) \in \Omega^{m+k} \otimes_\algA \modM$ and $T(\omega \eta) = (-1)^{nk} \omega T(\eta)$ for any $\eta \in \Omega^{m} \otimes_\algA \modM$ and any $\omega \in \Omega^n$). 

Let $\modM'$ be a left module such that $\modM \oplus \modM' = \algA^N$ and denote by $p : \algA^N \rightarrow \modM$ the projection and $\phi : \modM \rightarrow \algA^N$ the inclusion. Then $p\phi = \Id_\modM$.

Using right multiplication on $\widetilde{\algA^N}$, one can make the identification $\End^\grast_\Omega(\widetilde{\algA^N}) = M_N(\Omega^\grast)$.

\begin{Proposition}[General properties of noncommutative connections]
Any noncommutative connection $\nabla$ can be extended into a map of degree $+1$ on $\widetilde{\modM}$ such that for any $m \in \modM$ and $\omega \in \Omega^n$,
\begin{equation*}
\nabla(\omega \otimes_\algA m) = (d \omega) \otimes_\algA m + (-1)^n \omega (\nabla m)
\end{equation*}
The space of connections is an affine space over $\End^{1}_\Omega(\widetilde{\modM})$.
\end{Proposition}

\begin{Definition}[Curvature of a noncommutative connection]
The curvature of $\nabla$ is the map $\Theta = \nabla^2 = \nabla \circ \nabla$.
\end{Definition}

We define the linear map $\delta : \End^\grast_\Omega(\widetilde{\modM}) \rightarrow \End^\grast_\Omega(\widetilde{\modM})$ by the relation $\delta(T) = \nabla T - (-1)^{k} T \nabla$, where $T \in \End^k_\Omega(\widetilde{\modM})$. One can easily show that $\delta$ is a graded derivation of degree $+1$ on the graded algebra $\End^\grast_\Omega(\widetilde{\modM})$. 

\begin{Proposition}
One has $\Theta \in \End^{2}_\Omega(\widetilde{\modM})$, $\delta(T) = \Theta T - T \Theta = [ \Theta, T]_\gr$ and the Bianchi identity $\delta(\Theta) = 0$.
\end{Proposition}

\begin{Example}[Existence of connections]
\label{ex-existenceofconnections}
Let $e \in \End_\algA(\algA^N)$ be a projector, and define the left module $\modM = e (\algA^N)$. $e$ is also a projector in $\End^0_\Omega(\Omega^\grast \otimes_\algA \algA^N)$ which is naturally extended by $\Omega^\grast$-linearity. Then, if $\nabla$ is a connection on $\algA^N$, the map $m \mapsto e (\nabla m)$ is a connection on $\modM$.

On $\algA^N$ there is a natural connection given by the differential map of the differential calculus: $\algA^N \ni (a_1, \dots, a_N) \mapsto (d a_1, \dots, da_n) \in \Omega^1 \otimes_\algA \algA^N = (\Omega^1)^N$. Then any finite projective module on $\algA$ admits at least one connection.
\end{Example}

\begin{Example}[Direct sum of connections]
Let $(\modM, \nabla^\modM)$ and $(\modN, \nabla^\modN)$ be two finite projective modules over $\algA$ for the same differential calculus. Then $\nabla : \modM \oplus \modN \rightarrow \Omega^1 \otimes_\algA (\modM \oplus \modN)$ defined by $\nabla(m \oplus n) = (\nabla^\modM m) \oplus (\nabla^\modN n)$ is a connection on $\modM \oplus \modN$ which we denote by $\nabla^\modM \oplus \nabla^\modN$.
\end{Example}

\begin{Definition}[Graded trace]
Let $\evV^\grast$ be a graded vector space. A graded trace on $\Omega^\grast$ with values in $\evV^\grast$ is a linear morphism of degree $0$, $\tau : \Omega^\grast \rightarrow \evV^\grast$, such that $\tau(\omega \eta) = (-1)^{m n}\tau(\eta \omega)$ for any $\omega \in \Omega^m$ and $\eta \in \Omega^n$.
\end{Definition}

Notice that the restriction $\tau : \algA = \Omega^0 \rightarrow \evV^0$ is an ordinary trace on $\algA$.

\begin{Proposition}[The universal trace]
If we denote by $[\Omega^\grast, \Omega^\grast]_\gr$ the subspace of $\Omega^\grast$ lineary generated by the graded commutators, then the graded vector space $\widehat{\Omega}^\grast = \Omega^\grast/[\Omega^\grast, \Omega^\grast]_\gr$ inherits the differential of $\Omega^\grast$, which we denote by $\widehat{d}$, and the projection $\tau_\Omega : \Omega^\grast \rightarrow \widehat{\Omega}^\grast$ is a graded trace which commutes which the differentials.

For any graded trace $\tau : \Omega^\grast \rightarrow \evV^\grast$ there is a factorisation $\tau = \overline{\tau} \tau_\Omega$ for a morphism $\overline{\tau} : \widehat{\Omega}^\grast \rightarrow \evV^\grast$. This is why $\tau_\Omega$ is called the universal trace on $\Omega^\grast$.
\end{Proposition}

\begin{Example}[The trace on $\End^\grast_\Omega(\widetilde{\modM})$]
Because of the identification $\End^\grast_\Omega(\widetilde{\algA^N}) = M_N(\Omega^\grast)$, there is a natural trace on $\End^\grast_\Omega(\widetilde{\algA^N})$ with values in $\Omega^\grast$ induced by the trace on the matrix algebra $M_N(\gC)$, which we denote by $\tr$. For any $T \in \End^\grast_\Omega(\widetilde{\modM})$, one has $\widehat{T} = \phi T p \in \End^\grast_\Omega(\widetilde{\algA^N})$, so that we can define $\tau_\Omega(\tr(\widehat{T})) \in \widehat{\Omega}^\grast$. This map is a graded trace which does not depend upon $p$, $\phi$ and $N$. The trace $\End^\grast_\Omega(\widetilde{\modM}) \rightarrow \widehat{\Omega}^\grast$ will be denoted by $\tr_\Omega$. It satisfies $\tr_\Omega \delta = \widehat{d} \tr_\Omega$.
\end{Example}

\begin{Definition}[Characteristic classes of $\modM$]
For any integer $k$, the cohomology class of $\tr_\Omega(\Theta^k)$ in $H^{2k}(\widehat{\Omega}^\grast, \widehat{d})$ is independent of the connection $\nabla$. This is the $k$-th characteristic class of $\modM$ for the differential calculus $(\Omega^\grast, d)$.
\end{Definition}

\begin{Definition}[The Chern character of $\modM$]
We define the Chern character $\ch(\modM) \in H^{\text{even}}(\widehat{\Omega}^\grast, \widehat{d})$ associated to $\modM$ by
\begin{align*}
\ch_k(\modM) &= \left[ \frac{(-1)^k}{(2i\pi)^k k!} \tr_\Omega(\Theta^k) \right] \in H^{2k}(\widehat{\Omega}^\grast, \widehat{d}) \\
\ch(\modM) &= \sum_{k \geq 0} \ch_k(\modM) 
= \tr_\Omega \circ \exp\left(\frac{i \Theta}{2\pi}\right)
\in H^{\text{even}}(\widehat{\Omega}^\grast, \widehat{d}) 
\end{align*}

\end{Definition}

Obviously, this definition is just an algebraic rephrasing of the expression given in Example~\ref{ex-theinvariantpolynomialoftheCherncharacter}.

\begin{Example}[The connection induced by a projector]
\label{ex-thecaseoftheconnectioninducedbyaprojector}
We saw in Example~\ref{ex-existenceofconnections} that there is a natural connection on $\modM = e (\algA^N)$ expressed in terms of the projector $e \in \End_\algA(\algA^N)$. From now on, $e$ will be identified with an element in the matrix algebra $M_N(\Omega^\grast)$, which acts on $\algA^N$ by multiplication on the right. One can compute explicitly the curvature of this connection using this matrix algebra, and then one finds, for any $a \in \modM \subset \algA^N$,
\begin{equation*}
\Theta(a) =  - a (d e) (d e)  e
\end{equation*}
This expression can be used to express the Chern character of $\modM$ in terms of the matrix $e$ :
\begin{equation*}
\ch(\modM) = \sum_{k \geq 0} \left[ \frac{1}{(2i\pi)^k k!} \tau_\Omega \tr(e (de)^{2k}) \right]
\end{equation*}

\end{Example}

\begin{Proposition}[Additive properties of $\ch$]
For any two finite projective left modules $\modM$ and $\modN$ on $\algA$, one has
\begin{equation*}
\ch(\modM \oplus \modN) = \ch(\modM) + \ch(\modN)
\end{equation*}
\end{Proposition}

\begin{Remark}[No product property!]
There is no product property which could satisfy this Chern character, because there is no possibility to define a tensor product of two finite projective left modules $\modM$ and $\modN$\dots
\end{Remark}

\begin{Example}[The geometric Chern Character]
In the case $\algA = C^\infty(M)$ and $\Omega^\grast = \Omega^\grast(M)$, the de~Rham differential calculus, one has $\widehat{\Omega}^\grast = \Omega^\grast$ because $\Omega^\grast(M)$ is graded commutative. Then the Chern character takes its values in the even de~Rham cohomology of $M$. By the Serre-Swan theorem in its algebraic version, Theorem~\ref{thm-serreswanalgebraicversion}, any finite projective module on $C^\infty(M)$ is the space of smooth sections of a vector bundle $E$ over $M$. It is easy to verify that a noncommutative connection is then an ordinary connection on $E$, seen as a covariant derivative maps on sections. This identification uses the natural isomorphism $\Omega^\grast \otimes_\algA \modM = \Omega^\grast(M, E)$, where $\Omega^\grast(M, E)$ is the space of differential forms on $M$ with values in $E$. The curvature is then an element in $\Omega^2(M, \End(E)) = \Omega^2(M) \otimes_{C^\infty(M)} \End_{C^\infty(M)}(\Gamma(E))$, the space of $2$-forms with values in the associated vector bundle $\End(E) = E \otimes E^\ast$. In this context, one has $\End^\grast_\Omega(\Omega^\grast \otimes_\algA \modM) = \Omega^\grast(M, \End(E))$ and the trace is the ordinary trace on the fibres of $\End(E)$.

Using these considerations and the explicit formulas defining them, the two definitions of the Chern characters coincide.

As an exercise, one can show that the relation $\delta(\Theta) = 0$ is indeed the Bianchi identity!
\end{Example}

\subsection{The Chern character from algebraic $K$-theory to periodic cyclic homology}

The definition of the (algebraic) Chern character we will use rests upon the two results concerning the algebras $\gC$ and $\gC[z,z^{-1}]$ given in Examples~\ref{ex-periodiccyclicC} and \ref{ex-thelaurentpolynomials}: $HP_0(\gC) = \gC$ and $HP_1(\gC[z,z^{-1}]) = \gC$. Recall that the trace map $\tr$ defined in Definition~\ref{def-tracemap} induces the Morita isomorphism in periodic cyclic homology.

Let $\algA$ be an associative unital algebra. Let $p \in M_N(\algA)$ be a projector. Then it defines a morphism of algebras $i_p : \gC \rightarrow M_N(\algA)$ by $\lambda \mapsto \lambda p$. Indeed, $1 \in \gC$ is mapped to $p$, and the relation $p^2 = p$ is the required compatibility with $1^2 = 1$. This morphism is not a morphism of unital algebras. 

Let $u \in M_N(\algA)$ be an invertible element. Then it defines a morphism of algebras $i_u : \gC[z,z^{-1}] \rightarrow M_N(\algA)$ completely given by $z \mapsto u$ and $1 \mapsto \bbbone_N$.

\begin{Definition}[The algebraic Chern character]
With the previous notations, the Chern character $[\ch_0(p)] \in HP_0(\algA)$ of the projector $p$ is the image of $1 \in HP_0(\gC)$ in the composite map
\begin{equation*}
HP_0(\gC) \xrightarrow{{i_p}_\indast} HP_0(M_N(\algA)) \xrightarrow{\tr_\indast} HP_0(\algA)
\end{equation*}

The Chern character $[\ch_1(u)] \in HP_1(\algA)$ of the invertible $u$ is the image of $1 \in HP_1(\gC[z,z^{-1}])$ in the composite map
\begin{equation*}
HP_1(\gC[z,z^{-1}]) \xrightarrow{{i_u}_\indast} HP_1(M_N(\algA)) \xrightarrow{\tr_\indast} HP_1(\algA)
\end{equation*}
\end{Definition}

\begin{Proposition}[The Chern character on algebraic $K$-theory]
The class $[\ch_0(p)] \in HP_0(\algA)$ (resp. $[\ch_1(u)] \in HP_1(\algA)$) depends only on the class of $p$ in $K^\text{alg}_0(\algA)$ (resp. on the class of $u$ in $K^\text{alg}_1(\algA)$).

The Chern character is a map $\ch : K^\text{alg}_\nu(\algA) \rightarrow HP_\nu(\algA)$ for $\nu= 0,1$.
\end{Proposition}

\begin{Example}[Explicit formula for $\ch_0(p)$ in $\Omega^\grast(\algA)$]
\label{ex-explicitformulach0}
In order to give an explicit formula for the representative $\ch_0(p)$ of the class of the Chern character in the mixed complex $(\Omega^\grast(\algA), b_H, B)$, one has to explicitly write down the generator $1 \in HP_0(\gC)=\gC$. It is convenient to do that in the same mixed bicomplex $(\Omega^\grast(\gC), b_H, B)$. In order to make notations clear, let us denote by $e \in \gC$ the unit element. Then one can show, using explicit formulas on $b_H$ and $B$, that
\begin{equation*}
e + \sum_{n \geq 1} (-1)^n \frac{(2n)!}{n!} \left(e - \frac{1}{2}\right) (de)^{2n}
\end{equation*}
is the generator of the class $1$, in the total complex of the mixed complex $(\Omega^\grast(\gC), b_H, B)$.

Using the composite map at the level of mixed bicomplexes ($i_p$ and $\tr$), one gets
\begin{equation*}
\ch_0(p) = \tr(p) + \sum_{n \geq 1} (-1)^n \frac{(2n)!}{n!} \tr\left( \left(p - \frac{1}{2}\right) (dp)^{2n} \right)
\end{equation*}
\end{Example}

\begin{Example}[Explicit formula for $\ch_1(u)$ in $\Omega_U^\grast(\algA)$]
\label{ex-explicitformulach1}
In the mixed bicomplex $(\Omega_U^\grast(\algA), b_H, B)$, we can give an explicit formula for the representative $\ch_1(u)$  using the following expression of the representative of $1 \in HP_1(\gC[z,z^{-1}]) = \gC$:
\begin{equation*}
\sum_{n \geq 0} n! z^{-1} dz (dz^{-1} dz)^{n}
\end{equation*}
Then the element $\ch_1(u)$ takes the form
\begin{equation*}
\ch_1(u) = \sum_{n \geq 0} n! \tr\left( u^{-1} du (du^{-1} du)^{n} \right)
\end{equation*}
\end{Example}

\begin{Remark}[What is really a representative of the Chern character?]
The Chern character is well defined only in the periodic cyclic homology of the algebra. But it is convenient to manipulate it as a cycle in the complex computing this homology. 

But which complex to consider? Indeed, as we saw before, there are many possibilities, at least as many mixed bicomplexes that are $b$-quasi-isomorphic (Proposition~\ref{prop-bquasiisomorphismHPHCneg}). So that one can expect some representative cycles in the complexes $CC_{\grast, \grast}^\text{per}(\algA)$, $(\Omega^\grast(\algA), b, B)$, $(\Omega_U^\grast(\algA), b, B)$, and even $(\Omega^{\grast}_{\algA|\gC}, 0, d_K)$ if the algebra is a smooth commutative algebra\dots

The representatives given in Examples~\ref{ex-explicitformulach0} and \ref{ex-explicitformulach1} are then only particular expressions. For instance, for the algebra $\algA = \symes \evV$, one can use the Kähler differential calculus, in which any element of degree $\geq \dim \evV$ is $0$. In that case, the Chern character is represented by a finite sum of differential forms of odd or even degrees. 

The expressions we gave above have the advantage that they are written in the universal differential calculi, in which all the degrees can be represented. Let us give another expression for the generator $1 \in HP_1(\gC[z,z^{-1}])$ in the bicomplex $CC_{\grast, \grast}^\text{per}(\gC[z,z^{-1}])$. In order to do that, define the family of elements
\begin{align*}
\alpha_n &= (n+1)! (z^{-1} - 1) \otimes (z -1) \otimes [(z^{-1} - 1) \otimes (z -1)]^{\otimes 2n} &\in \gC[z,z^{-1}]^{\otimes 2n+2}\\
\beta_n &= (n+1)! (z -1) \otimes [(z^{-1} - 1) \otimes (z -1)]^{\otimes 2n} &\in \gC[z,z^{-1}]^{\otimes 2n+1}
\end{align*}
then the representative of the generator is 
\begin{equation*}
c = \sum_{n\geq 0} \alpha_n \oplus \beta_n \in TCC_{1}^\text{per}(\gC[z,z^{-1}])
\end{equation*}
Using the identification $\Omega^{2n+1}(\gC[z,z^{-1}]) = \gC[z,z^{-1}]^{\otimes 2n+2} \oplus \gC[z,z^{-1}]^{\otimes 2n+1}$, this generator is also directly written as a generator in the mixed bicomplex $(\Omega^\grast(\gC[z,z^{-1}]), b, B)$.

Finally, notice that the explicit development of the Chern character in one of the complexes mentioned above is completely determined by the lowest degrees, in which a normalisation is imposed, and the condition to be a cycle in the periodic complex. Hence this object is a very canonical one.
\end{Remark}

\begin{Proposition}[Naturality of the Chern character]
For any short exact sequence of associative algebras $\xymatrix@1{{\algzero} \ar[r] & {\algI} \ar[r] & {\algA} \ar[r] & {\algA/\algI} \ar[r] & {\algzero}}$, one has the commutative diagram
\begin{equation}
\label{eq-caracteredechernentresuitesexacteslonguesktheorieethomologiecycliqueperiodique}
\xymatrix{ 
{K^\text{alg}_1(\algI)} \ar[d]^-{\ch} \ar[r] & {K^\text{alg}_1(\algA)} \ar[d]^-{\ch} \ar[r] & {K^\text{alg}_1(\algA/\algI)} \ar[d]^-{\ch} \ar[r]^-{\delta} & {K^\text{alg}_0(\algI)} \ar[d]^-{\ch} \ar[r] & {K^\text{alg}_0(\algA)} \ar[d]^-{\ch} \ar[r] & {K^\text{alg}_0(\algA/\algI)} \ar[d]^-{\ch} \\
{HP_1(\algI)} \ar[r] & {HP_1(\algA)} \ar[r] & {HP_1(\algA/\algI)} \ar[r]^-{\delta} & {HP_0(\algI)} \ar[r] & {HP_0(\algA)} \ar[r] & {HP_0(\algA/\algI)}
}
\end{equation}

\end{Proposition}

\begin{Remark}[The topological case]
When the algebra is a topological algebra, one can show that the Chern character is in fact a map from the $K$-groups defined on topological algebras and the continuous cyclic periodic homology. Indeed, one can show that it is homotopic invariant. 

Nevertheless, remember that in Remark~\ref{rmk-cyclichomologyandKtheory} we mentioned that $K$-theory is well adapted to $C^\ast$-algebras and continuous functional calculus in general, but that cyclic periodic homology is only useful for topological algebras underlying some smooth structures\dots

If one wants to connect $K$-theory and cyclic periodic homology directly at the level of representative cycles, one has to consider some intermediate algebras between ``algebraic'' and ``$C^\ast$'', for instance Fréchet or locally convex algebras. In these cases, unfortunately, the $K$-groups are not defined using projections and unitaries, so that the interpretation of the Chern character is not at all transparent whereas it looks so clear in the algebraic version\dots 

When the Bott periodicity takes place in $K$-theory, the commutative diagram \eqref{eq-caracteredechernentresuitesexacteslonguesktheorieethomologiecycliqueperiodique} connects in reality the two six term exact sequences of Propositions~\ref{prop-sixtermsexactsequencesK} and \ref{prop-sixtermsexactsequencesHP}. But there is a defect in this closed relation, a factor $2 \pi i$ is necessary in the morphism $\delta : K_0(\algA/\algI) \rightarrow K_{1}(\algI)$ to get a commutative diagram (see~\cite{CuntSkanTsyg:04}). 

\end{Remark}

\begin{Remark}[The Chern character as an isomorphism]
In Theorem~\ref{thm-thecherncharacterasanisomorphism}, we saw that the Chern character realizes an isomorphism between $K$-theory of topological spaces (in fact its torsion-free part) and the ordinary cohomology of the underlying topological space.

In \cite{Soll:05}, it is shown that the Chern character for topological algebras realizes an equivalent isomorphism for a large class of Fréchet algebras in the following form
\begin{equation*}
\ch \otimes \Id : K_\grast(\algA) \otimes \gC \xrightarrow{\simeq} HP_\grast(\algA)
\end{equation*}
In particular, the Fréchet algebras $C^\infty(M)$ for a locally compact manifold $M$ is in this class.
\end{Remark}

\begin{Remark}[Chern character and cyclic cohomology]
For a Banach algebra $\algA$, the Chern character can be realized as a pairing $K_\nu(\algA) \times HP^\nu(\algA) \rightarrow \gC$, using the natural pairing between periodic cyclic cohomology and periodic cyclic homology.

Let us consider the case $\nu=0$. In Example~\ref{ex-firstversionofcycliccohomology}, we defined cyclic cohomology using the Connes complex. Let $\phi \in C^{2n}_\lambda(\algA)$ be a cyclic cocycle and $p \in M_N(\algA)$ a projector. Define
\begin{equation*}
\langle [p], [\phi] \rangle = \frac{1}{(2 i \pi)^n n!} \sum_{i_1, \dots, i_N} \phi( p_{i_1 i_2}, p_{i_2 i_3}, \dots , p_{i_N i_1})
\end{equation*}
One can show that this pairing is well defined at the level of the $K_0$ group and at the level of $HC^{2n}(\algA)$, and that it satisfies $\langle [p], S[\phi] \rangle = \langle [p], [\phi] \rangle$. Because the periodic cyclic cohomology group $HP^{0}(\algA)$ can be defined as an inductive limit through the periodic operator $S$ on the $HC^{2n}(\algA)$ spaces, the previous pairing is indeed a pairing between $K_0(\algA)$ and $HP^0(\algA)$.

Using this construction, no extra structure is required. One then recovers that the Chern character is indeed a canonical object in the context of $K$-theory and periodic cyclic (co)homology. There is a similar expression for $\nu=1$.
\end{Remark}

\begin{Remark}[Comparing Chern characters]
The expressions in Examples~\ref{ex-theinvariantpolynomialoftheCherncharacter}, \ref{ex-thecherncharacterK1(m)rightarrowHodd(MQ)}, \ref{ex-explicitformulach0} and \ref{ex-explicitformulach1} look very similar. But there are differences which are important to be noted. In order to make them clear, we will call ``geometric Chern character''  the expressions given in Examples~\ref{ex-theinvariantpolynomialoftheCherncharacter}, \ref{ex-thecherncharacterK1(m)rightarrowHodd(MQ)} (and also \ref{ex-thecaseoftheconnectioninducedbyaprojector}), and ``algebraic Chern character'' the expressions given in Examples~\ref{ex-explicitformulach0} and \ref{ex-explicitformulach1}.

First, the spaces on which these Chern characters are expressed as ``differential forms'' are not the same in the two situations. In the geometric one, it is the de~Rham differential calculus. In the algebraic one, it is one of the universal differential calculi. 

In order to compare them, one has to take into account a situation in which they both make sense, the case of the algebra $\algA = C^\infty(M)$ for instance. In that case, one knows that the identification of the Hochschild homology with de~Rham forms can be expressed as in Example~\ref{ex-thefrechetalgebraCinfty(M)}. Using these expressions, one easily show that the following squares are commutative, where the vertical isomorphisms concerning $K$-theories express the Serre-Swan Theorem~\ref{thm-serreswanalgebraicversion},
\begin{equation*}
\xymatrix{ 
{K_{0}(C^\infty(M))} \ar[d]^-{\simeq} \ar[r]^-{\ch_{\text{alg}}} & {HP_0^{\text{cont}}(C^\infty(M))} \ar[d]^-{\phi}_{\simeq} \\
{K^{0}(M)} \ar[r]^-{\ch_{\text{geom}}} & {H^{\text{even}}(M; \gQ)} 
}
\qquad \qquad
\xymatrix{ 
{K_{1}(C^\infty(M))} \ar[d]^-{\simeq} \ar[r]^-{\ch_{\text{alg}}} & {HP_1^{\text{cont}}(C^\infty(M))} \ar[d]^-{\phi}_{\simeq} \\
{K^{-1}(M)} \ar[r]^-{\ch_{\text{geom}}} & {H^{\text{odd}}(M; \gQ)} 
}
\end{equation*}
This explains the extra factors used in the isomorphism $\phi$ in Example~\ref{ex-thefrechetalgebraCinfty(M)}. Notice that the two definitions of the Chern characters are constrained: the geometric case is normalised in such a way that it is a ring morphism, the algebraic one is expressed as an infinite cycle in cyclic periodic homology, so that all the terms are normalised by the first one. The only degree of freedom in this square is the isomorphism of vector spaces $\phi$ (and fortunately not an isomorphism of algebras since $HP_\nu(\algA)$ has no natural structure of algebra). 
\end{Remark}

\section{Conclusion}

There cannot be any conclusion to a subject that is still full of vivacity! Thousands of mathematicians try everyday to conquest some new landmarks in this extraordinary vast and rich world. In this lecture, only some selected aspects of this theory have been presented. For instance, no mention has been made about ``noncommutative measure theory'', in which von~Neumann algebras play the role of $C^\ast$-algebras for measurable spaces.

We have seen that one can reasonably manipulate ``noncommutative topological spaces'' using the $K$-theory of $C^\ast$-algebras. One can convince oneself that differentiable structures are available in the heart of cyclic homology.

Nevertheless this research project is facing a challenge which have not yet been solved: what is the noncommutative counterpart of smooth functions? Does it exist? We have made it clear that cyclic homology sees some smooth structures, but the right category of ``noncommutative smooth algebras'' has not yet been identified. Some paths have been investigated. For instance, Cuntz has considered $m$-algebras (see \cite{Frag:05}), some kind of locally convex algebras, on which he succeed to enrich $K$-theory and cyclic homology (see \cite{CuntSkanTsyg:04}). But what is still missing is a Gelfand-Neumark theorem for smooth functions. 

\nocite{Davi:96}
\bibliographystyle{alpha}
\bibliography{biblio-articles-perso,biblio-livre,biblio-articles}

\end{document}